\def\txi#1{\widetilde{\bxi}\phantom{x}\hspace{-1.2ex}^{#1}}
\def\phan{\left.\phantom{|^|_|}\!\!\!}
\def\begB{\setcounter{qq}{1}\renewcommand{\theequation}{\arabic{equation}\alph{qq}}\begin{equation}}
\def\endB#1#2{\label{#1}\end{equation}\vspace*{-\baselineskip}\phantom{ }\newcounter{#2}\setcounter{#2}{\value{equation}}}
\def\begI{\addtocounter{equation}{-1}\addtocounter{qq}{1}\begin{equation}}
\def\endI#1{\label{#1}\end{equation}}
\def\endL#1{\label{#1}\end{equation}\vspace*{-\baselineskip}\renewcommand{\theequation}{\arabic{equation}}\break}
\def\al{ {\it et al}.}
\def\eqn#1{\eqno({\rm #1})}
\def\mi{\vspace*{2mm}\noindent}
\def\lbd{\{\hskip-4pt\{}
\def\rbd{\}\hskip-4pt\}}
\def\ladb{\left\langle\phantom{|^|_|}\hskip-10pt\right\langle}
\def\radb{\left\rangle\phantom{|^|_|}\hskip-10pt\right\rangle}
\def\lbdb{\left\{\phantom{|^|_|}\hskip-11pt\right\{}
\def\rbdb{\left\}\phantom{|^|_|}\hskip-11pt\right\}}
\def\lpar{\left(\phantom{|^|_|}\hskip-9pt\right.}
\def\rpar{\left.\phantom{|^|_|}\hskip-9pt\right)}
\def\lad{\langle\hskip-3pt\langle}
\def\rad{\rangle\hskip-3pt\rangle}
\def\lb{\{}
\def\rb{\}}
\def\lsb{[}
\def\rsb{]}
\def\la{\langle}
\def\ra{\rangle}
\def\btau{\mbox{\boldmath$\tau$\unboldmath}}
\def\bxi{\mbox{\boldmath$\xi$\unboldmath}}
\def\bal{\mbox{\boldmath$\alpha$\unboldmath}}
\def\bgamma{\mbox{\boldmath$\gamma$\unboldmath}}
\def\bphi{\mbox{\boldmath$\varphi$\unboldmath}}
\def\bzeta{\mbox{\boldmath$\zeta$\unboldmath}}
\def\bOm{{\bf\Omega}}
\def\bom{\mbox{\boldmath$\omega$\unboldmath}}
\def\Alfa{\mbox{\boldmath$\cal A$\unboldmath}}
\def\rf#1{(\ref{#1})}
\def\xrf#1#2{(\ref{#1})--(\ref{#2})}
\documentstyle[12pt]{article}
\begin{document}
\newcounter{qq}
\oddsidemargin 5mm
{\center

\mi{\bf MEAN-FIELD EQUATIONS\\
FOR WEAKLY NONLINEAR TWO-SCALE PERTURBATIONS\\
OF FORCED HYDROMAGNETIC CONVECTION\\
IN A ROTATING LAYER\\

\mi
Vladislav Zheligovsky}

\mi
International Institute of Earthquake Prediction Theory\\
and Mathematical Geophysics\\
84/32 Profsoyuznaya St., 117997 Moscow, Russian Federation\\

\mi
Observatoire de la C\^ote d'Azur, CNRS\\
U.M.R. 6529, BP 4229, 06304 Nice Cedex 4, France\\

\mi
{\it Accepted in Geophysical Astrophysical Fluid Dynamics}

}

\medskip We consider stability of regimes of hydromagnetic thermal convection
in a rotating horizontal layer with free electrically conducting boundaries, to
perturbations involving large spatial and temporal scales. Equations governing
the evolution of weakly nonlinear mean perturbations are derived under the
assumption that the $\alpha-$effect is insignificant in the leading order
(e.g., due to a symmetry of the system). The mean-field equations generalise
the standard
equations of hydromagnetic convection: New terms emerge -- a second-order
linear operator representing the combined eddy diffusivity, and quadratic terms
associated with the eddy advection. If the perturbed CHM regime is non-steady
and insignificance of the $\alpha-$effect in the system does not rely on the
presence of a spatial symmetry, the combined eddy diffusivity operator also
involves a non-local pseudodifferential operator. If the perturbed CHM state
is almost symmetric, $\alpha-$effect terms appear in the mean-field equations
as well. Near a point of a symmetry-breaking bifurcation, cubic nonlinearity
emerges in the equations. All the new terms are in general anisotropic.
A method for evaluation of their coefficients is presented; it requires
solution of a significantly smaller number of auxiliary problems than in
a straightforward approach.

\mi{\bf 1. Introduction}

\mi
An attribute of convection in a melted planetary core is the presence of
interacting structures involving different spatial scales. One example of such
a structure is the Ekman layer in rotating convective flows, whose instability
can result in magnetic field generation (see Ponty\al, 2001a,b, 2003; Rotvig and
Jones, 2002). Core-mantle coupling, which is supposed to be responsible for the
length of day variation at time scales of decades, is another one: e.g.,
topographic coupling is due to topographic structures at the core-mantle
boundary, which do not exceed 5 km in size (Bowin, 1986;
Merrill\al, 1996); this is small compared to the radius of the liquid core
(see also a discussion of the influence of irregularities of the boundary of
the Earth's liquid core on the flow in it and on the Earth's magnetic field
in Anufriev and Braginsky, 1975, 1977a,b). While
in these two examples the small-scale features are located at the boundaries of
the liquid core, scale separation can occur in the entire volume, where
convection takes place. For instance, in geostrophic flows in rapidly rotating
spherical or cylindrical shells fluid moves in the so-called Taylor columns,
parallel to the axis of rotation; the columns have a smaller width than
the size of the container of the fluid. Narrow (in the horizontal direction)
cells emerge in thermal convection of fluids rapidly rotating about
the vertical axis (Bassom and Zhang, 1994; Julien and Knobloch, 1997,
1998, 1999; Julien\al, 1998) and in magnetoconvection in the
presence of strong magnetic fields (in the limit of large Chandrasekhar
numbers; see Matthews, 1999; Julien\al, 1999, 2000, 2003). A hierarchy of scales
and related phenomena, such as intermittency and direct and inverse energy
cascades, are characteristic features of turbulence (Frisch, 1995).

This suggests that analytical asymptotic methods can be used in combination with
numerical ones to study various convective and MHD regimes. This approach was
followed, in particular, in the well-known geodynamo studies by Braginsky
(1964a-d,\break 1967, 1975) and Soward (1972, 1974), who constructed asymptotic
expansions of solutions to dynamo problems to determine the values of the
magnetic $\alpha$-effect coefficients.

In multiscale MHD stability problems it is typically assumed that the
characteristic spatial and temporal scales of a perturbation are much larger
than the characteristic scales of the perturbed regime. We assume that
the perturbation depends on the so-called fast spatial $\bf x$ and temporal $t$
variables, and on the slow ones\footnote{The width $L$ of the layer is fixed,
and hence no slow spatial variable in the vertical direction is introduced.}:
${\bf X}=\varepsilon(x_1,x_2)$, $T=\varepsilon^2t$. The small parameter
$\varepsilon$ is the spatial scale ratio. In what follows, vector
fields which depend only on small-scale (fast) variables will be called
small-scale fields, and those depending as well on large-scale (slow) variables
will be called large-scale fields. In this terminology, we are concerned with
stability of small-scale regimes to large-scale perturbations.

The perturbation is expanded in power series in the small parameter
$\varepsilon$. General methods of the theory of homogenisation for equations in
partial derivatives for multiscale systems (see, e.g., the monographs
Bensoussan\al, 1978; Oleinik\al, 1992; Jikov\al, 1994; Cioranescu and Donato,
1999) are then applied to derive rigorously a closed system of equations for large-scale
structures in the perturbation, averaged over small scales. The influence of
the small-scale dynamics is represented by new terms (often called eddy
corrections) emerging in the equations. Computation of coefficients in these
terms requires numerical solution of systems of linear differential equations
in partial derivatives -- the so-called auxiliary problems. The multiscale
asymptotic techniques offer a tool to split analytically the large- and
small-scale dynamics in the complete problem, if the small-scale dynamics is in
a certain sense homogeneous. Consequently, a full resolution is not required for
the solution of the auxiliary problems, which is a significant advantage of the
combined approach. The techniques were applied to specific two- (Sivashinsky and Yakhot,
1985; Sivashinsky and Frenkel, 1992; Novikov and Papanicolaou, 2001; Novikov, 2004)
and general and specific three-dimensional equations of hydrodynamics (Dubrulle
and Frisch, 1991; Wirth, 1994, Wirth\al, 1995), passive scalar transport
(Biferale\al, 1995; Vergassola and Avellaneda, 1997; Majda and Kramer, 1999)
and kinematic magnetic dynamos (Lanotte\al, 1999; Zheligovsky\al, 2001;
Zheligovsky and Podvigina, 2003; Zheligovsky, 2005). The effect of eddy viscosity
was observed in direct numerical investigation (Murakami\al, 1995) of stability
of planar flows to large-scale three-dimensional perturbations (without a prior
recourse to asymptotic analysis).

Nepomnyashchy (1976) considered eddy viscosity in the Kolmogorov flow and found
that near the bifurcation where eddy viscosity passes through zero, the regime
satisfies an equation with a cubic nonlinearity of the Cahn-Hilliard type,
studied by She (1987). E and Shu (1993) studied numerically inverse
cascade in Kolmogorov flow, using homogenised nonlinear ``effective'' equations.
Mean-field equations for weakly nonlinear perturbations of two-dimensional
hydrodynamic steady flows were considered in the terms of the stream function by
Gama\al~(1994) and Frisch\al~(1996) (in the latter paper an additional term
describing the $\beta$-effect was introduced into the Navier-Stokes equation to
take into account rotation of fluid, and the case of marginally negative eddy
viscosity was assumed). Newell (1983) and Cross and Newell (1984)
investigated equations for perturbations, analogous by the method of derivation
to the mean-field equations under consideration; however, their derivation
was carried out for model equations describing approximately convective flows
in a layer in the form of deformed rolls. Newell\al~(1990a,b, 1993, 1996) and
Ponty\al~(1997) explored weakly nonlinear dynamics of convective patterns and
the developing defects in a system of rolls. They considered a complete system
of equations of the Boussinesq convection in a layer with rigid boundaries,
employing the variables the amplitude and the phase; the mean-field equation
for the slow phase was derived. (No magnetic field was considered in the papers
referred to in this paragraph.)

Using the multiscale asymptotic techniques, Zheligovsky studied linear
(Zheligovsky, 2003) and weakly nonlinear (Zheligovsky, 2006a) stability of
parity-invariant three-dimensional space-periodic MHD regimes, and
Baptista\al~(2007) considered linear stability of convective
hydromagnetic (CHM) steady states in a layer, symmetric about a vertical axis.
As originally found by Dubrulle and Frisch (1991), in this class of multiscale
problems a term manifesting the $\alpha$-effect appears in the mean-field
equations in the leading order, $\varepsilon$; no other terms, linear
or nonlinear, survive averaging at this
stage (whichever kind of stability -- linear or weakly nonlinear -- is
inspected). Thus, one obtains mean-field equations involving other ``eddy''
effects, which can give rise to non-trivial dynamics of mean fields, only if in
the leading order $\varepsilon$ the $\alpha$-effect is insignificant (e.g., if appropriate
components of the $\alpha$-tensor vanish). The symmetries of the perturbed
regimes assumed in the papers mentioned above in this paragraph guarantee that
the $\alpha$-effect vanishes, or it is insignificant. However, the symmetries
are not necessary for this: e.g., the AKA (kinematic $\alpha$)
effect does not emerge in ABC flows (Wirth\al, 1995). Zheligovsky (2006b)
considered weakly nonlinear stability of three-dimensional CHM regimes
in a layer just assuming that the $\alpha$-effect is insignificant.

In the linear stability problems mentioned above, the main terms of the series
expansion of the perturbation modes of steady states and their growth rates
are eigenvectors and eigenvalues, respectively, of the operator of combined
eddy diffusivity. This is a linear partial differential operator of the second
order, which is not, in general, isotropic or negatively defined. If it has
eigenvalues with a positive real part, one says that there is negative eddy
diffusivity (Starr, 1968). In the weakly nonlinear stability problem it is
assumed that the amplitude of the perturbation is of the same order as the
scale ratio, $\varepsilon$, and the evolution of such perturbation due to
unabridged nonlinear system of equations is considered. Then additional
quadratic terms emerge in the mean-field equations for the perturbation, which
are analogous to the advection terms of the original equations.

Lanotte\al~(1999), Zheligovsky\al~(2001), Zheligovsky and Podvigina (2003) and
Zheligovsky (2003, 2005) have shown that the presence of small scales
is beneficial for the action of kinematic magnetic dynamos and for the growth of
magnetohydrodynamic instabilities, and the effect of eddy viscosity can provide
a working mechanism for development of these phenomena. A question immediately
arises: What happens to this mechanism, when the initial stage of
development of an instability is over and nonlinear effects switch on? It is
natural to search for preliminary answers to this question in the context of
weakly nonlinear stability theory.

In the present paper we are concerned with weakly nonlinear stability of
regimes in Rayleigh-B\'enard convection in a layer in the presence of
magnetic field. We consider the so called ``forced'' convection, where external
forces (in addition to the buoyancy and Coriolis forces) are supposed
to act, and/or sources of heat and magnetic field are present
inside the layer (section 2). In their absence the system is translation
invariant in horizontal directions, and as a result the problem becomes more
complex; this CHM system will be considered in a sequel to the present paper.

The leading term in the expansion of perturbations involving large scales is
comprised of amplitude-modulated neutral modes of the operator of linearisation
of the equations, governing the behaviour of the CHM system:
$$({\bf v}_0,{\bf h}_0,\theta_0)=\sum_{k=1}^KC^k{\bf s}^\cdot_k,$$
where ${\bf s}^\cdot_k({\bf x},t)$ are the small-scale neutral modes, and
$C^k({\bf X},T)$ are amplitudes, slowly varying in space and time. Our goal is
to derive the equations, describing the evolution of the amplitudes $C^k$.
The derivation relies on existence of constant
vector fields in the kernel of the operator, adjoint to the operator of
linearisation near the perturbed regime. The homogenised equations are
solvability conditions for equations in fast variables, which by the Fredholm
alternative theorem (see section~4)
amounts to the orthogonality of the right-hand sides of the equations
to the kernel of the adjoint operator, and in our case is equivalent to
averaging of the respective components of the equations. Direct substitution
shows that any combination of a constant flow and magnetic field with a zero
temperature component belongs to the kernel, if they satisfy boundary
conditions. Thus the boundary conditions determine the number of the
homogenised equations, and the auxiliary problems to be solved in order to
evaluate the coefficients of the eddy terms in these equations.
For electrically conducting free boundaries held at fixed temperatures,
considered here, the dimension $K$ of the kernels of the linearisation and
of the adjoint operator is at least 4.

In the problem under consideration, generically $K=4$ and mean horizontal
components of the flow velocity and magnetic field of the neutral
small-scale modes are non-zero. Hence we consider the neutral modes
${\bf s}^\cdot_k=({\bf S}^{v,v}_k+{\bf e}_k,{\bf S}^{v,h}_k,{\bf S}^{v,\theta}_k)$ and
${\bf s}^\cdot_{k+2}=({\bf S}^{h,v}_k,{\bf S}^{h,h}_k+{\bf e}_k,{\bf S}^{h,\theta}_k)$,
$k=1,2$, where horizontal flow velocity and magnetic field components of all
vector fields ${\bf S}_k^{\cdot,\cdot}$ vanish. Let $\lad{\bf f}\rad_h$ denote
the average over fast variables of the horizontal component of a vector field $\bf f$. Clearly,
$$\ladb\sum_{k=1}^2C^k{\bf s}^\cdot_k\radb_h=(\lad{\bf v}_0\rad_h,0,0),\qquad
\ladb\sum_{k=3}^4C^k{\bf s}^\cdot_k\radb_h=(0,\lad{\bf h}_0\rad_h,0),$$
i.e., $C^k$ are the mean horizontal components of perturbations of the flow,
$\lad{\bf v}_0\rad_h$, and magnetic field, $\lad{\bf h}_0\rad_h$. Consequently,
we call ``mean-field'' the homogenised equations for $C^k$, that we construct.
(Mean vertical flow and magnetic field components of perturbation vanish
due to solenoidality and the boundary conditions.)

In general, the $\alpha$-effect
emerges in the system (section 6). Then the mean-field equations turn out to be
linear, resulting in instability of the CHM regime to large-scale
perturbations. It can be, however, insignificant in the leading order, for
instance, due to the presence of certain spatial or spatio-temporal symmetries
in the system (section 7). Assuming it to be insignificant, as in Zheligovsky
(2006b), we derive (sections 8 and 9) the mean-field equations \rf{eq59} and
\rf{eq62} for the evolution of the mean part of a perturbation,
which generalise the standard
equations of hydromagnetic convection. The linear operator of combined eddy
diffusivity and quadratic terms associated with eddy advection emerge in
the mean-field equations, like in the studies cited above. If the perturbed
CHM regime is non-steady and the symmetry responsible for the
$\alpha$-effect insignificance in the leading order is spatio-temporal,
non-local operators representing new physical effects emerge in the mean-field
equations. We show (section 10) that if the perturbed CHM flow is
asymptotically close to a symmetric one, and its antisymmetric part is
of the order of the scale ratio $\varepsilon$ (in particular, this happens
when a branch of CHM regimes emerges in a symmetry breaking Hopf bifurcation),
then $\alpha$-effect terms also appear in the mean-field equations, which
now take the form \rf{eq71} and \rf{eq73}.

Finally, in section 11 we consider a symmetric CHM system close to a point of
a symmetry-breaking pitchfork bifurcation for overcriticality of the order of
$\varepsilon^2$. In this case the kernel of the operator of linearisation is
five-dimensional ($K=5$); in addition to neutral small-scale modes with non-zero
horizontal components of the flow velocity and magnetic field, there exist
a mean-free neutral small-scale mode, whose amplitude (denoted $c_0$ in section
11) is not associated with any mean field. The mean-field equations \rf{eq83}
and \rf{eq86} now constitute a closed system together with the equation
for the evolution of this amplitude (since this equation is bulky, it is moved
to appendix~F). Eddy diffusivity, eddy advection
and $\alpha$-effect terms and non-local operators can still be found
in the mean-field equations. Cubic nonlinearity emerges in the equation for
the amplitude of the neutral mean-free mode, and if the perturbed
CHM regime is non-steady and the symmetry responsible
for the $\alpha$-effect insignificance in the leading order is spatio-temporal,
the equation involves a variety of linear and nonlinear non-local operators.

Pairs of the mean-field equations \rf{eq59}, \rf{eq62}; \rf{eq71}, \rf{eq73};
and \rf{eq83}, \rf{eq86} combined with the amplitude equation presented
in appendix F, arising in the analysis of weakly nonlinear stability of
a single CHM regime and branches of regimes emerging in the Hopf and pitchfork
bifurcations, respectively, constitute the main results of the present paper.
Following Zheligovsky (2006b), we do not assume that the perturbed CHM regime
is periodic in horizontal directions (although this geometry is
convenient for numerical solution of the auxiliary problems). This condition
must be relaxed in order to consider stability of semi-regular structures,
such as spiral-defect chaos in thermal convection (see
Figs.~5b,d,f in Bodenschatz\al, 2000). Also, like Zheligovsky (2006a,b), we
do not assume that the perturbed CHM regime is steady. Consequently, in
derivation of the mean-field equations averaging must be performed over the
entire domain of the fast (small-scale) spatial and temporal variables.
The coefficients in the newly emerging terms are constants, like this is the
case if a steady CHM state is perturbed. (In particular, temporal dependence of
the eddy diffusivity tensor, studied by Gama and Chaves, 2000, is consequential
only on time scales, which are not involved in the derivation of the mean-field
equations.)

In appendix G a method for evaluation of the coefficients is presented, which
requires to solve a significantly smaller number of auxiliary problems than in
the straightforward approach. This method relies on solution of auxiliary
problems for the adjoint operator, like in Zheligovsky (2005, 2006a,b).

Like Zheligovsky (2006b), we consider convection of fluid, rotating about
a vertical axis (which is more astrophysically sound than the absence of
rotation). This gives rise to the following algebraic difficulty. In the
presence of the Coriolis force, the kernel of the adjoint operator includes
constant vector fields with a non-zero constant flow only, if a linear growth
in horizontal directions of the potential of the subtracted gradient is
allowed. However, in this case surface integrals appear from the pressure
gradient when averaging of the equations over fast variables is performed, which
is technically inconvenient. To overcome this difficulty we consider (like
Zheligovsky, 2006b) the Navier-Stokes equation for vorticity. As a result,
the equation for the perturbation of the flow emerges as the solvability
condition for equations in fast variables at the order of $\varepsilon^3$, and
not $\varepsilon^2$, as usual. Moreover, solution of equations at the order of
$\varepsilon^n$ is preceded by derivation of the spatial mean
$\la{\bf v}_n\ra_h$ from the spatial mean of the vorticity equation
at the order of $\varepsilon^{n+1}$.

\mi{\bf 2. Equations of hydromagnetic thermal convection, boundary conditions

and the linearisation operator}

\mi
A CHM regime ${\bf V,H},\cal T$, the weakly nonlinear stability of which we are
considering, satisfies in the Boussinesq approximation the equations
\begB
{\partial\bOm\over\partial t}=\nu\nabla^2\bOm+\nabla\times({\bf V}\times\bOm
-{\bf H}\times(\nabla\times{\bf H}))+\nabla\times({\bf V\times\tau e}_3
+\beta{\cal T}{\bf e}_3+{\bf F}),
\endB{eq1p1}{CHMB}\begI
{\partial{\bf H}\over\partial t}=
\eta\nabla^2{\bf H}+\nabla\times({\bf V}\times{\bf H})+{\bf J},
\endI{eq1p2}
$${\partial{\cal T}\over\partial t}=\kappa\nabla^2{\cal T}
-({\bf V}\cdot\nabla){\cal T}+S,$$
\begI
\nabla\cdot{\bf V}=\nabla\cdot{\bf H}=0,
\endI{eq1p3}\begI
\nabla\times{\bf V}=\bOm.
\endL{eq1p4}
Here ${\bf V(x},t)$ denotes the velocity and
$\bOm({\bf x},t)$ the vorticity of an electrically conducting fluid flow,
${\bf H}({\bf x},t)$ magnetic field, ${\cal T}({\bf x},t)$ temperature,
$t$ time, $\nu$, $\eta$ and $\kappa$ kinematic, magnetic and thermal molecular
diffusivities, respectively, $\tau/2$ angular velocity of the rotation,
$\beta{\cal T}{\bf e}_3$ the Archimedes (buoyancy) force,
${\bf e}_k$ is the unit vector along the coordinate axis $x_k$,
${\bf F(x},t)$ a body force, ${\bf J(x},t)$ reflects the presence
of externally induced currents in the layer and $S({\bf x},t)$ of heat sources.
To simplify the notation we will occasionally use 10-dimensional vectors
$(\bom,{\bf v,h},\theta)$ and 7-dimensional vectors $(\bom,{\bf h},\theta)$
or $({\bf v,h},\theta)$.

The following conditions are assumed on the horizontal boundaries
of the layer:
\\$\bullet$ free boundaries:
\begB
\left.{\partial V_1\over\partial x_3}\right|_{x_3=\pm L/2}=
\left.{\partial V_2\over\partial x_3}\right|_{x_3=\pm L/2}=0,\quad
\left.\phantom{|_|}V_3\right|_{x_3=\pm L/2}=0
\endB{eq2p1}{BCs}

\begI
\Rightarrow\left.\phantom{|_|}\Omega_1\right|_{x_3=\pm L/2}
=\left.\phantom{|_|}\!\!\Omega_2\right|_{x_3=\pm L/2}=0,\quad
\left.{\partial\Omega_3\over\partial x_3}\right|_{x_3=\pm L/2}=0;
\endI{eq2p2}

\noindent
$\bullet$ electrically conducting boundaries:
\begI
\left.{\partial H_1\over\partial x_3}\right|_{x_3=\pm L/2}=
\left.{\partial H_2\over\partial x_3}\right|_{x_3=\pm L/2}=0,\quad
\left.\phantom{|_|}H_3\right|_{x_3=\pm L/2}=0;
\endL{eq2p3}

\noindent
$\bullet$ boundaries at fixed temperatures:
$$\left.\phantom{|_|}{\cal T}\right|_{x_3=-L/2}={\cal T}_1,\quad
\left.\phantom{|_|}{\cal T}\right|_{x_3=L/2}={\cal T}_2.$$
(Vector components are enumerated by the subscript.) It is convenient
to introduce a new variable $\Theta={\cal T}-{\cal T}_1+\delta(x_3+L/2)$,
where $\delta=({\cal T}_1-{\cal T}_2)/L$ (convection is only
possible for $\delta>0$). It satisfies the equation
\begin{equation}
{\partial\Theta\over\partial t}=\kappa\nabla^2\Theta
-({\bf V}\cdot\nabla)\Theta+\delta V_3+S
\label{eq3}\end{equation}
and homogeneous boundary conditions
\begin{equation}
\left.\phantom{|_|}\Theta\right|_{x_3=\pm L/2}=0.
\label{eq4}\end{equation}

Linearisation of (\theCHMB) in the vicinity of the CHM regime ${\bf V,H},\Theta$
is the operator ${\cal L}=({\cal L}^\omega,{\cal L}^h,{\cal L}^\theta)$, where
$${\cal L}^\omega(\bom,{\bf v,h},\theta)\equiv-{\partial\bom\over\partial t}
+\nu\nabla^2\bom+\nabla\times\lpar\bf V\times\bom+v\times\bOm$$
\begB
{\bf-H\times(\nabla\times h)-h\times(\nabla\times H})\rpar
+\tau{\partial{\bf v}\over\partial x_3}+\beta\nabla\theta\times{\bf e}_3,
\endB{eq6p1}{defL}

\begI
{\cal L}^h({\bf v,h})\equiv-{\partial{\bf h}\over\partial t}+\eta\nabla^2{\bf h}
+\nabla\times({\bf v\times H+V\times h}),
\endI{eq6p2}\begI
{\cal L}^\theta({\bf v},\theta)\equiv-{\partial\theta\over\partial t}+\kappa\nabla^2\theta
-({\bf V}\cdot\nabla)\theta-({\bf v}\cdot\nabla)\Theta+\delta v_3,
\endL{eq6p3}

\noindent
and $\bom=\nabla\times\bf v$.
Evidently, ${\cal L}^\omega=(\nabla\times){\cal L}^v$, where
$${\cal L}^v({\bf v,h},\theta,p)\equiv-{\partial{\bf v}\over\partial t}
+\nu\nabla^2{\bf v}+{\bf V}\times(\nabla\times{\bf v})+{\bf v}\times\bOm$$
$$-{\bf H}\times(\nabla\times{\bf h})-{\bf h}\times(\nabla\times{\bf H})
+\tau{\bf v}\times{\bf e}_3+\beta\theta{\bf e}_3-\nabla p$$
is the linearisation of the Navier-Stokes equation for the flow velocity:
$${\partial{\bf V}\over\partial t}=\nu\nabla^2{\bf V}+{\bf V}\times\bOm
-{\bf H}\times(\nabla\times{\bf H})
+\tau{\bf V}\times{\bf e}_3+\beta\Theta{\bf e}_3-\nabla P.$$
Here $P=P'/\rho+|{\bf V}|^2/2+\tau^2(x_1^2+x_2^2)/8$ is a modified pressure,
$P'$ pressure and $\rho$ fluid density.

We assume that in a weakly nonlinear regime the amplitude of a perturbation
is of the order of $\varepsilon$. The perturbed state $\bOm+\varepsilon\bom$,
$\bf V+\varepsilon v$, $\bf H+\varepsilon h$, $\Theta+\varepsilon\theta$
satisfies equations (\theCHMB)--\rf{eq4}, and hence the profiles of perturbations
$\bom,{\bf v,h},\theta$ (which we henceforth call perturbations) satisfy
the equations
\begB
{\cal L}^\omega(\bom,{\bf v,h},\theta)
+\varepsilon\nabla\times({\bf v}\times\bom-{\bf h}\times(\nabla\times{\bf h}))=0,
\endB{eq5p1}{EQpert}\begI
{\cal L}^h({\bf v,h})+\varepsilon\nabla\times({\bf v\times h})=0,
\endI{eq5p2}\begI
{\cal L}^\theta({\bf v},\theta)-\varepsilon({\bf v}\cdot\nabla)\theta=0,
\endI{eq5p3}\begI
\nabla\cdot{\bf h}=0,
\endI{eq5p4}\begI
\nabla\cdot\bom=\nabla\cdot{\bf v}=0,
\endI{eq5p5}\begI
\nabla\times{\bf v}=\bom.
\endI{eq5p6}

Define the spatial mean over fast variables and fluctuating parts of a scalar
or vector field $f$ as
$$\la f\ra\equiv\lim_{\ell\to\infty}{1\over L\ell^2}\int_{-L/2}^{L/2}
\int_{-\ell/2}^{\ell/2}\int_{-\ell/2}^{\ell/2}f({\bf x})\,dx_1\,dx_2\,dx_3,
\qquad\lb f\rb\equiv f-\la f\ra,$$
and the spatio-temporal mean and the fluctuating part of $f$ as
$$\lad f\rad\equiv\lim_{\hat{t}\to\infty}
{1\over\hat{t}}\int_0^{\hat{t}}\la f({\bf x},t)\ra\,dt,\qquad
\lbd f\rbd\equiv f-\lad f\rad,$$
respectively. Denote by $\la f\ra_k$ and $\lad f\rad_k$ the $k-$th
components of the means $\la\bf f\ra$ and $\lad\bf f\rad$, respectively,
$$\la{\bf f}\ra_v\equiv\la f\ra_3\,{\bf e}_3,\qquad
\lb{\bf f}\rb_v\equiv{\bf f}-\la{\bf f}\ra_v;$$
$$\lad{\bf f}\rad_v\equiv\lad f\rad_3\,{\bf e}_3,\qquad
\lbd{\bf f}\rbd_v\equiv{\bf f}-\lad{\bf f}\rad_v;$$
$$\la{\bf f}\ra_h\equiv\la f\ra_1{\bf e}_1+\la f\ra_2\,{\bf e}_2,\qquad
\lb{\bf f}\rb_h\equiv{\bf f}-\la{\bf f}\ra_h;$$
$$\lad{\bf f}\rad_h\equiv\lad f\rad_1{\bf e}_1+\lad f\rad_2\,{\bf e}_2,\qquad
\lbd{\bf f}\rbd_h\equiv{\bf f}-\lad{\bf f}\rad_h.$$
(Thus the subscripts $v$ and $h$ are used to denote vertical and horizontal
components of three-dimensional vector fields; we will also use the
superscripts $v$ and $h$ to denote the flow velocity and magnetic field
components of 7- or 10-dimensional vector fields.)

Averaging the horizontal component of the equation for the flow perturbation,
\pagebreak
$${\cal L}^v({\bf v,h},\theta,p)+\varepsilon({\bf v}\times(\nabla\times{\bf v})-{\bf h}\times(\nabla\times{\bf h}))=0,$$
over the layer of the fluid, find
$${\partial\la{\bf v}\ra_h\over\partial t}=
\la{\bf v}\ra_h\times\tau{\bf e}_3-\la\nabla p\ra_h.$$
Thus, the mean horizontal component of the perturbation of the flow
is controlled by the mean of the gradient of the pressure perturbation $p$,
which must be specified. We assume that there is no pumping of
fluid through the layer due to pressure variation at the infinity, i.e.
the growth of pressure at the infinity is bounded so that $\la\nabla p\ra_h=0$.
Under this condition, if initially the horizontal component of the perturbation
of the velocity vanishes, then at any time
\begI
\la{\bf v}\ra_h=0.
\endL{eq5p7}

\mi{\bf 3. Asymptotic expansions of large-scale weakly nonlinear perturbations

of CHM regimes}

\mi
We introduce the slow spatial, ${\bf X}=\varepsilon(x_1,x_2)$, and temporal,
$T=\varepsilon^2t$, variables. The exponent in the temporal scale ratio is
tailored for CHM regimes, where the $\alpha-$effect is insignificant in the
leading order. A solution to the problem (\theEQpert) is sought in the form of power series:
\begB
\bom=\sum_{n=0}^\infty\bom_n({\bf x},t,{\bf X},T)\varepsilon^n,
\endB{eq16p1}{pertser}

\begI
{\bf v}=\sum_{n=0}^\infty{\bf v}_n({\bf x},t,{\bf X},T)\varepsilon^n,
\endI{eq16p2}\begI
{\bf h}=\sum_{n=0}^\infty{\bf h}_n({\bf x},t,{\bf X},T)\varepsilon^n,
\endI{eq16p3}\begI
\theta=\sum_{n=0}^\infty\theta_n({\bf x},t,{\bf X},T)\varepsilon^n.
\endL{eq16p4}

Consider the series in $\varepsilon$ resulting from substitution of \rf{eq16p2}
and \rf{eq16p3} into \rf{eq5p4} and \rf{eq5p5}. The mean and fluctuating
parts of equations at order $n\ge0$ are
\begB
\nabla_{\bf X}\cdot\la{\bf v}_n\ra_h=\nabla_{\bf X}\cdot\la{\bf h}_n\ra_h=0,
\endB{eq17p1}{termsol}\begI
\nabla_{\bf x}\cdot\lb{\bf v}_n\rb_h+\nabla_{\bf X}\cdot\lb{\bf v}_{n-1}\rb_h=0,
\endI{eq17p2}\begI
\nabla_{\bf x}\cdot\lb{\bf h}_n\rb_h+\nabla_{\bf X}\cdot\lb{\bf h}_{n-1}\rb_h=0.
\endI{eq17p3}
(In differential operators with the indices $\bf x$ and $\bf X$
differentiation in the respective spatial fast and slow variables is performed;
$\nabla_{\bf X}=(\partial/\partial X_1,\partial/\partial X_2,0)$. Henceforth
differentiation in fast variables is assumed in the definition (\thedefL)
of the operator $\cal L$. All quantities for $n<0$ are zero by definition.) Clearly,
\begI
\nabla_{\bf X}\cdot\la\bom_n\ra_v=0,
\endI{eq17p4}
and $\nabla\cdot\bom=0$ implies
\begI
\nabla_{\bf x}\cdot\lb\bom_n\rb_v+\nabla_{\bf X}\cdot\lb\bom_{n-1}\rb_v=0
\endL{eq17p5}
for all $n\ge0$.

Evidently, $\la\bom_n\ra_v$, $\la{\bf v}_n\ra_h$ and $\la{\bf h}_n\ra_h$ satisfy
the boundary conditions (\theBCs). Due to \rf{eq17p1}, we can introduce a stream function
$\psi_n(X_1,X_2,t,T)$ for $\la{\bf v}_n\ra_h$:
$$\la{\bf v}_n\ra_h=\left(-{\partial\psi_n\over\partial X_2},
{\partial\psi_n\over\partial X_1},0\right).$$
Substitution of the series \rf{eq16p1} and \rf{eq16p2} into \rf{eq5p6} yields
\begin{equation}
\nabla_{\bf x}\times\lb{\bf v}_n\rb_h=\bom_n-\nabla_{\bf X}\times{\bf v}_{n-1}.
\label{eq18}\end{equation}
Consider this as an equation in $\lb{\bf v}_n\rb_h$. It can be shown using
\rf{eq18} after the change of index $n\to n-1$ and \rf{eq17p5}, that the right-hand
side of \rf{eq18} is solenoidal in fast variables. Consider the operator curl
which acts from the space of solenoidal vector fields satisfying \rf{eq2p1} into
the space of solenoidal vector fields. The domain of the adjoint operator,
also a curl, is the space of solenoidal vector fields satisfying \rf{eq2p2}; its
kernel consists of constant vector fields $(0,0,C)$. Thus \rf{eq18} has a solution
as long as the average of the vertical component of the right-hand side
vanishes\footnote{This condition is sufficient for a space-periodic in horizontal
directions CHM system. In general a sufficiently fast decay near zero
of the spectrum of the field, to which the inverse curl is applied.
We assume that the examined CHM state is such that the inverse curl can be
applied to any field, whenever the space average of the vertical
component of the field vanishes.}:
\begin{equation}
\la\bom_n\ra_v=\nabla_{\bf X}\times\la{\bf v}_{n-1}\ra_h.
\label{eq19}\end{equation}
(For $n=0$ this reduces to $\la\bom_0\ra_v=0$.) In the terms of the stream
function this equation is $\nabla^2_{\bf X}\psi_{n-1}=\la\bom_n\ra_3$.
One can uniquely determine $\la{\bf v}_{n-1}\ra_h$ from this equation under the
condition that $\psi_{n-1}$ is globally bounded (so that \rf{eq5p7} is satisfied),
if the mean of $\la\bom_n\ra_v$ over the plane of slow variables vanishes.

In view of \rf{eq19}, \rf{eq18} reduces to
\begin{equation}
\nabla_{\bf x}\times\lb{\bf v}_n\rb_h=\lb\bom_n\rb_v
-\nabla_{\bf X}\times\lb{\bf v}_{n-1}\rb_h.
\label{eq30r}\end{equation}
After a substitution $\lb{\bf v}_n\rb_h={\bf v}+\nabla_{\bf x}B$,
where $B$ is a globally bounded solution to the Neumann problem
$$\nabla^2_{\bf x}B=-\nabla_{\bf X}\cdot\lb{\bf v}_{n-1}\rb_h,\qquad
\left.{\partial B\over\partial x_3}\right|_{x_3=\pm L/2}=0,$$
the system of equations \rf{eq30r} and \rf{eq17p2} takes the form of \xrf{eq5p5}{eq5p7}.

A solution to the system \xrf{eq5p5}{eq5p7} is
${\bf v}={\cal P}\lb{\bf A}\rb_h$, where $\bf A$ solves the Poisson's equation
$$\nabla^2{\bf A}=-\nabla\times\bom,\qquad
\left.{\partial A_1\over\partial x_3}\right|_{x_3=\pm L/2}=
\left.{\partial A_2\over\partial x_3}\right|_{x_3=\pm L/2}=0,\qquad
\left.A_3\right|_{x_3=\pm L/2}=0,$$
and $\cal P$ is the projection of a three-dimensional vector field into the
subspace of solenoidal fields: ${\cal P}{\bf A\equiv A}-\nabla a$,
$a$ being a solution to the Neumann problem
$$\nabla^2a=\nabla\cdot{\bf A},\qquad
\left.{\partial a\over\partial x_3}\right|_{x_3=\pm L/2}=0.$$
This defines the operator $\cal R$, inverse to the curl, which
acts from the space of solenoidal fields satisfying $\la\bom\ra_v=0$
and boundary conditions for vorticity, into the space of solenoidal fields
satisfying \rf{eq5p7} and boundary conditions for flows.

We substitute (\thepertser) into \xrf{eq5p1}{eq5p3} and transform them into power series in
$\varepsilon$. The equations emerging at different orders $\varepsilon^n$
constitute a hierarchy (see (A1)--(A3) in appendix A); they can be solved
together with (\thetermsol) and \rf{eq18} successively, by considering separately their mean
and fluctuating parts. When considering the equations at order $n$, the mean
vorticity equation at order $n+1$ is also used to find $\la\bom_{n+1}\ra_v$ and
$\la{\bf v}_n\ra_h$.
A closed system of nonlinear equations for the averaged leading terms of (\thepertser)
(which we call the mean-field equations) will be derived as solvability
conditions for equations in fast variables at the orders $n=2$ and 3.

\mi{\bf 4. Solvability of auxiliary problems}

\mi
Vector fields $\bOm,{\bf V,H},\Theta$ constituting the CHM regime, whose
stability is examined, depend only on fast variables. We assume that they are
smooth and globally bounded together with their derivatives\footnote{This
is understood in the following sense: a field ${\bf f(x},t,{\bf X},T)$
is globally bounded together with its derivatives, if the field and all
required partial derivatives are bounded, with the bounds
depending only on the order of the derivative, slow variables and fast
time.} and all average quantities in fast variables, considered below,
are correctly defined. This condition is satisfied, for instance, if
the perturbed CHM regime is periodic in horizontal directions and time (since
then the domain, in which all the fields are defined, is compact).

Auxiliary problems of the following structure will be considered:
\begB
{\cal L}^\omega(\bom,{\bf v,h},\theta)={\bf f}^\omega,\qquad{\cal L}^h({\bf v,h})={\bf f}^h,
\qquad{\cal L}^\theta({\bf v},\theta)=f^\theta,
\endB{eq9p1}{linePRO}\begI
\nabla_{\bf x}\cdot\bom=d^\omega,\qquad\nabla_{\bf x}\cdot{\bf v}=d^v,\qquad\nabla_{\bf x}\cdot{\bf h}=d^h,
\endI{eq9p2}\begI
\bom-\nabla_{\bf x}\times{\bf v}={\bf a},
\endI{eq9p3}
and the means $\lad\bom\rad_v$, $\lad{\bf h}\rad_h$ and $\la{\bf v}\ra_h$
are specified.
Here the right-hand sides are known and the compatibility relations, obtained
by combining \rf{eq9p2} with the divergences of \rf{eq9p3} and the first two
equations in \rf{eq9p1}, must be satisfied:
\begI
\nabla_{\bf x}\cdot{\bf a}-d^\omega=0,
\endI{eq9p6}\begI
-{\partial d^\omega\over\partial t}+\nu\nabla_{\bf x}^2d^\omega
+\tau{\partial d^\omega\over\partial x_3}=\nabla_{\bf x}\cdot{\bf f}^\omega,
\endI{eq9p4}\begI
-{\partial d^h\over\partial t}+\eta\nabla_{\bf x}^2d^h=\nabla_{\bf x}\cdot{\bf f}^h;
\endI{eq9p5}
$\bf a$ satisfies the boundary conditions for vorticity,
\begI
\la d^v\ra=\la d^h\ra=0.
\endL{eq9p7}

\pagebreak\noindent
Due to \rf{eq9p4} and \rf{eq9p5}, it suffices that vorticity and magnetic field
satisfy \rf{eq9p2} initially.

The fields $\bom,{\bf v,h}$ and $\theta$
must be globally bounded with their derivatives and satisfy the same boundary
conditions as $\bOm,{\bf V,H}$ and $\Theta$, respectively.
For vector fields from this class, the identities
\begB
\la{\cal L}^\omega(\bom,{\bf v,h},\theta)\ra_v=-{\partial\la\bom\ra_v\over\partial t},
\endB{eq11p1}{weakdercond}

\begI
\la{\cal L}^h({\bf v,h})\ra_h=-{\partial\la{\bf h}\ra_h\over\partial t},
\endL{eq11p2}

\noindent
hold true, and hence
\begin{equation}
\lad{\cal L}^\omega(\bom,{\bf v,h},\theta)\rad_v=0,\qquad
\lad{\cal L}^h({\bf v,h})\rad_h=0.
\label{eq11p3}\end{equation}
\rf{eq11p3} implies that the conditions
\begB
\lad{\bf f}^\omega({\bf x},t)\rad_v=0,
\endB{eq13p1}{weaksolco}\begI
\lad{\bf f}^h({\bf x},t)\rad_h=0
\endL{eq13p2}
are necessary for existence of a solution to equations (\thelinePRO).
Integrating (\theweakdercond) in fast time, find
\begin{equation}
\left.\phantom{|^|_|}\!\!\!\!\!\!
\la\bom\ra_v\right|_{t=0}-\la\bom\ra_v=\int_0^t\la{\bf f}^\omega\ra_v\,dt,
\qquad\left.\phantom{|^|_|}
\la{\bf h}\ra_h\right|_{t=0}-\la{\bf h}\ra_h=\int_0^t\la{\bf f}^h\ra_h\,dt,
\label{initave}\end{equation}
implying
\begin{equation}
\la\bom\ra_v=\lad\bom\rad_v-\lbdb\int_0^t\la{\bf f}^\omega\ra_v\,dt\rbdb,\qquad
\la{\bf h}\ra_h=\lad{\bf h}\rad_h-\lbdb\int_0^t\la{\bf f}^h\ra_h\,dt\rbdb
\label{solvealpha}\end{equation}
(although the integrals in the right-hand sides fluctuate only in time,
the notation $\lbd\cdot\rbd$ is still applicable).
Thus, $\la\bom\ra_v$ and $\la{\bf h}\ra_h$ are well-defined, if and only if
the means $\lad\int_0^t\la{\bf f}^\omega\ra_vdt\rad$ and
$\lad\int_0^t\la{\bf f}^h\ra_hdt\rad$ are. The latter conditions are stronger
than (\theweaksolco), since if $\lad{\bf f}^\omega\rad_v$ and
$\lad{\bf f}^h\rad_h$ exist, (\theweaksolco) follows from existence of
$\lad\int_0^t\la{\bf f}^\omega\ra_vdt\rad$ and $\lad\int_0^t\la{\bf f}^h\ra_hdt\rad$.
Averaging the vertical component of \rf{eq9p3} over fast spatial variables,
find its solvability condition (see section 3):
\begin{equation}
\la{\bf a}\ra_v=\lad\bom\rad_v-\lbdb\int_0^t\la{\bf f}^\omega\ra_v\,dt\rbdb.
\label{solveomega}\end{equation}

In what follows we assume that for arbitrary globally bounded together with
their derivatives smooth solenoidal zero-mean (\theweaksolco) fields
${\bf f}^\omega({\bf x},t)$, ${\bf f}^h({\bf x},t)$ and $f^\theta({\bf x},t)$
the system (\thelinePRO) has a solution $\bom,{\bf v,h},\theta$ in the
considered class at least for some smooth initial conditions, satisfying
\rf{solvealpha} for $t=0$. This is necessary to ensure that the auxiliary
problems, stated in sections 5 and 8, have solutions with the required
properties. A smooth solution to (\thelinePRO) can be constructed
as a solution to a parabolic equation for smooth initial conditions and
right-hand sides in \rf{eq11p1}. However, it is not guaranteed that it is
globally bounded, since the region occupied by the fluid is not compact.

\pagebreak
By virtue of (\theweakdercond), the spatial means
$\la\bom\ra_v$ and $\la{\bf h}\ra_h$ of solutions from this class to the system
\begin{equation}
{\cal L}(\bom,{\bf v,h},\theta)=0,
\label{eq15}\end{equation}
complemented by \xrf{eq5p4}{eq5p7}, are time-independent. A CHM regime is
{\it linearly stable to small-scale perturbations}, if a solution
to \rf{eq15}, \xrf{eq5p4}{eq5p7} exponentially decays in time for any
smooth globally bounded initial conditions satisfying $\la\bom\ra_v=\la{\bf h}\ra_h=0$.
It is shown in appendix B, that if a CHM state is linearly stable to
small-scale perturbations, then for any smooth globally bounded right-hand
sides in \rf{eq11p1} and initial conditions satisfying \rf{eq9p2}, \rf{eq9p3}
and \rf{solvealpha}, the solution to (\thelinePRO) is globally bounded.
However, we do not demand that the perturbed CHM regime
${\bf V,H},\Theta$ is linearly stable to small-scale perturbations
(despite any small-scale instability has a larger, order zero,
growth rate, than a large-scale instability), so that the formalism which is
developed here were applicable to analyse weakly nonlinear instability of
chaotic CHM attractors.

In the remaining part of this section we show that, for a generic space-periodic
steady or time-periodic CHM regimes, (\theweaksolco) are sufficient conditions
for existence of a solution to (\thelinePRO), steady and/or having the same
periods as the perturbed CHM regime. Define the operator
of linearisation not involving the flow velocity explicitly:
\begB
{\cal M}'^\omega(\bom',{\bf h}',\theta)\equiv
{\cal L}^\omega(\bom',{\cal R}\bom',{\bf h}',\theta),
\endB{eq7p1}{Mdef}\begI
{\cal M}'^h(\bom',{\bf h}')\equiv{\cal L}^h({\cal R}\bom',{\bf h}'),
\endI{eq7p2}\begI
{\cal M}'^\theta(\bom',\theta)\equiv{\cal L}^\theta({\cal R}\bom',\theta).
\endL{eq7p3}
The operator ${\cal M}'=({\cal M}'^\omega,{\cal M}'^h,{\cal M}'^\theta)$ acts in the space
of 7-dimensional vector fields $(\bom',{\bf h}',\theta)$, where
vorticity and magnetic field are solenoidal, $\la\bom'\ra_v=0$, and the boundary
conditions of the kind of \rf{eq2p2}, \rf{eq2p3} and \rf{eq4} are satisfied.
Let $\cal M$ denote a restriction of ${\cal M}'$ on the subspace in the
domain of $\cal M$ defined by the condition $\la{\bf h}'\ra_h=0$. Substituting
$$\bom=\bom'+{\bf a},\qquad
{\bf v}={\bf v}'+\nabla_{\bf x}A^v+\la{\bf v}\ra_h,$$
$${\bf h}={\bf h}'+\nabla_{\bf x}A^h+\lad{\bf h}\rad_h-\lbdb\int_0^t\la{\bf f}^h\ra_h\,dt\rbdb,$$
where $A^v$ and $A^h$ are solutions to the Neumann problems
$$\nabla_{\bf x}^2A^v=d^v,\quad
\left.{\partial A^v\over\partial x_3}\right|_{x_3=\pm L/2}=0;\qquad
\nabla_{\bf x}^2A^h=d^h,\quad
\left.{\partial A^h\over\partial x_3}\right|_{x_3=\pm L/2}=0,$$
transform the problem (\thelinePRO) into an equivalent one:
\begB
{\cal M}^\omega(\bom',{\bf h}',\theta)={\bf f}'^\omega,
\endB{MEQ1}{Msys}
\vspace*{-2mm}
\begI
{\cal M}^h(\bom',{\bf h}',\theta)={\bf f}'^h,
\endI{MEQ2}\begI
{\cal M}^\theta(\bom',{\bf h}',\theta)=f'^\theta,
\endI{MEQ3}\begI
\nabla_{\bf x}\cdot\bom'=\nabla_{\bf x}\cdot{\bf h}'=0,
\endI{MEQ4}\begI
\la\bom'\ra_v=0,\qquad\la{\bf v}'\ra_h=0,\qquad\la{\bf h}'\ra_h=0,
\endI{MEQ5}\begI
\nabla_{\bf x}\cdot{\bf f}'^\omega=\nabla_{\bf x}\cdot{\bf f}'^h=0,
\endI{MEQ6}\begI
\la{\bf f}'^\omega\ra_v=\la{\bf f}'^h\ra_h=0.
\endL{MEQ7}

The operator
$\widetilde{\cal L}^*=((\widetilde{\cal L}^*)^v,(\widetilde{\cal L}^*)^h,(\widetilde{\cal L}^*)^\theta)$
adjoint to $\widetilde{\cal L}=({\cal L}^v,{\cal L}^h,{\cal L}^\theta)$
can be derived as usual by integration by parts in the defining identity
$$\la\widetilde{\cal L}^*({\bf v,h},\theta)\cdot({\bf v',h'},\theta')\ra\equiv
\la({\bf v,h},\theta)\cdot\widetilde{\cal L}({\bf v',h'},\theta')\ra:$$
$$(\widetilde{\cal L}^*)^v({\bf v,h},\theta)={\partial{\bf v}\over\partial t}+\nu\nabla^2{\bf v}
-\nabla\times({\bf V}\times{\bf v})$$
$$+\,{\cal P}\lb{\bf H}\times(\nabla\times{\bf h})-{\bf v}\times\bOm
-\tau{\bf v\times e}_3+\delta\theta{\bf e}_3-\theta\nabla\Theta\rb_h,$$
$$(\widetilde{\cal L}^*)^h({\bf v,h})={\partial{\bf h}\over\partial t}+\eta\nabla^2{\bf h}
+\nabla\times({\bf H}\times{\bf v})
+{\cal P}({\bf v}\times(\nabla\times{\bf H})-{\bf V}\times(\nabla\times{\bf h})),$$
$$(\widetilde{\cal L}^*)^\theta({\bf v},\theta)={\partial\theta\over\partial t}
+\kappa\nabla^2\theta+({\bf V}\cdot\nabla)\theta+\beta v_3.$$
Conditions on horizontal boundaries for vector fields in the
domain\footnote{Vector fields in the domains of $\widetilde{\cal L}$ and
$\widetilde{\cal L}^*$ are assumed to be globally bounded with their
derivatives; in this class the bilinear form $\lad\bf a\cdot b\rad$ is not
a genuine scalar product, since, for instance, $\lad|{\bf a}|^2\rad=0$ for any
smooth field $\bf a$ with a compact support.} of $\widetilde{\cal L}^*$
can be obtained demanding as usual that the boundary surface integrals
arising in this integration are zero. For (\theBCs) and \rf{eq4} defining
the domains of $\cal L$ and $\widetilde{\cal L}$ (together with
the solenoidality condition for the flow, vorticity and magnetic components),
the domain of $\widetilde{\cal L}^*$ and $\widetilde{\cal L}$ is the same.

The operator ${\cal M}'^*=(({\cal M}'^*)^\omega,\,({\cal M}'^*)^h,\,({\cal M}'^*)^\theta)$,
adjoint to ${\cal M}'$, can be found as an adjoint to a composition
of $\widetilde{\cal L}$ with the curl:
$$({\cal M}'^*)^\omega(\bom,{\bf h},\theta)={\partial\bom\over\partial t}
+\nu\nabla^2\bom+{\cal R}'\lb-\nabla\times({\bf V}\times(\nabla\times\bom))$$
$$+{\bf H}\times(\nabla\times{\bf h})+\bOm\times(\nabla\times\bom)
-\tau{\partial\bom\over\partial x_3}+\delta\theta{\bf e}_3-\theta\nabla\Theta\rb_h,$$
$$({\cal M}'^*)^h(\bom,{\bf h})={\partial{\bf h}\over\partial t}+\eta\nabla^2{\bf h}
+\nabla\times({\bf H}\times(\nabla\times\bom))+{\cal P}((\nabla\times\bom)\times(\nabla\times{\bf H})
-{\bf V}\times(\nabla\times{\bf h})),$$
$$({\cal M}'^*)^\theta(\bom,{\bf h},\theta)={\partial\theta\over\partial t}
+\kappa\nabla^2\theta+({\bf V}\cdot\nabla)\theta+\beta{\bf e}_3\cdot(\nabla\times\bom).$$
Here the operator ${\cal R}'$ is also ``an inverse curl'', like $\cal R$,
but it is defined for different boundary conditions:
If $\bom$ is a solution to the problem
$$\nabla\times\bom={\bf v}-\nabla p\quad\Leftrightarrow\quad
\nabla^2\bom=-\nabla\times\bf v,$$
$$\nabla\cdot\bom=0,\qquad\la\bom\ra_v=0,$$
satisfying \rf{eq2p2}, then ${\cal R}'{\bf v}=\bom$. Since $\cal M$ is a
composition of ${\cal M}'$ with the projection onto the subspace of fields with
a zero spatial mean of the horizontal magnetic component,
\begin{equation}
{\cal M}^*(\bom,{\bf h},\theta)=(({\cal M}'^*)^\omega(\bom,{\bf h},\theta),
\lb({\cal M}'^*)^h(\bom,{\bf h},\theta)\rb_h,({\cal M}'^*)^\theta(\bom,{\bf h},\theta)).
\label{Madj}\end{equation}
(${\cal M}'^*\!$ and ${\cal M}^*$ have the same domains, as ${\cal M}'$ and
$\cal M$, respectively.)

\pagebreak
Clearly, $(0,(C_1,C_2,0),0)\in\ker{\cal M}'^*$ for any constant $C_1$ and $C_2$.
We consider the generic case, where the kernel of ${\cal M}'^*$ consists of such
constant vectors. (The case, where no external source terms $\bf F,J$ and $S$
are present in (\theCHMB), is not generic.) Then $\ker{\cal M}^*$ is trivial:
for $(\bom,{\bf h},\theta)\in\ker{\cal M}^*$
$${\cal M}^*(\bom,{\bf h},\theta)=(0,(c_1(t),c_2(t),0),0)\ \Leftrightarrow
\ {\cal M}^*\!\left(\!\bom,{\bf h}-\sum_{k=1}^2\int_0^tc_k(t')\,dt'\,{\bf e}_k,\theta\right)\!=0,$$
by virtue of \rf{Madj}. Hence the property of $\ker{\cal M}'^*$ implies
$$\bom=0,\quad{\bf h}-\sum_{k=1}^2\int_0^tc_k(\tilde{t})\,d\tilde{t}\,{\bf e}_k
=(C_1,C_2,0),\quad\theta=0,$$
and thus ${\bf h}=0$, since vector fields from the domain of $\cal M$ and
${\cal M}^*$ satisfy $\la{\bf h}\ra_h=0$.

When the kernel of ${\cal M}'^*$ is two-dimensional, the conditions (\theweaksolco)
are sufficient for existence of a steady or time-periodic solution to
(\theMsys), which can be shown as in Zheligovsky (2003). This follows from
the theorem on Fredholm alternative (see, e.g., Liusternik and Sobolev, 1961),
stating that a solution to a problem
$({\cal I}+{\cal K}){\cal X}=\cal F$ in a Hilbert space exists if and only if
$\cal F$ is orthogonal to ker$({\cal I}+{\cal K}^*)$. Here $\cal I$ is the
identity operator, $\cal K$ is compact and ${\cal K}^*$ is its adjoint.
Apply the operators $(-\partial/\partial t+\nu\nabla^2)^{-1}$,
$(-\partial/\partial t+\eta\nabla^2)^{-1}$ and
$(-\partial/\partial t+\kappa\nabla^2)^{-1}$ to \xrf{MEQ1}{MEQ3}, respectively.
(By virtue of \rf{MEQ7} and the assumed boundary conditions they can be applied
to the right-hand sides of \rf{MEQ1} and \rf{MEQ2}; $\partial/\partial t=0$
in a problem with the steady data.) The system (\theMsys) is reduced thereby
to an equivalent problem of the form
${\cal M}^\circ(\bom',{\bf h}',\theta)={\bf f}''$, where
${\cal M}^\circ$ acts in the domain of $\cal M$ and has a trivial kernel,
as long as $\ker\cal M$ is trivial. This problem is of the type considered
in the Fredholm theorem. Thus application of the theorem guarantees existence
of a unique solution with the required steadiness or time-periodicity.

In the absence of space periodicity the condition (\theweaksolco) may be insufficient
for existence of a globally bounded solution to (\thelinePRO), equivalent to (\theMsys).
For instance, this can occur for CHM regimes, quasi-periodic in a horizontal
direction: In this case, the Laplacian $\nabla^2$, acting in the space of
vector fields the appropriate averages of which vanish, does not have a bounded
inverse, and it is impossible to make a reduction of the problem, for
which the Fredholm theorem can be readily applied.

In sections 5--10 we assume that the problem (\theMsys) has a bounded solution
for any right-hand sides ${\bf f}'^\omega,{\bf f}'^h,f'^\theta$ (not imposing
any periodicity conditions). Then the system of mean-field equations, that we
derive, is comprised of equations for mean horizontal components of
perturbations of the flow and magnetic field.
In section 11 we consider the case, where ${\cal M}^*$ has an one-dimensional
kernel; then the system of mean-field equations is expanded by a scalar
equation for the amplitude of the respective mean-free neutral mode.

\mi{\bf 5. Order $\varepsilon^0$ equations}

\mi
The leading terms of the series \xrf{eq5p1}{eq5p3} yield the equation
\begin{equation}
{\cal L}(\bom_0,{\bf v}_0,{\bf h}_0,\theta_0)=0.
\label{eq20}\end{equation}
Averaging over fast spatial variables the vertical vorticity component of \rf{eq20},
the horizontal magnetic component of \rf{eq20} and the vertical component of
(A1) for $n=1$ obtain, respectively,
$${\partial\la\bom_0\ra_v\over\partial t}=0\quad\Rightarrow\quad
\la\bom_0\ra_v=0$$
(provided $\la\bom_0\ra_v|_{t=0}=0$ to agree with \rf{eq19} for $n=0$),
\begB
{\partial\la{\bf h}_0\ra_h\over\partial t}=0\quad\Rightarrow\quad
\la{\bf h}_0\ra_h=\lad{\bf h}_0\rad_h,
\endB{eq21p1}{initmean}\begI
{\partial\la\bom_1\ra_v\over\partial t}=0\quad\Rightarrow\quad
\la\bom_1\ra_v=\lad\bom_1\rad_v.
\endI{eq21p2}
The last equation together with \rf{eq17p1} for $n=0$ and \rf{eq19} for $n=1$ implies
$$\la{\bf v}_0\ra_h=\lad{\bf v}_0\rad_h+{\bf v}'_0(t,T).$$
It is shown in appendix C that
$\la\,\lb{\bf v}_0\rb_h|_{{\bf X}=\varepsilon(x_1,x_2)}\ra$ is asymptotically
smaller than any power of $\varepsilon$. Thus \rf{eq5p7} asymptotically holds
true to any order of $\varepsilon^n$ at any time, if ${\bf v}'_0=0$, i.e.
\begI
\la{\bf v}_0\ra_h=\lad{\bf v}_0\rad_h,
\endL{zerofluc0}
and the average of
$\lad{\bf v}_0\rad_h$ over the plane of slow spatial variables vanishes.

Separating the mean and fluctuating parts of the unknown vector fields,
transform \rf{eq20} into an equivalent equation
$${\cal M}(\lb\bom_0\rb_v,\lb{\bf h}_0\rb_h,\theta_0)=
-\lpar\nabla_{\bf x}\times(\lad{\bf v}_0\rad_h\times\bOm
-\lad{\bf h}_0\rad_h\times(\nabla_{\bf x}\times{\bf H})),$$
\begin{equation}
(\lad{\bf h}_0\rad_h\cdot\nabla_{\bf x}){\bf V}
-(\lad{\bf v}_0\rad_h\cdot\nabla_{\bf x}){\bf H},
\quad-(\lad{\bf v}_0\rad_h\cdot\nabla_{\bf x})\Theta\rpar.
\label{eq22}\end{equation}
$\lad{\bf v}_0\rad_h$ and $\lad{\bf h}_0\rad_h$ are independent of fast
variables, and slow variables are not involved in the definition of
the operator $\cal M$; thus, by linearity, a solution to equations
\rf{eq22}, \rf{eq17p2}, \rf{eq17p3} and \rf{eq18} for $n=0$ can be expressed as
\begin{equation}
(\lb\bom_0\rb_v,\lb{\bf v}_0\rb_h,\lb{\bf h}_0\rb_h,\theta_0)
=\bxi_0^\cdot+\sum_{k=1}^2\left({\bf S}^{v,\cdot}_k\lad v_0\rad_k
+{\bf S}^{h,\cdot}_k\lad h_0\rad_k\right)
\label{eq23}\end{equation}
(interpreted as an equality of 10-dimensional vectors).
${\bf S}^{\cdot,v}_k$ can be determined from the equations
\begB
\nabla_{\bf x}\times{\bf S}^{v,v}_k={\bf S}^{v,\omega}_k,\quad
\nabla_{\bf x}\cdot{\bf S}^{v,v}_k=0,\quad\la{\bf S}^{v,v}_k\ra_h=0;
\endB{eq24p1}{omegaSv}\begI
\nabla_{\bf x}\times{\bf S}^{h,v}_k={\bf S}^{h,\omega}_k,\quad
\nabla_{\bf x}\cdot{\bf S}^{h,v}_k=0,\quad\la{\bf S}^{h,v}_k\ra_h=0.
\endL{eq24p2}
${\bf S}^{\cdot,\cdot}_k({\bf x},t)=({\bf S}^{\cdot,\omega}_k,{\bf S}^{\cdot,v}_k,{\bf S}^{\cdot,h}_k,
S^{\cdot,\theta}_k)$ are solutions to {\it auxiliary problems of type} I,
satisfying the boundary conditions similar to (\theBCs) and \rf{eq4}:
\begB
{\cal M}({\bf S}^{v,\omega}_k,{\bf S}^{v,h}_k,S^{v,\theta}_k)=\left(
{\partial\bOm\over\partial x_k},{\partial{\bf H}\over\partial x_k},
{\partial\Theta\over\partial x_k}\right),
\endB{eq25p1}{Iauxiv}\vspace*{2mm}\begI
\nabla_{\bf x}\cdot{\bf S}^{v,\omega}_k=\nabla_{\bf x}\cdot{\bf S}^{v,h}_k=0,
\endI{eq25p2}\begI
\la{\bf S}^{v,\omega}_k\ra_v=0,\qquad\la{\bf S}^{v,h}_k\ra_h=0,
\endL{eq25p3}
together with \rf{eq24p1};

\pagebreak
\begB
{\cal M}({\bf S}^{h,\omega}_k,{\bf S}^{h,h}_k,S^{h,\theta}_k)=
\left(-\nabla_{\bf x}\times{\partial{\bf H}\over\partial x_k},
-{\partial{\bf V}\over\partial x_k},0\right),
\endB{eq26p1}{Iauxih}\vspace*{2mm}\begI
\nabla_{\bf x}\cdot{\bf S}^{h,\omega}_k=\nabla_{\bf x}\cdot{\bf S}^{h,h}_k=0,
\endI{eq26p2}\begI
\la{\bf S}^{h,\omega}_k\ra_v=0,\qquad\la{\bf S}^{h,h}_k\ra_h=0,
\endL{eq26p3}
together with \rf{eq24p2}.

The problems (\theIauxiv), \rf{eq24p1} and (\theIauxih), \rf{eq24p2} are
particular cases of the problem (\thelinePRO); consistency relations
\xrf{eq9p6}{eq9p7} and solvability conditions (\theweaksolco) and
\rf{solveomega} are trivially satisfied for them (~(\theweaksolco) is
satisfied due to the assumption that $\bOm$ and $\bf H$ are globally bounded).
Relations \rf{eq25p2}, \rf{eq25p3}, \rf{eq26p2} and \rf{eq26p3} hold true
at any time, provided they do at $t=0$. Therefore, any
smooth vector fields, globally bounded together with their derivatives and
satisfying \rf{eq25p2}, \rf{eq25p3}, \rf{eq26p2}, \rf{eq26p3} and the
boundary conditions similar to (\theBCs) and \rf{eq4},
can serve as initial conditions for the problems (\theIauxiv) and (\theIauxih),
as long as the respective solutions remain globally bounded in time. For CHM
states ${\bf V,H},\Theta$, linearly stable to small-scale perturbations,
the boundedness for any initial data is demonstrated in appendix B.
Existence of steady or time-periodic solutions for generic perturbed CHM
states, periodic in horizontal directions, is shown in section 3.

Equations \rf{eq25p1} and \rf{eq26p1} are equivalent to
$${\cal L}({\bf S}^{v,\omega}_k,{\bf S}^{v,v}_k+{\bf e}_k,{\bf S}^{v,h}_k,S^{v,\theta}_k)=0;\qquad
{\cal L}({\bf S}^{v,\omega}_k,{\bf S}^{v,v}_k,{\bf S}^{v,h}_k+{\bf e}_k,S^{v,\theta}_k)=0,$$
respectively. The second of these equations is an eigenvalue problem for
eigenfunctions from $\ker{\cal M}'$ with non-zero spatio-temporal means of
horizontal magnetic components. Since $(0,(C_1,C_2,0),0)\in\ker{\cal M}'^*$
for constant $C_1$ and $C_2$, this problem has a solution regardless of whether
ker$\cal M$ is trivial, as assumed in sections 5--10, or not, provided
${\cal M}'$ does not have a Jordan cell associated with the zero eigenvalue.
In computations it may be natural to consider, instead of vorticity components
of \rf{eq25p1} and \rf{eq26p1}, equations involving their vector potentials:
$${\cal L}^v({\bf S}^{v,v}_k,{\bf S}^{v,h}_k,S^{v,\theta}_k,S^{v,p}_k)=
{\partial{\bf V}\over\partial x_k};\qquad
{\cal L}^v({\bf S}^{h,v}_k,{\bf S}^{h,h}_k,S^{h,\theta}_k,S^{h,p}_k)=
-{\partial{\bf H}\over\partial x_k},$$
where $\la\nabla_{\bf x}S^{\cdot,p}_k\ra_h=0$, and to
solve numerically auxiliary problems of type I in the terms of
${\bf S}^{\cdot,v}_k,{\bf S}^{\cdot,h}_k,S^{\cdot,\theta}_k$ and $S^{\cdot,p}_k$.

$\bxi^\cdot_0({\bf x},t,{\bf X},T)=(\bxi_0^\omega,\bxi_0^v,\bxi_0^h,\xi_0^\theta)$
satisfies the equations
\begB
{\cal L}(\bxi_0^\omega,\bxi_0^v,\bxi_0^h,\xi_0^\theta)=0,
\endB{eq28p1}{Ixi}\begI
\nabla_{\bf x}\cdot\bxi_0^\omega=\nabla_{\bf x}\cdot\bxi_0^h=0,\qquad
\endI{eq28p2}\begI
\la\bxi_0^\omega\ra_v=0,\qquad\la\bxi_0^h\ra_h=0,
\endI{eq28p3}\begI
\nabla_{\bf x}\times\bxi_0^v=\bxi_0^\omega,\qquad
\nabla_{\bf x}\cdot\bxi_0^v=0,\qquad\la\bxi_0^v\ra_h=0.
\endL{eq28p4}
Relations (\theomegaSv), \rf{eq25p2}, \rf{eq26p2}, \rf{eq28p2} and
\rf{eq28p4} guarantee that conditions \rf{eq17p2}, \rf{eq17p3} and \rf{eq18}
hold true for $n=1$. Again, it is sufficient to demand that \rf{eq28p2} and
\rf{eq28p3} are satisfied at $t=0$. Initial
conditions for (\theIxi) can be found from \rf{eq23} at $t=0$.
(From \rf{eq21p1}, $\lad{\bf h}_0\rad_h|_{T=0}=\la{\bf h}_0\ra_h|_{t=0}$.)
Initial conditions for $\bxi^\cdot_0$ must belong to the stable manifold of
the perturbed CHM state ${\bf V,H},\Theta$, i.e. $\bxi^\cdot_0$ must
exponentially decay in time. A permissible change of initial conditions for
${\bf S}^{\cdot,\cdot}_k$ in the considered class implies a change of initial
conditions for $\bxi^\cdot_0$, but then the respective changes in
${\bf S}^{\cdot,\cdot}_k$ and $\bxi^\cdot_0$ exponentially decay in time.

\mi{\bf 6. Order $\varepsilon^1$ and $\varepsilon^2$ equations: $\alpha$-effect
in the leading order}

\mi
Averaging the vertical component of (A1) for $n=2$ over fast spatial variables
and taking into account \rf{eq11p1}, \xrf{eq17p1}{eq17p3}, \rf{eq18} for $n=0$
and 1, \rf{eq19} for $n=0$, the boundary conditions for $\bf V$, $\bf H$,
${\bf v}_i$ and ${\bf h}_i$, and solenoidality of ${\bf v}_0$ and ${\bf h}_0$, find
$$-{\partial\la\bom_2\ra_v\over\partial t}+\nabla_{\bf X}\times\la
{\bf V}\times(\nabla_{\bf X}\times{\bf v}_0)-{\bf V}\nabla_{\bf X}\cdot\lb{\bf v}_0\rb_h$$
\begin{equation}
-{\bf H}\times(\nabla_{\bf X}\times{\bf h}_0)
+{\bf H}\nabla_{\bf X}\cdot\lb{\bf h}_0\rb_h\ra_h=0.
\label{eq29}\end{equation}
Substitution of the flow and magnetic field \rf{eq23} now yields
\begin{equation}
{\partial\la\bom_2\ra_v\over\partial t}=\nabla_{\bf X}\times
\left(\,\sum_{k=1}^2\sum_{m=1}^2\left(\alpha^v_{m,k}
{\partial\lad v_0\rad_k\over\partial X_m}+\alpha^h_{m,k}
{\partial\lad h_0\rad_k\over\partial X_m}\right){\bf e}_k+\txi{\omega}\right).
\label{eq30}\end{equation}
Here it is denoted
$$\txi{\omega}=\la{\bf V}\times(\nabla_{\bf X}\times\bxi^v_0)
-{\bf V}\nabla_{\bf X}\cdot\bxi^v_0-{\bf H}\times(\nabla_{\bf X}\times\bxi^h_0)
+{\bf H}\nabla_{\bf X}\cdot\bxi^h_0\ra_h;$$
\begB
{\bf a}^v_{m,k}=\la{\bf V}\times({\bf e}_m\times({\bf S}^{v,v}_k+{\bf e}_k))
-{\bf H}\times({\bf e}_m\times{\bf S}^{v,h}_k)
-{\bf V}(S^{v,v}_k)_m+{\bf H}(S^{v,h}_k)_m\ra_h,
\endB{eq32p1}{AKAco}\vspace*{-1mm}\begI
{\bf a}^h_{m,k}=\la{\bf V}\times({\bf e}_m\times{\bf S}^{h,v}_k)
-{\bf H}\times({\bf e}_m\times({\bf S}^{h,h}_k+{\bf e}_k))
-{\bf V}(S^{h,v}_k)_m+{\bf H}(S^{h,h}_k)_m\ra_h;
\endI{eq32p2}\begI
\alpha^v_{1,1}=(a^v_{1,1})_1-(a^v_{2,1})_2-(a^v_{2,2})_1,\qquad
\alpha^v_{2,1}=(a^v_{2,1})_1,
\endI{eq32p4}\begI
\alpha^v_{1,2}=(a^v_{1,2})_2,\qquad
\alpha^v_{2,2}=(a^v_{2,2})_2-(a^v_{1,2})_1-(a^v_{1,1})_2,
\endL{eq32p5}
where $(f)_m$ is the $m-$th component of a three-dimensional vector $\bf f$;
$\alpha^h_{m,k}$ are also expressed in the terms of ${\bf a}^h_{j,n}$
by the last four formulae, where the superscript $v$ is changed to $h$.
The differential operator in the right-hand side of \rf{eq30} represents
the AKA--effect (anisotropic kinematic $\alpha-$effect) operator
in the leading order.

Applying \rf{solvealpha}, obtain from \rf{eq30}
\begin{equation}
\la\bom_2\ra_v=\lad\bom_2\rad_v
\label{eq39}\end{equation}
$$+\,\nabla_{\bf X}\!\times\!\left(\,\sum_{k=1}^2\sum_{m=1}^2\!\!\left(\!\lbdb\int_0^t\alpha^v_{m,k}\,dt\rbdb
{\partial\lad v_0\rad_k\over\partial X_m}+\lbdb\int_0^t\alpha^h_{m,k}\,dt\rbdb
{\partial\lad h_0\rad_k\over\partial X_m}\right)\!{\bf e}_k+\lbdb\int_0^t\txi{\omega}dt\rbdb\right).$$
This expression is well-defined, if and only if
the means $\lad\int_0^t\alpha^\cdot_{m,k}\,dt\rad$ for $m,k=1,2$ are.
This is {\it the condition of insignificance of the AKA-effect} in the leading
order. It is stronger than the solvability condition \rf{eq13p1}
for the system (A1)--(A3) for $n=2$:
\begin{equation}
\lad\alpha^\cdot_{m,k}\rad=0.
\label{eq34}\end{equation}

\pagebreak
A similar procedure reveals a possible presence of magnetic $\alpha-$effect
in the CHM system. Averaging of the horizontal component of (A2) for $n=1$
over fast spatial variables with the use of \rf{eq11p2} and
substitution of the flow and magnetic component of \rf{eq23} yields
\begin{equation}
{\partial\la{\bf h}_1\ra_h\over\partial t}=\nabla_{\bf X}\times\left(\,\sum_{k=1}^2
\left(\bal^v_k\lad v_0\rad_k+\bal^h_k\lad h_0\rad_k\right)+\txi{h}\right).
\label{eq35}\end{equation}
Here it is denoted
\begin{equation}
\bal^v_k=\la{\bf V}\times{\bf S}^{v,h}_k+({\bf S}^{v,v}_k+{\bf e}_k)\times{\bf H}\ra_v,\qquad
\bal^h_k=\la{\bf V}\times({\bf S}^{h,h}_k+{\bf e}_k)+{\bf S}^{h,v}_k\times{\bf H}\ra_v;
\label{eq36}\end{equation}
$$\txi{h}=\la\bxi^v_0\times{\bf H}+{\bf V}\times\bxi^h_0\ra_v.$$
The differential operator in the right-hand side of \rf{eq35} represents
the magnetic $\alpha-$effect. \rf{solvealpha} applied to \rf{eq35} yields
\begin{equation}
\la{\bf h}_1\ra_h=\lad{\bf h}_1\rad_h\!+\!\nabla_{\bf X}\!\times\!\left(\,
\sum_{k=1}^2\left(\lbdb\int_0^t\!\bal^v_k\,dt\rbdb\lad v_0\rad_k
\!+\!\lbdb\int_0^t\!\bal^h_k\,dt\rbdb\lad h_0\rad_k\right)
\!+\!\lbdb\int_0^t\!\txi{h}dt\rbdb\right)\!.
\label{Malpha}\end{equation}
{\it Magnetic $\alpha-$effect is insignificant} in the leading order, if
the means $\lad\int_0^t\bal^\cdot_kdt\rad$ exist and as a result
$\la{\bf h}_1\ra_h$ is well-defined by \rf{Malpha}. The condition of
insignificance of the magnetic $\alpha-$effect is stronger than the solvability
condition \rf{eq13p2} for the problem (A1)--(A3) for $n=1$:
\begin{equation}
\lad\bal^v_k\rad=\lad\bal^h_k\rad=0.
\label{eq37}\end{equation}

If the perturbed CHM state is steady or periodic in time, the 8 scalar
relations \rf{eq34} are sufficient for insignificance of kinematic
$\alpha-$effect, and the 4 scalar relations \rf{eq37} are sufficient for
insignificance of the magnetic $\alpha-$effect. Note that vanishing of
vertical components of the kinematic $\alpha-$tensor in the vorticity equation
or of horizontal components of the magnetic $\alpha-$tensor in the magnetic
induction equation is not required. For this reason we are speaking about
{\it insignificance} and not about the absence of the $\alpha-$effect.

The terms $\lbd\int_0^t\txi{\omega}dt\rbd$ in \rf{eq39} and
$\lbd\int_0^t\txi{h}dt\rbd$ in \rf{Malpha} are not problematic, because
$\txi{\omega}$ and $\txi{h}$ exponentially decay and thus the averages
$\lad\int_0^t\txi{\omega}dt\rad$ and $\lad\int_0^t\txi{h}dt\rad$ are
well-defined. It is shown in appendix D that $\lbd\int_0^t\txi{h}dt\rbd$ and
$\lbd\int_0^t\txi{h}dt\rbd$ also exponentially decay.

The multiscale approach remains feasible even, if \rf{eq34} or \rf{eq37} do not
hold true. It is well-known (see Dubrulle and Frisch, 1991), that in this case
another slow time scale is appropriate, $T=\varepsilon t$. Then the new terms,
$\partial\la\bom_1\ra_v/\partial T$ and $\partial\la{\bf h}_0\ra_h/\partial T$,
emerging in the left-hand sides of \rf{eq30} and \rf{eq35}, balance the
$\alpha-$effect terms. \rf{eq30} and \rf{eq35} then constitute a closed system
of equations (together with \rf{eq17p1} and \rf{eq19} for $n=1$). The
mean-field equations turn out to be linear first-order PDE's; solutions to such
equations generically exhibit unbounded exponential growth. Therefore, from
now on we focus on a potentially more interesting case of an insignificant
$\alpha$-effect, where possible growth of perturbations may saturate due to
nonlinearity.

\pagebreak
\mi{\bf 7. Symmetries, guaranteeing insignificance of the $\alpha-$effect\\
in the leading order}

\mi
In this section we consider the symmetries, in the presence of which the
$\alpha-$effect is insignificant in the leading order. They are compatible with
the equations of thermal hydromagnetic convection and boundary conditions
considered in this paper.

A CHM regime $\bOm,{\bf V,H},\Theta$ is called {\it parity-invariant with the
time shift $\widetilde T$}, if the fields $\bf V$ and $\bf H$ are
parity-invariant with the time shift $\widetilde T$, and $\bOm$ and $\Theta$
are parity-antiinvariant. A three-dimensional vector field $\bf f$
is parity-invariant with the time shift $\widetilde T$, if
$${\bf f}(-{\bf x},t)=-{\bf f}({\bf x},t+\widetilde T),$$
and parity-antiinvariant, if
$${\bf f}(-{\bf x},t)={\bf f}({\bf x},t+\widetilde T);$$
a scalar field $f$ is parity-invariant with the time shift $\widetilde T$, if
$$f(-{\bf x},t)=f({\bf x},t+\widetilde T),$$
and parity-antiinvariant, if
$$f(-{\bf x},t)=-f({\bf x},t+\widetilde T).$$
(We have assumed here that the origin of the coordinate system is located
at the centre of symmetry on the mid-plane of the liquid layer.)
If the perturbed CHM regime is parity-invariant, then we call a set
of fields $(\bom,\bf v,h,\theta)$ {\it symmetric}, if $\bf v$ and $\bf h$
are parity-invariant, and $\bom$ and $\theta$ are parity-antiinvariant; it is
called {\it antisymmetric}, if $\bf v$ and $\bf h$ are parity-antiinvariant,
and $\bom$ and $\theta$ are parity-invariant.

We call a CHM regime $\bOm,{\bf V,H},\Theta$ {\it symmetric about the
vertical axis \hbox{$x_1=x_2=0$} with the time shift $\widetilde T$}, if all the
fields $\bf V,H,\bOm$ and $\Theta$ are symmetric. A three-dimensional vector
field $\bf f$ is symmetric about the vertical axis with the time shift
$\widetilde T$, if
$$f_1(-x_1,-x_2,x_3,t)=-f_1(x_1,x_2,x_3,t+\widetilde T),$$
$$f_2(-x_1,-x_2,x_3,t)=-f_2(x_1,x_2,x_3,t+\widetilde T),$$
$$f_3(-x_1,-x_2,x_3,t)=f_3(x_1,x_2,x_3,t+\widetilde T),$$
and antisymmetric, if
$$f_1(-x_1,-x_2,x_3,t)=f_1(x_1,x_2,x_3,t+\widetilde T),$$
$$f_2(-x_1,-x_2,x_3,t)=f_2(x_1,x_2,x_3,t+\widetilde T),$$
$$f_3(-x_1,-x_2,x_3,t)=-f_3(x_1,x_2,x_3,t+\widetilde T);$$
a scalar field $f$ is symmetric about the vertical axis with the time shift
$\widetilde T$, if
$$f(-x_1,-x_2,x_3,t)=f(x_1,x_2,x_3,t+\widetilde T),$$
and antisymmetric, if
$$f(-x_1,-x_2,x_3,t)=-f(x_1,x_2,x_3,t+\widetilde T).$$
(Without any loss of generality the origin of the coordinate system is assumed
to reside at the vertical axis.) When this symmetry is concerned, we call a set
of fields $(\bom,\bf v,h,\theta)$ {\it symmetric}, if all the four fields are,
and {\it antisymmetric}, if they are all antisymmetric.

Symmetries without a time shift (for $\widetilde T=0$) are spatial,
they can be possessed by CHM regimes of arbitrary time dependence.
We will call such symmetries {\it parity invariance} and {\it symmetry
about the vertical axis}, respectively. By contrast, only regimes, periodic
in time with the period $2\widetilde T$ can have spatio-temporal
symmetries with a time shift $\widetilde T\ne0$.
Travelling waves, for instance, can have such symmetries.

Suppose the symmetry is defined by means of the operator $\cal S$, i.e.
symmetric and antisymmetric fields satisfy the conditions ${\cal S}\bf f=f$ and
${\cal S}\bf f=-f$, respectively. Since all the symmetries considered in this
section are of the second order, an arbitrary field $\bf f$ can be decomposed
in a sum of a symmetric field $({\bf f+{\cal S}f})/2$ and antisymmetric
$({\bf f-{\cal S}f})/2$.

If a CHM regime is symmetric, then symmetric and antisymmetric sets of fields
are invariant subspaces for the linearisation operators $\cal L$ and $\cal M$.
In this case the right-hand sides of equations \rf{eq25p1} and \rf{eq26p1}
are antisymmetric sets, and hence essentially ${\bf S}^{\cdot,\cdot}_k$ are
antisymmetric sets (they are antisymmetric if initial conditions are; by
construction, for any permissible initial conditions a symmetric part of a
solution to auxiliary problems of type I exponentially decays). Consequently,
(\theAKAco) and \rf{eq36} imply that in the presence of a symmetry the AKA--
and magnetic $\alpha-$effects are insignificant. Moreover, since
$$\alpha^\cdot_{m,k}(t+\widetilde T)=-\alpha^\cdot_{m,k}(t)$$
for a symmetric perturbed CHM state, the function
$$\vartheta^\cdot(t)\equiv\lbdb\int_0^t\alpha^\cdot_{m,k}\,dt\rbdb$$
satisfies a similar relation $\vartheta^\cdot(t+\widetilde T)=-\vartheta^\cdot(t)$:
$$\vartheta(t)=\int_0^t\alpha^\cdot_{m,k}\,dt-{1\over2\widetilde T}
\int_0^{2\widetilde T}\int_0^{\tilde{t}}\alpha^\cdot_{m,k}(t')\,dt'\,d\tilde{t}
={1\over2\widetilde T}\int_0^{2\widetilde T}\int_{\tilde{t}}^t\alpha^\cdot_{m,k}(t')\,dt'\,d\tilde{t},$$
whereby
$$\vartheta^\cdot(t)+\vartheta^\cdot(t+\widetilde T)=
{1\over2\widetilde T}\int_0^{2\widetilde T}
\left(\int_{\tilde{t}}^t\alpha^\cdot_{m,k}(t')\,dt'
+\int_{\tilde{t}}^{t+\widetilde{T}}\alpha^\cdot_{m,k}(t')\,dt'\right)d\tilde{t}$$
$$={1\over2\widetilde T}\int_0^{2\widetilde T}
\!\!\left(\int_{\tilde{t}}^t\alpha^\cdot_{m,k}(t')\,dt'-
\int_{\tilde{t}-\widetilde{T}}^t\alpha^\cdot_{m,k}(t')\,dt'\right)d\tilde{t}
=-{1\over2\widetilde T}\int_0^{2\widetilde T}
\int_{\tilde{t}-\widetilde{T}}^{\tilde{t}}\alpha^\cdot_{m,k}(t')\,dt'd\tilde{t}$$
$$=-{1\over2\widetilde T}\int_0^{\widetilde T}
\left(\int_{\tilde{t}-\widetilde{T}}^{\tilde{t}}\alpha^\cdot_{m,k}(t')\,dt'
+\int_{\tilde{t}}^{\tilde{t}+\widetilde{T}}\alpha^\cdot_{m,k}(t')\,dt'\right)d\tilde{t}=0.$$

\mi{\bf 8. Order $\varepsilon^1$ equations: $\alpha$-effect, insignificant in the leading order}

\mi
In this section we solve equations (A1)--(A3) for $n=1$, assuming from now
on that the AKA-- and magnetic $\alpha-$effects are insignificant.

The solution to equations \rf{eq19} for $n=2$, \rf{eq17p1} for $n=1$ and \rf{eq39} is
$$\la{\bf v}_1\ra_h=\lad{\bf v}_1\rad_h+\sum_{k=1}^2\sum_{m=1}^2\!\left(\!
\lbdb\int_0^t\alpha^v_{m,k}\,dt\rbdb{\partial\lad v_0\rad_k\over\partial X_m}+
\lbdb\int_0^t\alpha^h_{m,k}\,dt\rbdb{\partial\lad h_0\rad_k\over\partial X_m}\right){\bf e}_k$$
\begB
-\nabla_{\bf X}\Pi+\txi{v},
\endB{eq40}{avevI}\begI
\Pi=\!\sum_{m=1}^2\!\left(\!
\lbdb\int_0^t(\alpha^v_{m,1}-\alpha^v_{m,2})\,dt\rbdb{\partial^2\nabla^{-2}_{\bf X}\lad v_0\rad_1\over\partial X_m\partial X_1}
\!+\lbdb\int_0^t(\alpha^h_{m,1}-\alpha^h_{m,2})\,dt\rbdb{\partial^2\nabla^{-2}_{\bf X}\lad h_0\rad_1\over\partial X_m\partial X_1}\!\right)\!,
\endL{eq41}
where the operator $\nabla^{-2}_{\bf X}$ is inverse
to the Laplacian in slow variables: for $\bf f(X)$, whose average over
$\bf X$ is zero, ${\bf g(X})=\nabla^{-2}_{\bf X}\bf f$ is the mean-free
solution to the equation $\nabla_{\bf X}^2\bf g=f$, which is globally bounded
together with the derivatives. In \rf{eq40}
\begin{equation}
\txi{v}=\lbdb\int_0^t\bxi'dt\rbdb,\qquad
\bxi'({\bf X},t,T)=\txi{\omega}-\nabla_{\bf X}\nabla^{-2}_{\bf X}(\nabla_{\bf X}\cdot\txi{\omega}).
\label{eq42}\end{equation}
$\bxi'$ exponentially decays in fast time (inheriting this property from
$\bxi^\cdot_0$). It is shown in appendix D that $\txi{v}$ also exponentially
decays in fast time.

Thus $\la{\bf v}_1\ra_h$ is expressed in the terms of
$\lad v_0\rad_k$ and $\lad h_0\rad_k$. Since the solvability condition
\rf{eq19} for the equation \rf{eq18} for $n=1$ is satisfied (see section 5),
$\lb{\bf v}_1\rb_h$ can be found from \rf{eq18}. As shown in the previous section,
vanishing of the average of $\la{\bf v}_1\ra_h$ over slow spatial variables
guarantees that the condition \rf{eq5p7} for ${\bf v}_1|_{{\bf X}=\varepsilon(x_1,x_2)}$
is asymptotically satisfied to any power of $\varepsilon$.

Transform the equations (A1)--(A3) for $n=1$ using in (A1) the identity
$\bom_1=\lb\bom_1\rb_v+\nabla_{\bf X}\times\lad{\bf v}_0\rad_h$:
$${\cal L}^\omega(\lb\bom_1\rb_v,\lb{\bf v}_1\rb_h,\lbd{\bf h}_1\rbd_h,\theta_1)
+(\nabla_{\bf X}\times\lad{\bf v}_0\rad_h)_3{\partial{\bf V}\over\partial x_3}
-(\la{\bf v}_1\ra_h\cdot\nabla_{\bf x})\bOm$$
$$+(\lad{\bf h}_1\rad_h\cdot\nabla_{\bf x})(\nabla_{\bf x}\times{\bf H})
+2\nu(\nabla_{\bf x}\cdot\nabla_{\bf X})\lb\bom_0\rb_v
-\nabla_{\bf x}\times({\bf H}\times(\nabla_{\bf X}\times{\bf h}_0))$$
$$+\nabla_{\bf X}\times\left({\bf V}\times\bom_0+{\bf v}_0\times\bOm
-{\bf H\times(\nabla_x\times h}_0)-{\bf h}_0\times(\nabla_{\bf x}\times{\bf H})\right)$$
$$+\nabla_{\bf x}\times({\bf v}_0\times\bom_0-{\bf h}_0\times(\nabla_{\bf x}\times{\bf h}_0))
+\beta\nabla_{\bf X}\theta_0\times{\bf e}_3=0;$$
$${\cal L}^h(\lb{\bf v}_1\rb_h,\lbd{\bf h}_1\rbd_h)
-(\la{\bf v}_1\ra_h\cdot\nabla_{\bf x}){\bf H}
+(\lad{\bf h}_1\rad_h\cdot\nabla_{\bf x}){\bf V}
+2\eta(\nabla_{\bf x}\cdot\nabla_{\bf X})\lb{\bf h}_0\rb_h$$
$$+\nabla_{\bf X}\times({\bf v}_0\times{\bf H}+{\bf V}\times{\bf h}_0)
+\nabla_{\bf x}\times({\bf v}_0\times{\bf h}_0)=0;$$
$${\cal L}^\theta(\lb{\bf v}_1\rb_h,\theta_1)
-(\la{\bf v}_1\ra_h\cdot\nabla_{\bf x})\Theta
+2\kappa(\nabla_{\bf x}\cdot\nabla_{\bf X})\theta_0
-({\bf V}\cdot\nabla_{\bf X})\theta_0
-({\bf v}_0\cdot\nabla_{\bf x})\theta_0=0.$$

In view of relations \rf{eq23} and (\theavevI), (\thetermsol) and \rf{eq18} for $n=1$,
and due to linearity of this problem, it has solutions of the following structure:
\pagebreak
$$(\lb\bom_1\rb_v,\lb{\bf v}_1\rb_h,\lbd{\bf h}_1\rbd_h,\theta_1)
=\bxi_1^\cdot+\left.\sum_{k=1}^2\right({\bf S}^{v,\cdot}_k\lad v_1\rad_k
+{\bf S}^{h,\cdot}_k\lad h_1\rad_k$$
$$\left.+\sum_{m=1}^2\right({\bf G}^{v,\cdot}_{m,k}{\partial\lad v_0\rad_k\over\partial X_m}
+{\bf G}^{h,\cdot}_{m,k}{\partial\lad h_0\rad_k\over\partial X_m}
+{\bf Y}^{v,\cdot}_{m,k}
{\partial^3\nabla^{-2}_{\bf X}\lad v_0\rad_1\over\partial X_k\partial X_m\partial X_1}
+{\bf Y}^{h,\cdot}_{m,k}
{\partial^3\nabla^{-2}_{\bf X}\lad h_0\rad_1\over\partial X_k\partial X_m\partial X_1}$$
\begin{equation}
\left.\left.\phantom{\partial\over\partial}
+{\bf Q}^{vv,\cdot}_{m,k}\lad v_0\rad_k\lad v_0\rad_m
+{\bf Q}^{vh,\cdot}_{m,k}\lad v_0\rad_k\lad h_0\rad_m
+{\bf Q}^{hh,\cdot}_{m,k}\lad h_0\rad_k\lad h_0\rad_m\right)\right).
\label{eq43}\end{equation}
Here
${\bf G}^{\cdot,\cdot}_{m,k}=({\bf G}^{\cdot,\omega}_{m,k},
{\bf G}^{\cdot,v}_{m,k},{\bf G}^{\cdot,h}_{m,k},G^{\cdot,\theta}_{m,k})$
solve {\it auxiliary problems of type II}:
$${\cal L}^\omega({\bf G}^{v,\cdot}_{m,k})=
-\epsilon_{m,k,3}{\partial{\bf V}\over\partial x_3}
-2\nu{\partial{\bf S}^{v,\omega}_k\over\partial x_m}
-{\bf e}_m\times\left({\bf V}\times{\bf S}_k^{v,\omega\phantom{\theta}}\right.$$
$$\left.+({\bf S}^{v,v}_k+{\bf e}_k)\times\bOm
-{\bf H}\times(\nabla_{\bf x}\times{\bf S}^{v,h}_k)
-{\bf S}^{v,h}_k\times(\nabla_{\bf x}\times{\bf H})
+\beta S^{v,\theta}_k{\bf e}_3\right)$$
\begB
+\nabla_{\bf x}\times({\bf H}\times({\bf e}_m\times{\bf S}^{v,h}_k))
+\lbdb\int_0^t\alpha^v_{m,k}dt\rbdb{\partial\bOm\over\partial x_k}
\endB{eq44p1}{auxIIv}

\vspace{1mm}\noindent
($\epsilon_{m,k,j}$ denotes the standard unit antisymmetric tensor),
\begI
{\cal L}^h({\bf G}^{v,\cdot}_{m,k})=
-2\eta{\partial{\bf S}^{v,h}_k\over\partial x_m}
-{\bf e}_m\times\left({\bf V}\times{\bf S}^{v,h}_k
+({\bf S}^{v,v}_k+{\bf e}_k)\times{\bf H}\right)
+\lbdb\int_0^t\alpha^v_{m,k}dt\rbdb{\partial{\bf H}\over\partial x_k},
\endI{eq44p5}\begI
{\cal L}^\theta({\bf G}^{v,\cdot}_{m,k})=
-2\kappa{\partial S^{v,\theta}_k\over\partial x_m}+V_mS^{v,\theta}_k
+\lbdb\int_0^t\alpha^v_{m,k}dt\rbdb{\partial\Theta\over\partial x_k},
\endI{eq44p7}\begI
\nabla_{\bf x}\times{\bf G}^{v,v}_{m,k}={\bf G}^{v,\omega}_{m,k}
-{\bf e}_m\times{\bf S}^{v,v}_k,
\endI{eq44p3}\begI
\nabla_{\bf x}\cdot{\bf G}^{v,\omega}_{m,k}=-(S^{v,\omega}_k)_m,\quad
\nabla_{\bf x}\cdot{\bf G}^{v,v}_{m,k}=-(S^{v,v}_k)_m,\quad
\nabla_{\bf x}\cdot{\bf G}^{v,h}_{m,k}=-(S^{v,h}_k)_m;
\endL{eq44sol}
$${\cal L}^\omega({\bf G}^{h,\cdot}_{m,k})=
-2\nu{\partial{\bf S}^{h,\omega}_k\over\partial x_m}
-{\bf e}_m\times\left({\bf V}\times{\bf S}^{h,\omega}_k\right.$$
$$\left.+{\bf S}^{h,v}_k\times\bOm
-{\bf H}\times(\nabla_{\bf x}\times{\bf S}^{h,h}_k)
-({\bf S}^{h,h}_k+{\bf e}_k)\times(\nabla_{\bf x}\times{\bf H})
+\beta S^{h,\theta}_k{\bf e}_3\right)$$
\begB
+\nabla_{\bf x}\times({\bf H}\times({\bf e}_m\times({\bf S}^{h,h}_k+{\bf e}_k))
+\lbdb\int_0^t\alpha^h_{m,k}dt\rbdb{\partial\bOm\over\partial x_k},
\endB{eq45p1}{auxIIh}\begI
{\cal L}^h({\bf G}^{h,\cdot}_{m,k})=
-2\eta{\partial{\bf S}^{h,h}_k\over\partial x_m}
-{\bf e}_m\times\left({\bf V}\times({\bf S}^{h,h}_k+{\bf e}_k)
+{\bf S}^{h,v}_k\times{\bf H}\right)
+\lbdb\int_0^t\alpha^h_{m,k}dt\rbdb{\partial{\bf H}\over\partial x_k},
\endI{eq45p5}\begI
{\cal L}^\theta({\bf G}^{h,\cdot}_{m,k})=
-2\kappa{\partial S^{h,\theta}_k\over\partial x_m}+V_mS^{h,\theta}_k
+\lbdb\int_0^t\alpha^h_{m,k}dt\rbdb{\partial\Theta\over\partial x_k};
\endI{eq45p7}\begI
\nabla_{\bf x}\times{\bf G}^{h,v}_{m,k}={\bf G}^{h,\omega}_{m,k}
-{\bf e}_m\times{\bf S}^{h,v}_k,
\endI{eq45p3}\begI
\nabla_{\bf x}\cdot{\bf G}^{h,\omega}_{m,k}=-(S^{h,\omega}_k)_m,\quad
\nabla_{\bf x}\cdot{\bf G}^{h,v}_{m,k}=-(S^{h,v}_k)_m,\quad
\nabla_{\bf x}\cdot{\bf G}^{h,h}_{m,k}=-(S^{h,h}_k)_m;
\endL{eq45sol}

\pagebreak\noindent
${\bf Q}^{\cdot\cdot,\cdot}_{m,k}=({\bf Q}^{\cdot\cdot,\omega}_{m,k},
{\bf Q}^{\cdot\cdot,v}_{m,k},{\bf Q}^{\cdot\cdot,h}_{m,k},Q^{\cdot\cdot,\theta}_{m,k})$
solve {\it auxiliary problems of type III}:
$$\left.{\cal M}^\omega({\bf Q}^{vv,\omega}_{m,k},{\bf Q}^{vv,h}_{m,k},Q^{vv,\theta}_{m,k})=
\rho_{m,k}\nabla_{\bf x}\times\right(
-({\bf S}^{v,v}_k+{\bf e}_k)\times{\bf S}^{v,\omega}_m$$
\begB
\left.+{\bf S}^{v,h}_k\times(\nabla_{\bf x}\times{\bf S}^{v,h}_m)
-({\bf S}^{v,v}_m+{\bf e}_m)\times{\bf S}^{v,\omega}_k
+{\bf S}^{v,h}_m\times(\nabla_{\bf x}\times{\bf S}^{v,h}_k)\right),
\endB{eq46p1}{auxIIIvv}\begI
{\cal M}^h({\bf Q}^{vv,\omega}_{m,k},{\bf Q}^{vv,h}_{m,k})=-\rho_{m,k}\nabla_{\bf x}\times
\left(({\bf S}^{v,v}_k+{\bf e}_k)\times{\bf S}^{v,h}_m
+({\bf S}^{v,v}_m+{\bf e}_m)\times{\bf S}^{v,h}_k\right),
\endI{eq46p3}\begI
{\cal M}^\theta({\bf Q}^{vv,\omega}_{m,k},Q^{vv,\theta}_{m,k})=\rho_{m,k}\left(
(({\bf S}^{v,v}_k+{\bf e}_k)\cdot\nabla_{\bf x})S^{v,\theta}_m
+(({\bf S}^{v,v}_m+{\bf e}_m)\cdot\nabla_{\bf x})S^{v,\theta}_k\right),
\endI{eq46p4}\begI
\nabla_{\bf x}\times{\bf Q}^{vv,v}_{m,k}={\bf Q}^{vv,\omega}_{m,k};
\endL{eq46p2}
$$\left.{\cal M}^\omega({\bf Q}^{vh,\omega}_{m,k},{\bf Q}^{vh,h}_{m,k},Q^{vh,\theta}_{m,k})=
-\nabla_{\bf x}\times\right(({\bf S}^{v,v}_k+{\bf e}_k)\times{\bf S}^{h,\omega}_m
+{\bf S}^{h,v}_m\times{\bf S}^{v,\omega}_k$$
\begB
\left.-{\bf S}^{v,h}_k\times(\nabla_{\bf x}\times{\bf S}^{h,h}_m)
-({\bf S}^{h,h}_m+{\bf e}_m)\times(\nabla_{\bf x}\times{\bf S}^{v,h}_k)\right),
\endB{eq47p1}{auxIIIvh}\begI
{\cal M}^h({\bf Q}^{vh,\omega}_{m,k},{\bf Q}^{vh,h}_{m,k})=-\nabla_{\bf x}\times
\left(({\bf S}^{v,v}_k+{\bf e}_k)\times({\bf S}^{h,h}_m+{\bf e}_m)+
{\bf S}^{h,v}_m\times{\bf S}^{v,h}_k\right),
\endI{eq47p3}\begI
{\cal M}^\theta({\bf Q}^{vh,\omega}_{m,k},Q^{vh,\theta}_{m,k})=
(({\bf S}^{v,v}_k+{\bf e}_k)\cdot\nabla_{\bf x})S^{h,\theta}_m
+({\bf S}^{h,v}_m\cdot\nabla_{\bf x})S^{v,\theta}_k,
\endI{eq47p4}\begI
\nabla_{\bf x}\times{\bf Q}^{vh,v}_{m,k}={\bf Q}^{vh,\omega}_{m,k};
\endL{eq47p2}
$${\cal M}^\omega({\bf Q}^{hh,\omega}_{m,k},{\bf Q}^{hh,h}_{m,k},Q^{hh,\theta}_{m,k})=
\rho_{m,k}\nabla_{\bf x}\times\left(-{\bf S}^{h,v}_k\times{\bf S}^{h,\omega}_m\right.$$
\begB
\left.+({\bf S}^{h,h}_k+{\bf e}_k)\times(\nabla_{\bf x}\times{\bf S}^{h,h}_m)
-{\bf S}^{h,v}_m\times{\bf S}^{h,\omega}_k
+({\bf S}^{h,h}_m+{\bf e}_m)\times(\nabla_{\bf x}\times{\bf S}^{h,h}_k)\right),
\endB{eq48p1}{auxIIIhh}\begI
{\cal M}^h({\bf Q}^{hh,\omega}_{m,k},{\bf Q}^{hh,h}_{m,k})=-\rho_{m,k}\nabla_{\bf x}\times
\left({\bf S}^{h,v}_k\times({\bf S}^{h,h}_m+{\bf e}_m)
+{\bf S}^{h,v}_m\times({\bf S}^{h,h}_k+{\bf e}_k)\right),
\endI{eq48p3}\begI
{\cal M}^\theta({\bf Q}^{hh,\omega}_{m,k},Q^{hh,\theta}_{m,k})=\rho_{m,k}\left(
({\bf S}^{h,v}_k\cdot\nabla_{\bf x})S^{h,\theta}_m
+({\bf S}^{h,v}_m\cdot\nabla_{\bf x})S^{h,\theta}_k\right),
\endI{eq48p4}\begI
\nabla_{\bf x}\times{\bf Q}^{hh,v}_{m,k}={\bf Q}^{hh,\omega}_{m,k};
\endL{eq48p2}
\begin{equation}
\nabla_{\bf x}\cdot{\bf Q}^{\cdot\cdot,\cdot}_{m,k}=0
\label{eq49}\end{equation}
($\rho_{m,k}=1$ for $m<k$, $\rho_{m,k}=1/2$ for $m=k$, and $\rho_{m,k}=0\,
\Rightarrow\,{\bf Q}^{vv,\cdot}_{m,k}={\bf Q}^{hh,\cdot}_{m,k}=0$ for $m>k$);
${\bf Y}^{\cdot,\cdot}_{m,k}=({\bf Y}^{\cdot,\omega}_{m,k},
{\bf Y}^{\cdot,v}_{m,k},{\bf Y}^{\cdot,h}_{m,k},Y^{\cdot,\theta}_{m,k})$
solve {\it auxiliary problems of type IV}:
$${\cal M}({\bf Y}^{v,\omega}_{m,k},{\bf Y}^{v,h}_{m,k},Y^{v,\theta}_{m,k})
=-\rho_{m,k}\left(\lbdb\int_0^t(\alpha^v_{m,1}-\alpha^v_{m,2})dt\rbdb
{\partial\over\partial x_k}\right.$$
\begB
\left.+\lbdb\int_0^t(\alpha^v_{k,1}-\alpha^v_{k,2})dt\rbdb
{\partial\over\partial x_m}\right)(\bOm,{\bf H},\Theta),
\endB{eq50p1}{auxIVv}

\begI
\nabla_{\bf x}\times{\bf Y}^{v,v}_{m,k}={\bf Y}^{v,\omega}_{m,k};
\endL{eq50p2}
$${\cal M}({\bf Y}^{h,\omega}_{m,k},{\bf Y}^{h,h}_{m,k},Y^{h,\theta}_{m,k})
=-\rho_{m,k}\left(\lbdb\int_0^t(\alpha^h_{m,1}-\alpha^h_{m,2})dt\rbdb
{\partial\over\partial x_k}\right.$$
\begB
\left.+\lbdb\int_0^t(\alpha^h_{k,1}-\alpha^h_{k,2})dt\rbdb
{\partial\over\partial x_m}\right)(\bOm,{\bf H},\Theta),
\endB{eq51p1}{auxIVh}

\begI
\nabla_{\bf x}\times{\bf Y}^{h,v}_{m,k}={\bf Y}^{h,\omega}_{m,k};
\endL{eq51p2}
\begin{equation}
\nabla_{\bf x}\cdot{\bf Y}^{\cdot,\cdot}_{m,k}=0
\label{eq52}\end{equation}
(${\bf Y}^{\cdot,\cdot}_{m,k}=0$ for $m>k$);
and $\bxi^\cdot_1=(\bxi^\omega_1,\bxi^v_1,\bxi^h_1,\xi^\theta_1)$ solve the problem
$${\cal L}^\omega(\bxi^\omega_1,\bxi^v_1,\bxi^h_1,\xi^\theta_1)
=-2\nu(\nabla_{\bf x}\cdot\nabla_{\bf X})\bxi^\omega_0
+\nabla_{\bf x}\times({\bf H}\times(\nabla_{\bf X}\times\bxi^h_0))
-\nabla_{\bf X}\times\lpar{\bf V}\times\bxi^\omega_0$$
$$+\bxi^v_0\times\bOm
-{\bf H}\times(\nabla_{\bf x}\times\bxi^h_0)-\bxi^h_0\times(\nabla_{\bf x}\times{\bf H})\rpar
-\nabla_{\bf x}\times\lpar{\bf v}_0\times\bxi^\omega_0+\bxi^v_0\times(\bom_0-\bxi^\omega_0)$$
\begB
-{\bf h}_0\times(\nabla_{\bf x}\times\bxi^h_0)
-\bxi^h_0\times(\nabla_{\bf x}\times({\bf h}_0-\bxi^h_0))\rpar
-\beta\nabla_{\bf X}\xi^\theta_0\times{\bf e}_3,
\endB{eq53p1}{IIxi}
$${\cal L}^h(\bxi^v_1,\bxi^h_1)=-2\eta(\nabla_{\bf x}\cdot\nabla_{\bf X})\bxi^h_1
-\nabla_{\bf X}\times(\bxi^v_0\times{\bf H}+{\bf V}\times\bxi^h_0)$$
\begI
-\nabla_{\bf x}\times({\bf v}_0\times\bxi^h_0+\bxi^v_0\times({\bf h}_0-\bxi^h_0))
\endI{eq53p5}\begI
{\cal L}^\theta(\bxi^v_1,\xi^\theta_1)
=-2\kappa(\nabla_{\bf x}\cdot\nabla_{\bf X})\xi^\theta_0
+({\bf V}\cdot\nabla_{\bf X})\xi^\theta_0+({\bf v}_0\cdot\nabla_{\bf x})\xi^\theta_0
+(\bxi^v_0\cdot\nabla_{\bf x})(\theta_0-\xi^\theta_0),
\endI{eq53p7}\begI
\nabla_{\bf x}\times\bxi^v_1-\bxi^\omega_1=-\nabla_{\bf X}\times\bxi^v_0,
\endI{eq53p3}\begI
\nabla_{\bf x}\cdot\bxi_1^\omega=-\nabla_{\bf X}\cdot\bxi_0^\omega,\quad
\nabla_{\bf x}\cdot\bxi_1^v=-\nabla_{\bf X}\cdot\bxi_0^v,\quad
\nabla_{\bf x}\cdot\bxi_1^h=-\nabla_{\bf X}\cdot\bxi_0^h.
\endL{eq53sol}
${\bf G}^{\cdot,\cdot}_{m,k},{\bf Q}^{\cdot\cdot,\cdot}_{m,k},
{\bf Y}^{\cdot,\cdot}_{m,k}$ and $\bxi^\cdot_1$ must satisfy the boundary
conditions similar to (\theBCs) and \rf{eq4}. Means over fast spatial variables
of their horizontal flow components and of vertical vorticity components must
vanish, as well as spatio-temporal means of horizontal magnetic components.
Clearly, the auxiliary problems are particular cases of the problem (\thelinePRO).

Equations \rf{eq46p1}, \rf{eq47p1} and \rf{eq48p1} are equivalent to
$$\left.{\cal L}^v({\bf Q}^{vv,v}_{m,k},{\bf Q}^{vv,h}_{m,k},Q^{vv,\theta}_{m,k},Q^{vv,p}_{m,k})=
\rho_{m,k}\right(-({\bf S}^{v,v}_k+{\bf e}_k)\times{\bf S}^{v,\omega}_m$$
\begin{equation}
\left.+{\bf S}^{v,h}_k\times(\nabla_{\bf x}\times{\bf S}^{v,h}_m)
-({\bf S}^{v,v}_m+{\bf e}_m)\times{\bf S}^{v,\omega}_k
+{\bf S}^{v,h}_m\times(\nabla_{\bf x}\times{\bf S}^{v,h}_k)\right),
\label{EQ26p1p}\end{equation}
$${\cal L}^v({\bf Q}^{vh,v}_{m,k},{\bf Q}^{vh,h}_{m,k},Q^{vh,\theta}_{m,k},Q^{vh,p}_{m,k})=
-({\bf S}^{v,v}_k+{\bf e}_k)\times{\bf S}^{h,\omega}_m
-{\bf S}^{h,v}_m\times{\bf S}^{v,\omega}_k$$
\begin{equation}
+{\bf S}^{v,h}_k\times(\nabla_{\bf x}\times{\bf S}^{h,h}_m)
+({\bf S}^{h,h}_m+{\bf e}_m)\times(\nabla_{\bf x}\times{\bf S}^{v,h}_k),
\label{EQ27p1p}\end{equation}
$${\cal L}^v({\bf Q}^{hh,v}_{m,k},{\bf Q}^{hh,h}_{m,k},Q^{hh,\theta}_{m,k},Q^{hh,p}_{m,k})=
\rho_{m,k}\left(-{\bf S}^{h,v}_k\times{\bf S}^{h,\omega}_m\right.
+({\bf S}^{h,h}_k+{\bf e}_k)\times(\nabla_{\bf x}\times{\bf S}^{h,h}_m)$$
\begin{equation}
\left.-{\bf S}^{h,v}_m\times{\bf S}^{h,\omega}_k
+({\bf S}^{h,h}_m+{\bf e}_m)\times(\nabla_{\bf x}\times{\bf S}^{h,h}_k)\right),
\label{EQ28p1p}\end{equation}
respectively, where $\la\nabla_{\bf x}Q^{\cdot\cdot,p}_{m,k}\ra_h=0$.

Together, \rf{eq44sol}, \rf{eq45sol}, \rf{eq49}, \rf{eq52} and \rf{eq53sol} are
equivalent to \rf{eq17p2}, \rf{eq17p3} and \rf{eq17p5} for $n=1$. It suffices
to impose these conditions for the vorticity and magnetic parts of the
\pagebreak
solutions at $t=0$, since the compatibility conditions \rf{eq9p4}, \rf{eq9p5}
evidently hold true (for the auxiliary problems of type II this can be
deduced from \rf{eq25p1}, \rf{eq25p2}, \rf{eq26p1} and \rf{eq26p2}~).
Similarly, equations \rf{eq44p3}, \rf{eq45p3}, \rf{eq46p2}, \rf{eq47p2},
\rf{eq48p2}, \rf{eq50p2}, \rf{eq51p2} and \rf{eq53p3} are equivalent to
equation \rf{eq30r} for $n=1$.

The spatial means of the initial conditions for the problems
(\theauxIIv)-(\theIIxi) can be determined averaging \rf{initave} over fast time:
\begin{equation}
\phan\la{\bf G}^{v,h}_{m,k}\ra_h\right|_{t=0}
=-\ladb{\bf e}_m\times\int_0^t\bal^v_k\,dt\radb_{\!h}\!,\qquad
\phan\la{\bf G}^{h,h}_{m,k}\ra_h\right|_{t=0}
=-\ladb{\bf e}_m\times\int_0^t\bal^h_k\,dt\radb_{\!h}\!,
\label{EQbegin0}\end{equation}
$$\phan\la\bxi^h_1\ra_h\right|_{t=0}=-\ladb\int_0^t\nabla_{\bf X}\times\txi{h}\,dt\radb_{\!h}$$
(the means exist by the assumption that the magnetic $\alpha-$effect is
insignificant); at any $t\ge0$
\begin{equation}
\la{\bf G}^{\cdot,\omega}_{m,k}\ra_v=\la{\bf Q}^{\cdot\cdot,\omega}_{m,k}\ra_v
=\la{\bf Y}^{\cdot,\omega}_{m,k}\ra_v=\la\bxi^\omega_1\ra_v
=\la{\bf Q}^{\cdot\cdot,h}_{m,k}\ra_h=\la{\bf Y}^{\cdot,h}_{m,k}\ra_h=0.
\label{EQbegin2}\end{equation}

An initial condition for mean vorticity is determined by \rf{eq21p2}:
$$\phan\lad\bom_1\rad_v\right|_{T=0}\phan=\la\bom_1\ra_v\right|_{t=0}.$$
Averaging the magnetic part of \rf{eq43} over spatial variables at $t=0$, find
$$\phan\lad{\bf h}_1\rad_h\right|_{T=0}
\phan=\la{\bf h}_1\ra_h\right|_{t=0}\phan-\la\bxi^h_1\ra_h\right|_{t=0}$$
$$-\sum_{k=1}^2\sum_{m=1}^2\left(
\phan\la{\bf G}^{v,h}_{m,k}\ra_h\right|_{t=0}{\partial\lad v_0\rad_k\over\partial X_m}\!\phan\right|_{T=0}\!\!
\phan+\la{\bf G}^{h,h}_{m,k}\ra_h\right|_{t=0}{\partial\lad h_0\rad_k\over\partial X_m}\!\phan\right|_{T=0}\right).$$
Now initial conditions for the problem (\theIIxi) can be determined
from \rf{eq43} at $t=0$.

The choice of initial data must ensure that all solutions to the auxiliary
problems are globally bounded with their derivatives. Generically, such
solutions do exist, if the perturbed CHM state ${\bf V,H},\Theta$ is periodic
in horizontal directions and steady or time-periodic (see section 4); solutions
are globally bounded for any smooth initial conditions, if the perturbed CHM
state ${\bf V,H},\Theta$ is linearly stable to
small-scale perturbations (see appendix C). Modification of initial
conditions for ${\bf G}^{\cdot,\cdot}_{m,k}$,
${\bf Q}^{\cdot\cdot,\cdot}_{m,k}$ and ${\bf Y}^{\cdot,\cdot}_{m,k}$ to other
permissible ones implies the respective changes in initial conditions for
$\bxi^\cdot_1$. These changes must belong to the stable manifold of the perturbed
CHM state ${\bf V,H},\Theta$, so that the resultant changes in solutions to the
auxiliary problems and in $\bxi^\cdot_1$ decay exponentially in fast time.

Since $\bxi^\cdot_0$ decay exponentially in time together with derivatives
(see section 5), the right-hand sides of equations \xrf{eq53p1}{eq53sol} also
decay exponentially. As shown in appendices D and B for mean parts of the
relevant components of $\bxi^\cdot_0$ and for the complementary fluctuating
part, respectively, this implies an exponential decay of $\bxi^\cdot_1$, if the
CHM state ${\bf V,H},\Theta$ is linearly stable to small-scale perturbations.
Similarly, under this condition, any changes in
${\bf G}^{\cdot,\cdot}_{m,k}$, ${\bf Q}^{\cdot\cdot,\cdot}_{m,k}$ and
${\bf Y}^{\cdot,\cdot}_{m,k}$ due to a permissible variation of the initial
data for ${\bf S}^{\cdot,\cdot}_k$ exponentially decay. Otherwise, an exponential decay
of $\bxi^\cdot_1$ must be ensured by an appropriate choice of the initial data.

If the perturbed CHM regime has a symmetry considered in section 7,
${\bf S}^{\cdot,\cdot}_k$ are supposed to be antisymmetric sets of fields and
therefore the right-hand sides of equations defining auxiliary problems
\pagebreak
of types II--V are symmetric sets. Due to invariance for the
operator $\cal L$ of the subspaces of symmetric and antisymmetric sets in the
presence of a symmetry, essentially ${\bf G}^{\cdot,\cdot}_{m,k}$,
${\bf Q}^{\cdot\cdot,\cdot}_{m,k}$ and ${\bf Y}^{\cdot,\cdot}_{m,k}$ are
symmetric sets (by construction the antisymmetric part of any permissible
solution to the problems (\theauxIIv)-(\theauxIVh) exponentially decays).
Then conditions \rf{EQbegin0} and \rf{EQbegin2} are automatically satisfied
(except for vanishing of the mean of the vertical component of vorticity
is not implied by parity invariance of the perturbed state).
If the perturbed CHM state is steady or its symmetry is without a time shift,
then the right-hand sides of auxiliary problems of type IV are zero and hence
${\bf Y}^{\cdot,\cdot}_{m,k}=0$.

If the perturbed CHM regime is steady, periodic or quasiperiodic in fast time
and/or spatial variables, it is natural to demand (in particular, for
convenience of computation of spatio-temporal averages), that
${\bf S}^{\cdot,\cdot}_k$, ${\bf G}^{\cdot,\cdot}_{m,k}$,
${\bf Q}^{\cdot\cdot,\cdot}_{m,k}$ and ${\bf Y}^{\cdot,\cdot}_{m,k}$ are
steady, or have the same periodicity or the same set of basic frequencies,
respectively. Then $\bxi^\cdot_0$ and $\bxi^\cdot_1$ are decaying transients,
bringing perturbations of the CHM system (in fast time) to their regular regime.

\mi{\bf 9. Order $\varepsilon^2$ and $\varepsilon^3$ equations}

\mi
Averaging the vertical component of (A1) for $n=3$ over fast spatial variables
taking into account \rf{eq11p1}, \xrf{eq17p1}{eq17p3} and \rf{eq18} for $n=0,1,2$,
the boundary conditions for $\bf V,\ H$, ${\bf v}_i$ and ${\bf h}_i$
and solenoidality of ${\bf v}_0$ and ${\bf h}_0$, find
$$-{\partial\la\bom_3\ra_v\over\partial t}-{\partial\la\bom_1\ra_v\over\partial T}
+\nu\nabla^2_{\bf X}\la\bom_1\ra_v$$
$$+\,\nabla_{\bf X}\times\la{\bf V}\times(\nabla_{\bf X}\times{\bf v}_1)
-{\bf V}\nabla_{\bf X}\cdot\lb{\bf v}_1\rb_h
-{\bf H}\times(\nabla_{\bf X}\times{\bf h}_1)
+{\bf H}\nabla_{\bf X}\cdot\lb{\bf h}_1\rb_h$$
\begin{equation}
+\,{\bf v}_0\times(\nabla_{\bf X}\times{\bf v}_0)
-{\bf v}_0\nabla_{\bf X}\cdot\lb{\bf v}_0\rb_h
-{\bf h}_0\times(\nabla_{\bf X}\times{\bf h}_0)
+{\bf h}_0\nabla_{\bf X}\cdot\lb{\bf h}_0\rb_h\ra_h=0.
\label{eq58}\end{equation}
Averaging this equation over fast time, substituting the
flow and magnetic components of \rf{eq23} and \rf{eq43},
employing \rf{eq40}, \rf{eq19} for $n=1$ and \rf{eq34}, and recalling that
$\bxi^\cdot_0$ and $\bxi^\cdot_1$ decay exponentially, find
$$\nabla_{\bf X}\times\left(-{\partial\over\partial T}\lad{\bf v}_0\rad_h
+\nu\nabla^2_{\bf X}\lad{\bf v}_0\rad_h\right.
-(\lad{\bf v}_0\rad_h\cdot\nabla_{\bf X})\lad{\bf v}_0\rad_h
+(\lad{\bf h}_0\rad_h\cdot\nabla_{\bf X})\lad{\bf h}_0\rad_h$$
$$+\sum_{j=1}^2\sum_{m=1}^2\sum_{k=1}^2{\partial\over\partial X_j}\left({\partial\over\partial X_m}
\right({\bf D}^{v,v}_{m,k,j}\lad v_0\rad_k+{\bf D}^{h,v}_{m,k,j}\lad h_0\rad_k
+{\partial^2\over\partial X_k\partial X_1}\nabla^{-2}_{\bf X}
\left({\bf d}^{v,v}_{m,k,j}\lad v_0\rad_1\right.$$
\begin{equation}
\left.+\,{\bf d}^{h,v}_{m,k,j}\lad h_0\rad_1\right)\!\!\left)
\left.\phantom{\partial\over\partial_j}\!\!\!\!\!\!\!\!\!
+\!{\bf A}^{vv,v}_{m,k,j}\lad v_0\rad_k\lad v_0\rad_m
\!+\!{\bf A}^{vh,v}_{m,k,j}\lad v_0\rad_k\lad h_0\rad_m
\!+\!{\bf A}^{hh,v}_{m,k,j}\lad h_0\rad_k\lad h_0\rad_m\right)\!\right)\!=0,
\label{eq59}\end{equation}
where
\begB
{\bf D}^{v,v}_{m,k,j}=\lad-V_j\widetilde{\bf G}^{v,v}_{m,k}-{\bf V}(G^{v,v}_{m,k})_j
+H_j{\bf G}^{v,h}_{m,k}+{\bf H}(G^{v,h}_{m,k})_j\rad_h,
\endB{eq60p1}{eddyvisc}

\begI
{\bf d}^{v,v}_{m,k,j}=\lad-V_j\widetilde{\bf Y}^{v,v}_{m,k}-{\bf V}(Y^{v,v}_{m,k})_j
+H_j{\bf Y}^{v,h}_{m,k}+{\bf H}(Y^{v,h}_{m,k})_j\rad_h,
\endI{eq60p2}\vspace*{-1mm}\begI
\widetilde{\bf G}^{v,v}_{m,k}={\bf G}^{v,v}_{m,k}+\lbdb\int_0^t\alpha^v_{m,k}dt\rbdb{\bf e}_k,\quad
\widetilde{\bf Y}^{v,v}_{m,k}={\bf Y}^{v,v}_{m,k}-\lbdb\int_0^t(\alpha^v_{m,1}-\alpha^v_{m,2})\,dt\rbdb{\bf e}_k;
\endL{eq60p3}

\noindent
changing in (\theeddyvisc) every occurrence of the first superscript ``$v$'' to ``$h$'',
obtain expressions for ${\bf D}^{h,v}_{m,k,j}$ and ${\bf d}^{h,v}_{m,k,j}$;
\pagebreak
\begB
{\bf A}^{vv,v}_{m,k,j}=\lad-V_j{\bf Q}^{vv,v}_{m,k}-{\bf V}(Q^{vv,v}_{m,k})_j
+H_j{\bf Q}^{vv,h}_{m,k}+{\bf H}(Q^{vv,h}_{m,k})_j
-(S^{v,v}_k)_j{\bf S}^{v,v}_m+(S^{v,h}_k)_j{\bf S}^{v,h}_m\rad_h,
\endB{eq61p1}{eddyadv}

$${\bf A}^{vh,v}_{m,k,j}=\lad-V_j{\bf Q}^{vh,v}_{m,k}-{\bf V}(Q^{vh,v}_{m,k})_j
+H_j{\bf Q}^{vh,h}_{m,k}+{\bf H}(Q^{vh,h}_{m,k})_j$$
\begI
-(S^{v,v}_k)_j{\bf S}^{h,v}_m-(S^{h,v}_m)_j{\bf S}^{v,v}_k
+(S^{v,h}_k)_j{\bf S}^{v,h}_m+(S^{h,h}_m)_j{\bf S}^{v,h}_k\rad_h,
\endI{eq61p2}

\begI
{\bf A}^{hh,v}_{m,k,j}=\lad-V_j{\bf Q}^{hh,v}_{m,k}-{\bf V}(Q^{hh,v}_{m,k})_j
+H_j{\bf Q}^{hh,h}_{m,k}+{\bf H}(Q^{hh,h}_{m,k})_j
-(S^{h,v}_k)_j{\bf S}^{h,v}_m+(S^{h,h}_k)_j{\bf S}^{h,h}_m\rad_h.
\endL{eq61p3}

Averaging the horizontal component of (A2) for $n=2$ over fast variables,
substituting (\theavevI) and flow and magnetic components of \rf{eq23} and \rf{eq43}, recalling
that $\bxi^\cdot_0$ and $\bxi^\cdot_1$ decay exponentially and taking into
account relations \rf{eq11p3}, \rf{eq37} and the boundary conditions
for $\bf V,\ H$, ${\bf v}_n$ and ${\bf h}_n$, find
$$\left.-{\partial\over\partial T}\lad{\bf h}_0\rad_h+\eta\nabla^2_{\bf X}\lad{\bf h}_0\rad_h
+\nabla_{\bf X}\times\right(\lad{\bf v}_0\rad_h\times\lad{\bf h}_0\rad_h$$
$$+\!\sum_{m=1}^2\sum_{k=1}^2\!\left({\partial\over\partial X_m}\left(
{\bf D}^{v,h}_{m,k}\lad v_0\rad_k+{\bf D}^{h,h}_{m,k}\lad h_0\rad_k
+{\partial^2\over\partial X_k\partial X_1}\nabla^{-2}_{\bf X}\left(
{\bf d}^{v,h}_{m,k}\lad v_0\rad_1
+{\bf d}^{h,h}_{m,k}\lad h_0\rad_1\right)\!\right)\right.$$
\begin{equation}
\left.\left.\phantom{\partial\over\partial}
+{\bf A}^{vv,h}_{m,k}\lad v_0\rad_k\lad v_0\rad_m
+{\bf A}^{vh,h}_{m,k}\lad v_0\rad_k\lad h_0\rad_m
+{\bf A}^{hh,h}_{m,k}\lad h_0\rad_k\lad h_0\rad_m\right)\right)=0,
\label{eq62}\end{equation}
where
\begin{equation}
{\bf D}^{v,h}_{m,k}=\lad{\bf V}\times{\bf G}^{v,h}_{m,k}-{\bf H}\times\widetilde{\bf G}^{v,v}_{m,k}\rad_v,\qquad
{\bf d}^{v,h}_{m,k}=\lad{\bf V}\times{\bf Y}^{v,h}_{m,k}-{\bf H}\times
\widetilde{\bf Y}^{v,v}_{m,k}\rad_v;
\label{eq63p2}\end{equation}
changing here every occurrence of the first superscript ``$v$'' to ``$h$'',
obtain expressions for ${\bf D}^{h,h}_{m,k}$ and ${\bf d}^{h,h}_{m,k}$;
\begB
{\bf A}^{vv,h}_{m,k}=\lad{\bf V}\times{\bf Q}^{vv,h}_{m,k}-{\bf H}\times{\bf Q}^{vv,v}_{m,k}
+{\bf S}^{v,v}_k\times{\bf S}^{v,h}_m\rad_v,
\endB{eq64p1}{eddyadh}\begI
{\bf A}^{vh,h}_{m,k}=\lad{\bf V}\times{\bf Q}^{vh,h}_{m,k}-{\bf H}\times{\bf Q}^{vh,v}_{m,k}
+{\bf S}^{v,v}_k\times{\bf S}^{h,h}_m+{\bf S}^{h,v}_m\times{\bf S}^{v,h}_k\rad_v,
\endI{eq64p2}\begI
{\bf A}^{hh,h}_{m,k}=\lad{\bf V}\times{\bf Q}^{hh,h}_{m,k}-{\bf H}\times{\bf Q}^{hh,v}_{m,k}
+{\bf S}^{h,v}_k\times{\bf S}^{h,h}_m\rad_v.
\endL{eq64p3}

${\bf D}^{\cdot,\cdot}$ are coefficients of the second order operators
representing the so-called {\it aniso\-tropic combined eddy diffusivity}.
Simpler expressions for the operators are presented in appendix E.
${\bf d}^{\cdot,\cdot}$ are coefficients of pseudodifferential operators,
formally also of the second order, which can be regarded as representing an
unconventional {\it non-local anisotropic combined eddy diffusivity}. All
${\bf d}^{\cdot,\cdot}$ vanish, if the perturbed CHM state is steady or
possesses a symmetry without a time shift of a type discussed in section 7.
To the best of our knowledge, such a physical effect was not encountered
before. ${\bf A}^{\cdot\cdot,\cdot}$ are coefficients of quadratic terms
representing the so-called {\it combined eddy advection}.

Equations \rf{eq59} and \rf{eq62} are solvability conditions for the
system (A1)--(A3) for $n=2$. To solve the order $\varepsilon^2$ equations,
determine from \rf{eq58} an expression for $\la\bom_3\ra_v$ similar to
\rf{eq39}, then from \rf{eq17p1} and \rf{eq19} for $n=2$ deduce an expression
for $\la{\bf v}_2\ra_h$ similar to (\theavevI), and finally solve the remaining
parts of equations (A1)--(A3) for $n=2$ (the fluctuating $\lb\cdot\rb_v$
\pagebreak
part of (A1), fluctuating $\lbd\cdot\rbd_h$ part of (A2), and (A3)~) to
find expressions for $\lb\bom_2\rb_v$, $\lb{\bf v}_2\rb_h$,
$\lbd{\bf h}_2\rbd_h$ and $\theta_2$ analogous to \rf{eq43}. It is possible to
construct in a similar way solutions to the systems arising for higher orders
of $\varepsilon^n$, and thus to obtain subsequent terms in expansions
(\thepertser).

\rf{eq59} and \rf{eq62} constitute a closed system of equations for mean
perturbations (more precisely, for the leading terms in their expansions in
power series in the spatial scale ratio). The vertical component of \rf{eq62}
and the horizontal component of \rf{eq59} vanish identically. One can
``uncurl'' \rf{eq59}, introducing the large-scale pressure. However, since
$\lad{\bf v}_0\rad_h$ and $\lad{\bf h}_0\rad_h$ are solenoidal in slow
variables, it is more natural to solve this system numerically introducing
stream functions for the two-dimensional flow $\lad{\bf v}_0\rad_h$ and
magnetic field $\lad{\bf h}_0\rad_h$ and considering vertical components
of \rf{eq59} and of the curl of \rf{eq62}.

Note that the Coriolis force enters only in the statement of the auxiliary
problems and not in the mean-field equations \rf{eq59}. Derivation of these
equations remains unaffected, if additionally the fluid is rotating in slow
time, and the total angular velocity is $\tau{\bf e}_3+\varepsilon^2\btau$.
Then the mean-field equation \rf{eq59} for the flow involves an additional term
representing the Coriolis force in the fluid layer rotating with the angular
velocity $\btau$.

\mi{\bf 10. Mean-field equations with $\alpha-$effect terms:\\
large-scale perturbations of CHM regimes near a Hopf bifurcation}

\mi
As discussed in section 6, construction of the system of mean-field equations
requires that the kinematic and magnetic $\alpha-$effects are insignificant
in the leading order. However, in the considered problem this implies
the absence of $\alpha-$effect terms in \rf{eq59} and \rf{eq62}. In this
section we consider perturbations of CHM regimes constituting
branches parametrised by a small parameter, and show that if this parameter is
linked with the scale ratio $\varepsilon$, then the $\alpha-$effect can remain
insignificant in the leading order, but emerge in the mean-field equations.

Let a family of CHM regimes admit an expansion in asymptotic power series:
\begin{equation}
\lpar\bOm,{\bf V,H},\Theta\rpar=\sum_{n=0}^\infty\lpar
\bOm_n({\bf x},t),{\bf V}_n({\bf x},t),{\bf H}_n({\bf x},t),
\Theta_n({\bf x},t)\rpar\varepsilon^n,
\label{eq65}\end{equation}
$$\bOm_n=\nabla_{\bf x}\times{\bf V}_n.$$
It can stem from similar expansions for the source terms
$\bf F,\ J$ and $S$ in (\theCHMB). Perhaps, a more important example is
a branch emerging from the CHM regime ${\bf V}_0,{\bf H}_0,\Theta_0$ (not
necessarily steady) in a Hopf bifurcation, occurring\footnote{The quantity $\beta$,
proportional to the Rayleigh number, is chosen here as the bifurcation
para\-meter, because investigation of bifurcations in convection happening when
this number is increased is a popular research subject (see, e.g.,
Podvigina 2006). Variation of any other quantity resulting in a Hopf
bifurcation, e.g., $\tau$ proportional to the Taylor number, can be
considered; then the mean-field equations are not altered and there are minor
modifications in the statements of auxiliary problems.} at $\beta=\beta_0$
when two complex conjugate eigenvalues of $\cal L$ cross the real axis.
Then the family of CHM regimes admits\footnote{Unless the case is not generic,
e.g., if the CHM state ${\bf V}_0,{\bf H}_0,\Theta_0$ has a large group
of symmetries.} the expansion \rf{eq65} for
\begin{equation}
\beta=\beta_0+\beta_2\varepsilon^2
\label{eq66}\end{equation}
(see, e.g., Guckenheimer and Holmes 1990). The deviation of the emerging CHM
regimes from the one at the critical point $\beta=\beta_0$ is asymptotically
close to a sum of eigenvectors associated with the pair of imaginary
eigenvalues of $\cal M$ at this point, i.e.
$\bOm_1,{\bf V}_1,{\bf H}_1,\Theta_1$ belongs to the subspace spanned by
these eigenvectors (which is generically two-dimensional).

We assume \rf{eq66}, where positive and negative $\beta_2$ correspond to
a direct and reverse Hopf bifurcation, respectively, or, for $\beta_2=0$,
the expansion \rf{eq65} is due to dependence of the source terms $\bf F,\ J$
and $S$ on $\varepsilon$. Everywhere in this and next section we refer to
the operators $\cal L$ and $\cal M$ in which $\bOm_0$, ${\bf V}_0$, ${\bf H}_0$,
$\Theta_0$ and $\beta_0$ are substituted in place of $\bOm,{\bf V,H},\Theta$
and $\beta$ in (\thedefL) and (\theMdef). The same substitution is assumed
in any equation and definition of any quantity introduced in previous sections
and referred to here or in the next section. Large-scale perturbations satisfy
(\theEQpert) and are sought in the form of the series (\thepertser).
Construction of the mean-field equations differs from that for
$\varepsilon-$independent CHM regimes only in a slight complication of algebra:
new terms involving higher-order terms $\bOm_n$, ${\bf V}_n$, ${\bf H}_n$ and
$\Theta_n$ for $n\ge 1$ emerge in the right-hand sides of the equations to be
solved.

Unless specifically mentioned, the CHM regime ${\bf V}_0,{\bf H}_0,\Theta_0$
at the point of bifurcation is not required to be symmetric. If it has
a symmetry considered in section 7, then symmetric and antisymmetric sets of
vector fields constitute invariant subspaces of $\cal M$, and generically
$\bOm_1,{\bf V}_1,{\bf H}_1,\Theta_1$ is either a symmetric, or antisymmetric set.

Substitution of (\thepertser) into (\theEQpert) gives rise to equations
(A4)--(A6) (see appendix~A). Using (A4) for $n=0$ and (A5) for $n=1$,
(\theinitmean) can be verified. The equations at order $\varepsilon^0$ coincide
with \rf{eq20}. Their solutions have the structure \rf{eq23}, where vector
fields ${\bf S}^{\cdot,\cdot}_k({\bf x},t)$ satisfy (\theomegaSv) and solve the
auxiliary problems (\theIauxiv) and (\theIauxih), and exponentially decaying
mean-free transients $\bxi^\cdot_0({\bf x},t,{\bf X},T)$ solve (\theIxi).
If the CHM state ${\bf V}_0,{\bf H}_0,\Theta_0$ possesses a symmetry of a type
considered in section 7, then ${\bf S}^{\cdot,\cdot}_k$ are antisymmetric sets,
and $\alpha$-effect is automatically insignificant in the leading order.

Averaging of the vertical component of (A4) for $n=2$ yields \rf{eq29} and,
consequently, $\la{\bf v}_1\ra_h$ is determined by (\theavevI),
provided the $\alpha-$effect is insignificant in the leading order.
Therefore, solutions to equations (A4)--(A6) for $n=1$ have the structure
$$\left.(\lb\bom_1\rb_v,\lb{\bf v}_1\rb_h,\lbd{\bf h}_1\rbd_h,\theta_1)
=\bxi_1^\cdot+\sum_{k=1}^2\right(
{\bf S}^{v,\cdot}_k\lad v_1\rad_k+{\bf S}^{h,\cdot}_k\lad h_1\rad_k
+\widehat{\bf S}^{v,\cdot}_k\lad v_0\rad_k+\widehat{\bf S}^{h,\cdot}_k\lad h_0\rad_k$$
$$\left.+\sum_{m=1}^2\right({\bf G}^{v,\cdot}_{m,k}{\partial\lad v_0\rad_k\over\partial X_m}
+{\bf G}^{h,\cdot}_{m,k}{\partial\lad h_0\rad_k\over\partial X_m}
+{\bf Q}^{vv,\cdot}_{m,k}\lad v_0\rad_k\lad v_0\rad_m
+{\bf Q}^{vh,\cdot}_{m,k}\lad v_0\rad_k\lad h_0\rad_m$$
\begin{equation}
\left.\left.+{\bf Q}^{hh,\cdot}_{m,k}\lad h_0\rad_k\lad h_0\rad_m
+{\bf Y}^{v,\cdot}_{m,k}
{\partial^3\nabla^{-2}_{\bf X}\lad v_0\rad_1\over\partial X_k\partial X_m\partial X_1}
+{\bf Y}^{h,\cdot}_{m,k}
{\partial^3\nabla^{-2}_{\bf X}\lad h_0\rad_1\over\partial X_k\partial X_m\partial X_1}
\right)\right).
\label{eq67}\end{equation}
Here ${\bf G}^{\cdot,\cdot}_{m,k}$, ${\bf Q}^{\cdot\cdot,\cdot}_{m,k}$ and
${\bf Y}^{\cdot,\cdot}_{m,k}$ are solutions to the auxiliary problems of types
II--IV; $\widehat{\bf S}^{\cdot,\cdot}_k=(\widehat{\bf S}^{\cdot,\omega}_k,
\widehat{\bf S}^{\cdot,v}_k,\widehat{\bf S}^{\cdot,h}_k,\widehat{\bf S}^{\cdot,\theta}_k)$
satisfy the boundary conditions of the kind of (\theBCs), \rf{eq4} and solve
{\it auxiliary problems of type V}:
\pagebreak
$${\cal M}(\widehat{\bf S}^{v,\omega}_k,\widehat{\bf S}^{v,h}_k,\widehat{S}^{v,\theta}_k)$$
$$=\lpar-\nabla_{\bf x}\times({\bf V}_1\times{\bf S}^{v,\omega}_k
+({\bf S}^{v,v}_k+{\bf e}_k)\times\bOm_1
-{\bf H}_1\times(\nabla_{\bf x}\times{\bf S}^{v,h}_k)
-{\bf S}^{v,h}_k\times(\nabla_{\bf x}\times{\bf H}_1)),$$
\begB
-\nabla_{\bf x}\times(({\bf S}^{v,v}_k+{\bf e}_k)\times{\bf H}_1+{\bf V}_1\times{\bf S}^{v,h}_k),\quad
({\bf V}_1\cdot\nabla_{\bf x})S^{v,\theta}_k
+(({\bf S}^{v,v}_k+{\bf e}_k)\cdot\nabla_{\bf x})\Theta_1\rpar,
\endB{eq68p1}{IIpauxiv}\vspace*{2mm}\begI
\nabla_{\bf x}\times\widehat{\bf S}^{v,v}_k=\widehat{\bf S}^{v,\omega}_k,\qquad
\nabla_{\bf x}\cdot\widehat{\bf S}^{v,\omega}_k=
\nabla_{\bf x}\cdot\widehat{\bf S}^{v,v}_k=\nabla_{\bf x}\cdot\widehat{\bf S}^{v,h}_k=0;
\endI{eq68p2}\begI
\la\widehat{\bf S}^{v,\omega}_k\ra_v=
\la\widehat{\bf S}^{v,v}_k\ra_h=\la\widehat{\bf S}^{v,v}_k\ra_h=0,
\endL{eq68p3}
$${\cal M}(\widehat{\bf S}^{h,\omega}_k,\widehat{\bf S}^{h,h}_k,\widehat{S}^{h,\theta}_k)$$
$$=\lpar-\nabla_{\bf x}\times({\bf V}_1\times{\bf S}^{h,\omega}_k+{\bf S}^{h,v}_k\times\bOm_1
-{\bf H}_1\times(\nabla_{\bf x}\times{\bf S}^{h,h}_k)
-({\bf S}^{h,h}_k+{\bf e}_k)\times(\nabla_{\bf x}\times{\bf H}_1)),$$
\begB
-\nabla_{\bf x}\times({\bf S}^{h,v}_k\times{\bf H}_1
+{\bf V}_1\times({\bf S}^{h,h}_k+{\bf e}_k)),\quad
({\bf V}_1\cdot\nabla_{\bf x})S^{h,\theta}_k+({\bf S}^{h,v}_k\cdot\nabla_{\bf x})\Theta_1\rpar,
\endB{eq69p1}{IIpauxih}\vspace*{2mm}\begI
\nabla_{\bf x}\times\widehat{\bf S}^{h,v}_k=\widehat{\bf S}^{h,\omega}_k,\qquad
\nabla_{\bf x}\cdot\widehat{\bf S}^{h,\omega}_k=
\nabla_{\bf x}\cdot\widehat{\bf S}^{h,v}_k=\nabla_{\bf x}\cdot\widehat{\bf S}^{h,h}_k=0,
\endI{eq69p2}\begI
\la\widehat{\bf S}^{h,\omega}_k\ra_v=
\la\widehat{\bf S}^{h,v}_k\ra_h=\la\widehat{\bf S}^{h,v}_k\ra_h=0.
\endL{eq69p3}
Solvability (compatibility) conditions \xrf{eq9p6}{eq9p5} and \rf{solveomega}
for the problems (\theIIpauxiv) and (\theIIpauxih) are easily verified.

Averaging of the vertical component of the vorticity equation (A4) for $n=3$
over fast spatial variables yields
\begin{equation}
-{\partial\la\bom_3\ra_v\over\partial t}-{\partial\la\bom_1\ra_v\over\partial T}
+\nu\nabla^2_{\bf X}\la\bom_1\ra_v
\label{eq70}\end{equation}
$$+\nabla_{\bf X}\times\la{\bf V}_0\times(\nabla_{\bf X}\times{\bf v}_1)
-{\bf V}_0\nabla_{\bf X}\cdot\lb{\bf v}_1\rb_h
+({\bf V}_1+{\bf v}_0)\times(\nabla_{\bf X}\times{\bf v}_0)
-({\bf V}_1+{\bf v}_0)\nabla_{\bf X}\cdot\lb{\bf v}_0\rb_h$$
$$-{\bf H}_0\times(\nabla_{\bf X}\times{\bf h}_1)
+{\bf H}_0\nabla_{\bf X}\cdot\lb{\bf h}_1\rb_h
-({\bf H}_1+{\bf h}_0)\times(\nabla_{\bf X}\times{\bf h}_0)
+({\bf H}_1+{\bf h}_0)\nabla_{\bf X}\cdot\lb{\bf h}_0\rb_h\ra_h=0.$$

Averaging \rf{eq70} over fast time upon substitution of \rf{eq67}, obtain
an equation for the mean flow perturbation, generalising \rf{eq59}:
$$\nabla_{\bf X}\times\left(-{\partial\over\partial T}\lad{\bf v}_0\rad_h
+\nu\nabla^2_{\bf X}\lad{\bf v}_0\rad_h\right.
-(\lad{\bf v}_0\rad_h\cdot\nabla_{\bf X})\lad{\bf v}_0\rad_h
+(\lad{\bf h}_0\rad_h\cdot\nabla_{\bf X})\lad{\bf h}_0\rad_h$$
$$+\sum_{j=1}^2\sum_{m=1}^2\left.{\partial\over\partial X_j}\right(
\Alfa^{v,v}_{m,j}\lad v_0\rad_m+\Alfa^{h,v}_{m,j}\lad h_0\rad_m
+\sum_{k=1}^2\left({\partial\over\partial X_m}\right(
{\bf D}^{v,v}_{m,k,j}\lad v_0\rad_k$$
$$+\,{\bf D}^{h,v}_{m,k,j}\lad h_0\rad_k
\left.+{\partial^2\over\partial X_k\partial X_1}\nabla^{-2}_{\bf X}
\left({\bf d}^{v,v}_{m,k,j}\lad v_0\rad_1
+{\bf d}^{h,v}_{m,k,j}\lad h_0\rad_1\right)\right)$$
\begin{equation}
\left.\left.\left.\phantom{\partial\over\partial_j}
+{\bf A}^{vv,v}_{m,k,j}\lad v_0\rad_k\lad v_0\rad_m
+{\bf A}^{vh,v}_{m,k,j}\lad v_0\rad_k\lad h_0\rad_m
+{\bf A}^{hh,v}_{m,k,j}\lad h_0\rad_k\lad h_0\rad_m\right)\!\right)\!\right)=0.
\label{eq71}\end{equation}
Here ${\bf D}^{\cdot,\cdot}_{m,k,j},\ {\bf d}^{\cdot,\cdot}_{m,k,j}$
and ${\bf A}^{\cdot\cdot,\cdot}_{m,k,j}$ are determined by (\theeddyvisc) and (\theeddyadv),
$$\Alfa^{v,v}_{m,j}=\lad-(V_0)_j\,\widehat{\bf S}^{v,v}_m
-{\bf V}_0\,(\widehat{S}^{v,v}_m)_j
-(V_1)_j\,({\bf S}^{v,v}_m+{\bf e}_m)-{\bf V}_1\,(S^{v,v}_m)_j$$
\begB
+(H_0)_j\,\widehat{\bf S}^{v,h}_m+{\bf H}_0\,(\widehat{S}^{v,h}_j)_m
+(H_1)_j\,{\bf S}^{v,h}_m+{\bf H}_1\,(S^{v,h}_m)_j\rad_h,
\endB{eq72v}{Alfav}
\pagebreak
$$\Alfa^{h,v}_{m,j}=\lad-(V_0)_j\,\widehat{\bf S}^{h,v}_m
-{\bf V}_0\,(\widehat{S}^{h,v}_m)_j
-(V_1)_j\,{\bf S}^{h,v}_m-{\bf V}_1\,(S^{h,v}_m)_j$$
\begI
+(H_0)_j\,\widehat{\bf S}^{h,h}_m+{\bf H}_0\,(\widehat{S}^{h,h}_j)_m
+(H_1)_j\,({\bf S}^{h,h}_m+{\bf e}_m)+{\bf H}_1\,(S^{h,h}_m)_j\rad_h.
\endL{eq72h}

Averaging (A5) for $n=2$ over fast variables upon substitution of \rf{eq67},
obtain an equation for the mean magnetic field perturbation, generalising \rf{eq62}:
$$-{\partial\over\partial T}\lad{\bf h}_0\rad_h+\eta\nabla^2_{\bf X}\lad{\bf h}_0\rad_h
+\nabla_{\bf X}\times\left(\lad{\bf v}_0\rad_h\times\lad{\bf h}_0\rad_h
+\sum_{k=1}^2\right(\Alfa^{v,h}_k\lad v_0\rad_k+\Alfa^{h,h}_k\lad h_0\rad_k$$
$$+\sum_{m=1}^2\left({\partial\over\partial X_m}\left(
{\bf D}^{v,h}_{m,k}\lad v_0\rad_k+{\bf D}^{h,h}_{m,k}\lad h_0\rad_k
+{\partial^2\over\partial X_k\partial X_1}\nabla^{-2}_{\bf X}\left(
{\bf d}^{v,h}_{m,k}\lad v_0\rad_1
+{\bf d}^{h,h}_{m,k}\lad h_0\rad_1\right)\right)\right.$$
\begin{equation}
\left.\left.\left.\phantom{\partial\over\partial}
+{\bf A}^{vv,h}_{m,k}\lad v_0\rad_k\lad v_0\rad_m
+{\bf A}^{vh,h}_{m,k}\lad v_0\rad_k\lad h_0\rad_m
+{\bf A}^{hh,h}_{m,k}\lad h_0\rad_k\lad h_0\rad_m\right)\!\right)\!\right)=0.
\label{eq73}\end{equation}
Here ${\bf D}^{\cdot,\cdot}_{m,k,j},\ {\bf d}^{\cdot,\cdot}_{m,k,j}$
and ${\bf A}^{\cdot\cdot,\cdot}_{m,k}$ are determined by \rf{eq63p2} and (\theeddyadh),
\begB
\Alfa^{v,h}_k=\lad
\widehat{\bf S}^{v,v}_k\times{\bf H}_0+({\bf S}^{v,v}_k+{\bf e}_k)\times{\bf H}_1
+{\bf V}_0\times\widehat{\bf S}^{v,h}_k+{\bf V}_1\times{\bf S}^{v,h}_k\rad_v,
\endB{eq74v}{Alfah}\begI
\Alfa^{h,h}_k=\lad
\widehat{\bf S}^{h,v}_k\times{\bf H}_0+{\bf S}^{h,v}_k\times{\bf H}_1
+{\bf V}_0\times\widehat{\bf S}^{h,h}_k+{\bf V}_1\times({\bf S}^{h,h}_k+{\bf e}_k)\rad_v.
\endL{eq74h}

The new terms involving coefficients $\Alfa^{\cdot,\cdot}_{m,j}$
and $\Alfa^{\cdot,\cdot}_k$ represent the AKA-- and magnetic $\alpha-$effects,
respectively. If the CHM regime ${\bf V}_0,{\bf H}_0,\Theta_0$ is symmetric,
but $\bOm_1,{\bf V}_1,{\bf H}_1,\Theta_1$ do not constitute a symmetric set
(e.g., if a symmetry-breaking Hopf bifurcation is examined), then
$\widehat{\bf S}^{\cdot,\cdot}_j$ have non-vanishing symmetric parts and
generically the coefficients (\theAlfav) and (\theAlfah) are non-zero. Thus, like
in the almost axisymmetric models considered by Braginsky (1964a-d, 1967, 1975)
and Soward (1972, 1974), a small deviation of the perturbed CHM regime from
symmetry gives rise to the $\alpha-$effect in the mean-field equations. The
combined eddy diffusivity operator in \rf{eq71} and \rf{eq73} can be simplified
using the identities of appendix~E.

\mi{\bf 11. Mean-field equations with the $\alpha-$effect and cubic nonlinearity:\\
large-scale perturbations of CHM regimes near a pitchfork bifurcation}

\mi
In this section we consider again a family of CHM regimes represented by the power
series \rf{eq65} in a small parameter, which we link with the ratio $\varepsilon$
of spatial scales in large-scale perturbations like in the previous section.
We assume now that the operator $\cal M$ (considered in this section, as
$\cal L$ and any equation or quantity introduced previously, for
$\bOm=\bOm_0$, ${\bf V}={\bf V}_0$, ${\bf H}={\bf H}_0$, $\Theta=\Theta_0$ and
$\beta=\beta_0$) has a non-trivial kernel including a small-scale eigenvector
${\bf S}^\cdot({\bf x},t)=({\bf S}^\omega,{\bf S}^v,{\bf S}^h,{\bf S}^\theta)$:
$${\cal M}({\bf S}^\omega,{\bf S}^h,S^\theta)=0,\qquad
\nabla_{\bf x}\times{\bf S}^v={\bf S}^\omega,$$
$$\nabla_{\bf x}\cdot{\bf S}^\omega=\nabla_{\bf x}\cdot{\bf S}^v=\nabla_{\bf x}\cdot{\bf S}^v=0,
\qquad\la{\bf S}^\omega\ra_v=\la{\bf S}^v\ra_h=\la{\bf S}^h\ra_h=0.$$
Generically, when exists, ${\bf S}^\cdot$ spans $\ker\cal M$. (If the CHM regime
${\bf V}_0,{\bf H}_0,\Theta_0$ is non-steady, an eigenvector in the kernel of $\cal M$
is a globally bounded in space and time solution to the system of equations
\rf{eq15}, \xrf{eq5p4}{eq5p7}, which does not decay in time.)

Like in section 10, we assume \rf{eq66} and investigate stability of CHM regimes,
whose dependence on $\varepsilon$ \rf{eq65} is due to a similar dependence on
$\varepsilon$ of the source terms $\bf F,\ J$ and $S$ in (\theCHMB) (for
$\beta_2=0$), or a family of CHM regimes emerging from ${\bf V}_0,{\bf H}_0,\Theta_0$
(not necessa\-rily a steady state or a periodic orbit) in a pitchfork bifurcation
at $\beta=\beta_0$, occurring when the Rayleigh number is varied and
an eigenvalue of $\cal M$ passes through zero. A positive or negative
$\beta_2$ corresponds to a direct or reverse bifurcation, respectively.
The family of CHM regimes emerging in the pitchfork bifurcation
admits\footnote{See footnotes 5 and 6.} the expansion
\rf{eq65} (see, e.g., Guckenheimer and Holmes
1990). Their deviation from the regime at the critical point $\beta=\beta_0$
is asymptotically close to the mean-free eigenvector from ker$\,\cal M$:
\begin{equation}
(\bOm_1,{\bf H}_1,{\bf V}_1,\Theta_1)=\chi({\bf S}^\omega,{\bf S}^v,{\bf S}^h,{\bf S}^\theta).
\label{eq75}\end{equation}

Construction of mean-field equations differs from the previously
considered cases in the following aspects. Now, a new solvability
condition for the system (\theMsys) must be taken into account:
orthogonality of the right-hand side of the equations \xrf{MEQ1}{MEQ3}, derived
from (A4)--(A6) for $n=2$, to the eigenvector
${\bf S}_*^\cdot({\bf x},t)\in\ker{\cal M}^*$. Also, expressions for the first
terms in expansions of solutions consist now of more terms, involving solutions
to a larger number of auxiliary problems. Insignificance of the $\alpha-$effect
in the leading order incorporates now more conditions than we have derived
in section 6, and it is unlikely that they all can be satisfied by parameter
tuning. Under these circumstances, in this section we impose symmetries of
the leading terms in the expansion of the perturbed CHM regime \rf{eq65}.
We assume that the CHM regime ${\bf V}_0,{\bf H}_0,\Theta_0$ at the
point of bifurcation has a symmetry considered in section 7. Consequently,
symmetric and antisymmetric sets of vector fields constitute invariant subspaces
of $\cal M$. We assume in addition that ${\bf S}^\cdot({\bf x},t)$, and hence
${\bf S}^\cdot_*({\bf x},t)$ and $\bOm_1,{\bf V}_1,{\bf H}_1,\Theta_1$ are
antisymmetric sets. Then the $\alpha-$effect is insignificant in the leading
order and all solvability conditions for auxiliary problems are automatically satisfied.

Upon substitution of (\thepertser) into (\theEQpert), the leading terms of the
resultant equations, at order $\varepsilon^0$, yield equations \rf{eq20}.
Their solution can be expressed as
\begin{equation}
(\bom_0,{\bf v}_0,{\bf h}_0,\theta_0)=\bxi_0^\cdot+\sum_{k=1}^2
\left({\bf S}^{v,\cdot}_k\lad v_0\rad_k+{\bf S}^{h,\cdot}_k\lad h_0\rad_k\right)
+c_0({\bf X},T){\bf S}^\cdot+(0,\lad{\bf v}_0\rad_h,\lad{\bf h}_0\rad_h,0).
\label{eq76}\end{equation}
Here ${\bf S}^{\cdot,\cdot}_k({\bf x},t)$ are antisymmetric sets satisfying
(\theomegaSv)--(\theIauxih), and exponentially decaying mean-free transients
$\bxi^\cdot_0({\bf x},t,{\bf X},T)$ solve (\theIxi).

Higher order equations (A4)--(A6) in the hierarchy are the same as
in the previous section. Averaging the vertical component of (A4) for
$n=2$ obtain \rf{eq29}; upon substitution of the flow and magnetic components
of \rf{eq76}, find from \rf{eq19} for $n=2$ and \rf{eq17p1} for $n=1$
\vspace*{-4mm}
$$\la{\bf v}_1\ra_h=\lad{\bf v}_1\rad_h+\txi{v}-\nabla_{\bf X}\Pi
+\sum_{k=1}^2\sum_{m=1}^2\left(
\lbdb\int_0^t\alpha^c_{m,k}dt\rbdb\,\,{\partial c_0\over\partial X_m}\right.$$
\begB
\left.+\lbdb\int_0^t\alpha^v_{m,k}dt\rbdb\,\,{\partial\lad v_0\rad_k\over\partial X_m}
+\lbdb\int_0^t\alpha^h_{m,k}dt\rbdb\,\,{\partial\lad h_0\rad_k\over\partial X_m}
\right){\bf e}_k,
\endB{EE77p1}{vonemean}
\vspace*{2mm}

\pagebreak\noindent
where $\alpha^v_{m,k}$ and $\alpha^h_{m,k}$ are determined by (\theAKAco),
$$\alpha^c_{m,k}=\la-(S^v)_m(V_0)_k-(V_0)_m(S^v)_k+(S^h)_m(H_0)_k+(H_0)_m(S^h)_k\ra,$$
$$\Pi=\sum_{m=1}^2\left(\lbdb\int_0^t(\alpha^v_{m,1}-\alpha^v_{m,2})\,dt\rbdb\,\,
{\partial^2\nabla^{-2}_{\bf X}\lad v_0\rad_1\over\partial X_m\partial X_1}\right.$$
\begI
\left.+\lbdb\int_0^t(\alpha^h_{m,1}-\alpha^h_{m,2})\,dt\rbdb\,\,
{\partial^2\nabla^{-2}_{\bf X}\lad h_0\rad_1\over\partial X_m\partial X_1}
+\sum_{k=1}^2\lbdb\int_0^t\alpha^c_{m,k}dt\rbdb\,\,
{\partial^2\nabla^{-2}_{\bf X}c_0\over\partial X_m\partial X_k}\right).
\endL{EE77p2}

\vspace*{1pt}\noindent
Equations (A4)--(A6) for $n=1$ have a solution
$$(\bom_1,{\bf v}_1,{\bf h}_1,\theta_1)=\bxi_1^\cdot+{\bf S}^\cdot c_1({\bf X},T)
+\widehat{\bf S}^\cdot c_0+{\bf Q}^{cc,\cdot}c_0^2
+\left.\sum_{k=1}^2\right({\bf G}^{c,\cdot}_k{\partial c_0\over\partial X_k}$$
$$+{\bf S}^{v,\cdot}_k\lad v_1\rad_k+{\bf S}^{h,\cdot}_k\lad h_1\rad_k
+\widehat{\bf S}^{v,\cdot}_k\lad v_0\rad_k+\widehat{\bf S}^{h,\cdot}_k\lad h_0\rad_k
+{\bf Q}^{cv,\cdot}_k\lad v_0\rad_k c_0+{\bf Q}^{ch,\cdot}_k\lad h_0\rad_k c_0$$
$$\left.+\sum_{m=1}^2\right({\bf G}^{v,\cdot}_{m,k}{\partial\lad v_0\rad_k\over\partial X_m}
+{\bf G}^{h,\cdot}_{m,k}{\partial\lad h_0\rad_k\over\partial X_m}
+{\bf Q}^{vv,\cdot}_{m,k}\lad v_0\rad_k\lad v_0\rad_m
+{\bf Q}^{vh,\cdot}_{m,k}\lad v_0\rad_k\lad h_0\rad_m$$
$$+{\bf Q}^{hh,\cdot}_{m,k}\lad h_0\rad_k\lad h_0\rad_m
+{\bf Y}^{v,\cdot}_{m,k}
{\partial^3\nabla^{-2}_{\bf X}\lad v_0\rad_1\over\partial X_k\partial X_m\partial X_1}
+{\bf Y}^{h,\cdot}_{m,k}
{\partial^3\nabla^{-2}_{\bf X}\lad h_0\rad_1\over\partial X_k\partial X_m\partial X_1}$$
\begin{equation}
\left.\left.+\sum_{j=1}^2{\bf Y}^{c,\cdot}_{m,k,j}
{\partial^3\nabla^{-2}_{\bf X}c_0\over\partial X_k\partial X_m\partial X_j}\right)\right)
+(\nabla_{\bf X}\times\lad{\bf v}_0\rad_h,\la{\bf v}_1\ra_h,\lad{\bf h}_1\rad_h,0).
\label{eq77}\end{equation}
Here ${\bf G}^{v,\cdot}_{m,k},{\bf G}^{h,\cdot}_{m,k},{\bf Q}^{vv,\cdot}_{m,k},
{\bf Q}^{vh,\cdot}_{m,k},{\bf Q}^{hh,\cdot}_{m,k},{\bf Y}^{v,\cdot}_{m,k},
{\bf Y}^{h,\cdot}_{m,k}$ and $\widehat{\bf S}^{\cdot,\cdot}_k$
are solutions to the auxiliary problems of types II--V;
${\bf G}^{c,\cdot}_k=({\bf G}^{c,\omega}_k,{\bf G}^{c,v}_k,{\bf G}^{c,h}_k,G^{c,\theta^{\phantom{|\!\!}}}_k)$
solve {\it auxiliary problems of type II$\,'$}:
$${\cal L}^\omega({\bf G}^{c,\cdot}_k)=-2\nu{\partial{\bf S}^\omega\over\partial x_k}
-{\bf e}_k\times\left({\bf V}_0\times{\bf S}^{\omega\phantom{\theta}}
+{\bf S}^v\times\bOm_0-{\bf H}_0\times(\nabla_{\bf x}\times{\bf S}^h)\right.$$
\begB
\left.-{\bf S}^h\times(\nabla_{\bf x}\times{\bf H}_0)+\beta_0S^\theta{\bf e}_3\right)
-\nabla_{\bf x}\times({\bf H}_0\times({\bf e}_k\times{\bf S}^h))
+\sum_{m=1}^2\lbdb\int_0^t\alpha^c_{k,m}dt\rbdb{\partial\bOm_0\over\partial x_m},
\endB{eq44p1x}{auxIIc}\begI
{\cal L}^h({\bf G}^{c,\cdot}_k)=-2\eta{\partial{\bf S}^h\over\partial x_k}
-{\bf e}_k\times\left({\bf V}_0\times{\bf S}^h+{\bf S}^v\times{\bf H}_0\right)
+\sum_{m=1}^2\lbdb\int_0^t\alpha^c_{k,m}dt\rbdb{\partial{\bf H}_0\over\partial x_m},
\endI{eq44p5x}\begI
{\cal L}^\theta({\bf G}^{c,\cdot}_k)=-2\kappa{\partial S^\theta\over\partial x_k}+(V_0)_kS^\theta
+\sum_{m=1}^2\lbdb\int_0^t\alpha^c_{k,m}dt\rbdb{\partial\Theta_0\over\partial x_m},
\endI{eq44p7x}\begI
\nabla_{\bf x}\times{\bf G}^{c,v}_k={\bf G}^{c,\omega}_k-{\bf e}_k\times{\bf S}^v,
\endI{eq44p3x}\begI
\nabla_{\bf x}\cdot{\bf G}^{c,\omega}_k=-(S^\omega)_k,\quad
\nabla_{\bf x}\cdot{\bf G}^{c,v}_k=-(S^v)_k,\quad
\nabla_{\bf x}\cdot{\bf G}^{c,h}_k=-(S^h)_k;
\endL{eq44p2x}
${\bf Q}^{c\cdot,\cdot}_k=({\bf Q}^{c\cdot,\omega}_k,{\bf Q}^{c\cdot,v}_k,{\bf Q}^{c\cdot,h}_k,Q^{c\cdot,\theta}_k)$
solve {\it auxiliary problems of type III$\,'$}:
\pagebreak
$${\cal M}({\bf Q}^{cv,\omega}_k,{\bf Q}^{cv,h}_k,Q^{cv,\theta}_k)$$
$$=\lpar\nabla_{\bf x}\times\left(-({\bf S}^{v,v}_k+{\bf e}_k)\times{\bf S}^\omega
-{\bf S}^v\times{\bf S}^{v,\omega}_k
+{\bf S}^{v,h}_k\times(\nabla_{\bf x}\times{\bf S}^h)
+{\bf S}^h\times(\nabla_{\bf x}\times{\bf S}^{v,h}_k)\right)\!,$$
\begB
-\nabla_{\bf x}\times\left(({\bf S}^{v,v}_k+{\bf e}_k)\times{\bf S}^h
+{\bf S}^v\times{\bf S}^{v,h}_k\right),\quad
(({\bf S}^{v,v}_k+{\bf e}_k)\cdot\nabla_{\bf x})S^\theta
+({\bf S}^v\cdot\nabla_{\bf x})S^{v,\theta}_k\rpar\!,
\endB{eq78p1}{auxIIIvc}\begI
\nabla_{\bf x}\times{\bf Q}^{cv,v}_k={\bf Q}^{cv,\omega}_k;
\endL{eq78p2}
$${\cal M}({\bf Q}^{ch,\omega}_k,{\bf Q}^{ch,h}_k,Q^{ch,\theta}_k)$$
$$=\lpar\nabla_{\bf x}\times\left(-{\bf S}^{h,v}_k\times{\bf S}^\omega
-{\bf S}^v\times{\bf S}^{h,\omega}_k+({\bf S}^{h,h}_k
+{\bf e}_k)\times(\nabla_{\bf x}\times{\bf S}^h)
+{\bf S}^h\times(\nabla_{\bf x}\times{\bf S}^{h,h}_k)\right)\!,$$
\begB
-\nabla_{\bf x}\times\left({\bf S}^{h,v}_k\times{\bf S}^h+
{\bf S}^v\times({\bf S}^{h,h}_k+{\bf e}_k)\right),\quad
({\bf S}^{h,v}_k\cdot\nabla_{\bf x})S^\theta
+({\bf S}^v\cdot\nabla_{\bf x})S^{h,\theta}_k\rpar\!,
\endB{eq79p1}{auxIIIhc}\begI
\nabla_{\bf x}\times{\bf Q}^{ch,v}_k={\bf Q}^{ch,\omega}_k;
\endL{eq79p2}
$${\cal M}({\bf Q}^{cc,\omega},{\bf Q}^{cc,h},Q^{cc,\theta})=
\lpar\nabla_{\bf x}\times\left(-{\bf S}^v\times{\bf S}^\omega
+{\bf S}^h\times(\nabla_{\bf x}\times{\bf S}^h)\right)\!,$$
\begB
-\nabla_{\bf x}\times({\bf S}^v\times{\bf S}^h),
\quad({\bf S}^v\cdot\nabla_{\bf x})S^\theta\rpar;
\endB{eq81p1}{auxIIIcc}\begI
\nabla_{\bf x}\times{\bf Q}^{cc,v}={\bf Q}^{cc,\omega};
\endL{eq81p2}
\begin{equation}
\nabla_{\bf x}\cdot{\bf Q}^{c\cdot,\omega}_k=
\nabla_{\bf x}\cdot{\bf Q}^{c\cdot,v}_k=\nabla_{\bf x}\cdot{\bf Q}^{c\cdot,h}_k=0;
\label{eq80}\end{equation}
${\bf Y}^{c,\cdot}_{m,k,j}$ solve {\it auxiliary problems of type IV$\,'$}:
\begB
{\cal M}({\bf Y}^{c,\omega}_{m,k,j},{\bf Y}^{c,h}_{m,k,j},Y^{c,\theta}_{m,k,j})
=-\lbdb\int_0^t\alpha^c_{m,k}dt\rbdb{\partial\over\partial x_j}(\bOm_0,{\bf H}_0,\Theta_0),
\endB{EQYpp1}{auxIVc}\begI
\nabla_{\bf x}\times{\bf Y}^{c,v}_{m,k,j}={\bf Y}^{c,\omega}_{m,k,j},\qquad
\nabla_{\bf x}\cdot{\bf Y}^{c,\cdot}_{m,k,j}=0;
\endL{EQYpp6}
$\widehat{\bf S}^\cdot=(\widehat{\bf S}^\omega,\widehat{\bf S}^v,\widehat{\bf S}^h,\widehat{S}^\theta)$
solve {\it auxiliary problems of type V$\,'$}:
$$\left.{\cal M}(\widehat{\bf S}^\omega,\widehat{\bf S}^h,\widehat{S}^\theta)=
\lpar\nabla_{\bf x}\times\right(-{\bf V}_1\times{\bf S}^\omega-{\bf S}^v\times\bOm_1
\left.+{\bf H}_1\times(\nabla_{\bf x}\times{\bf S}^h)
+{\bf S}^h\times(\nabla_{\bf x}\times{\bf H}_1)\right)\!,$$
\begB
-\nabla_{\bf x}\times({\bf S}^v\times{\bf H}_1+{\bf V}_1\times{\bf S}^h),
\quad({\bf V}_1\cdot\nabla_{\bf x})S^\theta+({\bf S}^v\cdot\nabla_{\bf x})\Theta_1\rpar\!,
\endB{eq82p1}{auxVp}\begI
\nabla_{\bf x}\times\widehat{\bf S}^\omega=\widehat{\bf S}^v,\qquad
\nabla_{\bf x}\cdot\widehat{\bf S}^\omega=\nabla_{\bf x}\cdot\widehat{\bf S}^v=
\nabla_{\bf x}\cdot\widehat{\bf S}^h=0.
\endL{eq82p3}
Solutions to these auxiliary problems must satisfy the boundary conditions
of the kind of (\theBCs) and \rf{eq4}, and have zero means over fast spatial
variables of horizontal components of flow parts and of vertical
components of vorticity parts must vanish, and spatio-temporal
means over fast variables of horizontal components of magnetic parts.
Due to the assumed symmetry of the CHM regime
$({\bf V}_0,{\bf H}_0,\Theta_0)$ and antisymmetry of the sets of fields
$({\bf V}_1,{\bf H}_1,\Theta_1)$, ${\bf S}^\cdot$ and
${\bf S}^{\cdot,\cdot}_k$, the right-hand sides of the systems of equations
(\theauxIIc)--(\theauxVp) are symmetric sets. Thus all solvability conditions
for these problems are satisfied, and the solutions ${\bf Q}^{c\cdot,\cdot}$,
${\bf Y}^{c,\cdot}_{m,k,j}$ and $\widehat{\bf S}^\cdot$ are symmetric sets.

Averaging of the vertical component of the vorticity equation (A4) for $n=3$ over fast
spatial variables yields \rf{eq70}. Averaging it over fast time and substituting
\rf{eq77}, obtain an equation for the mean flow perturbation, analogous to \rf{eq71}:
\pagebreak
$$\nabla_{\bf X}\times\left(-{\partial\over\partial T}\lad{\bf v}_0\rad_h
+\nu\nabla^2_{\bf X}\lad{\bf v}_0\rad_h\right.
-(\lad{\bf v}_0\rad_h\cdot\nabla_{\bf X})\lad{\bf v}_0\rad_h
+(\lad{\bf h}_0\rad_h\cdot\nabla_{\bf X})\lad{\bf h}_0\rad_h$$
$$+\sum_{j=1}^2{\partial\over\partial X_j}\left(\Alfa^{c,v}_jc_0+{\bf A}^v_jc_0^2
+\sum_{m=1}^2\right({\partial\over\partial X_m}\left(\,
\sum_{k=1}^2\right({\bf D}^{v,v}_{m,k,j}\lad v_0\rad_k+{\bf D}^{h,v}_{m,k,j}\lad h_0\rad_k$$
$$+{\partial^2\over\partial X_k\partial X_1}\nabla^{-2}_{\bf X}
\left({\bf d}^{v,v}_{m,k,j}\lad v_0\rad_1+{\bf d}^{h,v}_{m,k,j}\lad h_0\rad_1\right)
+\sum_{i=1}^2{\bf d}^{c,v}_{m,k,j,i}\left.\left.{\partial^2\nabla^{-2}_{\bf X}c_0
\over\partial X_k\partial X_i}\right)+{\bf D}^{c,v}_{m,j}c_0\right)$$
\begin{equation}
+\Alfa^{v,v}_{m,j}\lad v_0\rad_m+\Alfa^{h,v}_{m,j}\lad h_0\rad_m
+{\bf A}^{cv,v}_{m,j}\lad v_0\rad_mc_0+{\bf A}^{ch,v}_{m,j}\lad h_0\rad_mc_0
\label{eq83}\end{equation}
$$\left.\left.\left.+\sum_{k=1}^2\left(
{\bf A}^{vv,v}_{m,k,j}\lad v_0\rad_k\lad v_0\rad_m
+{\bf A}^{vh,v}_{m,k,j}\lad v_0\rad_k\lad h_0\rad_m
+{\bf A}^{hh,v}_{m,k,j}\lad h_0\rad_k\lad h_0\rad_m\right)\right)\right)\right)=0.$$
Here ${\bf D}^{\cdot,v}_{m,k,j}$, ${\bf d}^{\cdot,v}_{m,k,j}$,
${\bf A}^{\cdot\cdot,v}_{m,k,j}$ and $\Alfa^{\cdot,v}_{m,j}$ are determined
by (\theeddyvisc), (\theeddyadv) and (\theAlfav), and
\begB
{\bf D}^{c,v}_{m,j}=\lad-(V_0)_j\widetilde{\bf G}^{c,v}_m-{\bf V}_0(G^{c,v}_m)_j
+(H_0)_j{\bf G}^{c,h}_m+{\bf H}_0(G^{c,h}_m)_j\rad_h.
\endB{eq85p4}{pitchcov}\begI
{\bf d}^{c,v}_{m,k,j,i}=\lad
-(V_0)_i\widetilde{\bf Y}^{c,v}_{m,k,j}-{\bf V}_0(Y^{c,v}_{m,k,j})_i
+(H_0)_i{\bf Y}^{c,h}_{m,k,j}+{\bf H}_0(Y^{c,h}_{m,k,j})_i\rad_h,
\endI{eq85p5}\begI
\widetilde{\bf G}^{c,v}_m={\bf G}^{c,v}_m+\sum_{k=1}^2\lbdb\int_0^t\alpha^c_{m,k}\,dt\rbdb{\bf e}_k,\quad
\widetilde{\bf Y}^{c,v}_{m,k,j}={\bf Y}^{c,v}_{m,k,j}-\lbdb\int_0^t\alpha^c_{m,k}\,dt\rbdb{\bf e}_j;
\endI{eq85p6}
$$\Alfa^{c,v}_j=\lad-(V_0)_j\widehat{\bf S}^v-{\bf V}_0(\widehat{S}^v)_j
+(H_0)_j\widehat{\bf S}^h+{\bf H}_0(\widehat{S}^h)_j$$
\begI
-(V_1)_j{\bf S}^v-{\bf V}_1(S^v)_j+(H_1)_j{\bf S}^h+{\bf H}_1(S^h)_j\rad_h;
\endI{eq84}\begI
{\bf A}^v_j=\lad-(V_0)_j{\bf Q}^{cc,v}-{\bf V}_0(Q^{cc,v})_j
+(H_0)_j{\bf Q}^{cc,h}+{\bf H}_0(Q^{cc,h})_j
-{\bf S}^v(S^v)_j+{\bf S}^h(S^h)_j\rad_h;
\endI{eq85p1}
$${\bf A}^{cv,v}_{m,j}=\lad-(V_0)_j{\bf Q}^{cv,v}_m-{\bf V}_0(Q^{cv,v}_m)_j
+(H_0)_j{\bf Q}^{cv,h}_m+{\bf H}_0(Q^{cv,h}_m)_j$$
\begI
-(S^v)_j{\bf S}_m^{v,v}-{\bf S}^v(S_m^{v,v})_j
+(S^h)_j{\bf S}_m^{v,h}+{\bf S}^h(S_m^{v,h})_j\rad_h,
\endI{eq85p2}
$${\bf A}^{ch,v}_{m,j}=\lad-(V_0)_j{\bf Q}^{ch,v}_m-{\bf V}_0(Q^{ch,v}_m)_j
+(H_0)_j{\bf Q}^{ch,h}_m+{\bf H}_0(Q^{ch,h}_m)_j$$
\begI
-(S^v)_j{\bf S}_m^{h,v}-{\bf S}^v(S_m^{h,v})_j
+(S^h)_j{\bf S}_m^{h,h}+{\bf S}^h(S_m^{h,h})_j\rad_h,
\endL{eq85p3}

Averaging (A4) for $n=2$ and substituting \rf{eq67} obtain an equation
for the mean magnetic field perturbation, analogous to \rf{eq73}:
$$\left.-{\partial\over\partial T}\lad{\bf h}_0\rad_h+\eta\nabla^2_{\bf X}\lad{\bf h}_0\rad_h
+\nabla_{\bf X}\times\right(\lad{\bf v}_0\rad_h\times\lad{\bf h}_0\rad_h
+\Alfa^{c,h}c_0+{\bf A}^hc_0^2$$
$$+\sum_{k=1}^2\left(\Alfa^{v,h}_k\lad v_0\rad_k+\Alfa^{h,h}_k\lad h_0\rad_k
+{\bf D}^{c,h}_k{\partial c_0\over\partial X_k}
+{\bf A}^{cv,h}_k\lad v_0\rad_kc_0+{\bf A}^{ch,h}_k\lad h_0\rad_kc_0\right)$$
$$+\sum_{m=1}^2\sum_{k=1}^2\left({\partial\over\partial X_m}\right(
{\bf D}^{v,h}_{m,k}\lad v_0\rad_k+{\bf D}^{h,h}_{m,k}\lad h_0\rad_k
+\sum_{j=1}^2{\bf d}^{c,h}_{m,k,j}{\partial^2\nabla^{-2}_{\bf X}c_0\over\partial X_k\partial X_j}$$
\begin{equation}
\left.+{\partial^2\over\partial X_k\partial X_1}\nabla^{-2}_{\bf X}
\left({\bf d}^{v,h}_{m,k}\lad v_0\rad_1+{\bf d}^{h,h}_{m,k}\lad h_0\rad_1\right)\right)
\label{eq86}\end{equation}
$$\left.\left.\phantom{\partial\over\partial}
+{\bf A}^{vv,h}_{m,k}\lad v_0\rad_k\lad v_0\rad_m
+{\bf A}^{vh,h}_{m,k}\lad v_0\rad_k\lad h_0\rad_m
+{\bf A}^{hh,h}_{m,k}\lad h_0\rad_k\lad h_0\rad_m\right)\right)=0.$$

\pagebreak\noindent
Here ${\bf D}^{\cdot,h}_{m,k,j}$, ${\bf d}^{\cdot,h}_{m,k}$ and
${\bf A}^{\cdot\cdot,h}_{m,k}$ are determined by \rf{eq63p2}
and (\theeddyadh), $\Alfa^{\cdot,h}_k$ by (\theAlfah), and
\begB
{\bf D}^{c,h}_k=\lad{\bf V}_0\times{\bf G}^{c,h}_k+\widetilde{\bf G}^{c,v}_k\times{\bf H}_0\rad_v;\qquad
{\bf d}^{c,h}_{m,k,j}=\lad{\bf V}_0\times{\bf Y}^{c,h}_{m,k,j}+
\widetilde{\bf Y}^{c,v}_{m,k,j}\times{\bf H}_0\rad_v;
\endB{eq88p2}{pitchcoh}\begI
\Alfa^{c,h}=\lad{\bf V}_0\times\widehat{\bf S}^h+{\bf V}_1\times{\bf S}^h
+\widehat{\bf S}^v\times{\bf H}_0+{\bf S}^v\times{\bf H}_1\rad_v;
\endI{eq87}\begI
{\bf A}^h=\lad{\bf V}_0\times{\bf Q}^{cc,h}
+{\bf Q}^{cc,v}\times{\bf H}_0+{\bf S}^v\times{\bf S}^h\rad_v;
\endI{eq88p1}\begI
{\bf A}^{cv,h}_k=\lad{\bf V}_0\times{\bf Q}^{cv,h}_k+{\bf Q}^{cv,v}_k\times{\bf H}_0
+{\bf S}^v\times{\bf S}^{v,h}_k+{\bf S}^{v,v}_k\times{\bf S}^h\rad_v,
\endI{eq88p3}\begI
{\bf A}^{ch,h}_k=\lad{\bf V}_0\times{\bf Q}^{ch,h}_k+{\bf Q}^{ch,v}_k\times{\bf H}_0
+{\bf S}^v\times{\bf S}^{h,h}_k+{\bf S}^{h,v}_k\times{\bf S}^h\rad_v.
\endL{eq88p4}
If a family of CHM regimes emerging in a pitchfork bifurcation is considered,
comparison of (\theauxIIIcc) and (\theauxVp) with the use of \rf{eq75}
reveals that $\widehat{\bf S}^\cdot=2\chi{\bf Q}^{cc,\cdot}$; consequently,
$\bal^v_j=2\chi{\bf A}^v_j$ and $\bal^h=2\chi{\bf A}^h$.

Equations \rf{eq83} and \rf{eq86} constitute a closed system together with the
equation for the amplitude $c_0$, which is the solvability condition for the
equations \xrf{MEQ1}{MEQ3} derived from (A4)--(A6) for $n=2$, consisting
of orthogonality of their right-hand sides to the eigenvector
${\bf S}_*^\cdot\in\ker{\cal M}^*$. As before,
orthogonality is understood in the terms of the scalar product
$\lad{\bf a^\cdot\cdot b^\cdot}\rad$, where ${\bf a}^\cdot=\{{\bf a}^\omega,{\bf a}^h,a^\theta\}$
and ${\bf b}^\cdot=\{{\bf b}^\omega,{\bf b}^h,b^\theta\}$ are 7-dimensional
vector fields in the layer.

To derive this equation we need to calculate
$\la{\bf v}_2\ra_h$. Using \rf{eq21p2}, find from \rf{eq70}
$$\la\bom_3\ra_v=\lad\bom_3\rad_v+\lbdb\!\nabla_{\bf X}\!\times\!\int_0^t
\!\left.\lbdb\!\la\right.{\bf V}_0\times(\nabla_{\bf X}\times{\bf v}_1)
-{\bf V}_0\nabla_{\bf X}\cdot\lb{\bf v}_1\rb_h
+{\bf V}_1\times(\nabla_{\bf X}\times{\bf v}_0)$$
$$-{\bf V}_1\nabla_{\bf X}\cdot\lb{\bf v}_0\rb_h
-{\bf H}_0\times(\nabla_{\bf X}\times{\bf h}_1)
+{\bf H}_0\nabla_{\bf X}\cdot\lb{\bf h}_1\rb_h
-{\bf H}_1\times(\nabla_{\bf X}\times{\bf h}_0)
+{\bf H}_1\nabla_{\bf X}\cdot\lb{\bf h}_0\rb_h$$
$$+{\bf v}_0\times(\nabla_{\bf X}\times{\bf v}_0)
-{\bf v}_0\nabla_{\bf X}\cdot\lb{\bf v}_0\rb_h
-{\bf h}_0\times(\nabla_{\bf X}\times{\bf h}_0)
+{\bf h}_0\nabla_{\bf X}\cdot\lb{\bf h}_0\rb_h\left\ra_h\!\rbdb\right.{\rm d}t\!\rbdb.$$
Substituting expressions \rf{eq76} and \rf{eq77} and
``uncurling'' this equation, using \rf{eq19} for $n=2$, obtain
\begB
\la{\bf v}_2\ra_h=\lad{\bf v}_2\rad_h+{\bf\Phi}
-\nabla_{\bf X}(\nabla^{-2}_{\bf X}(\nabla_{\bf X}\cdot{\bf\Phi}))+\widetilde\xi_2,
\endB{eq89p1}{vtwoave}

\vspace*{-1mm}\noindent
where $\widetilde\xi_2$ decays exponentially in fast time and
$${\bf\Phi(X},T,t)=\sum_{j=1}^2{\partial\over\partial X_j}
\left(\widetilde{\Alfa}^{c,v}_jc_0+\widetilde{\bf A}^v_jc_0^2
+\sum_{m=1}^2\right({\partial\over\partial X_m}\left(\widetilde{\bf D}^{c,v}_{m,j}c_0+
\,\sum_{k=1}^2\right(\widetilde{\bf D}^{v,v}_{m,k,j}\lad v_0\rad_k$$
$$+\,\widetilde{\bf D}^{h,v}_{m,k,j}\lad h_0\rad_k
+{\partial^2\over\partial X_k\partial X_1}\nabla^{-2}_{\bf X}
\left(\widetilde{\bf d}^{v,v}_{m,k,j}\lad v_0\rad_1
+\widetilde{\bf d}^{h,v}_{m,k,j}\lad h_0\rad_1\right)
+\sum_{i=1}^2\widetilde{\bf d}^{c,v}_{m,k,j,i}\left.\left.{\partial^2\nabla^{-2}_{\bf X}c_0
\over\partial X_k\partial X_i}\right)\right)$$
$$+\,\widetilde{\Alfa}^{v,v}_{m,j}\lad v_0\rad_m+\widetilde{\Alfa}^{h,v}_{m,j}\lad h_0\rad_m
+\widetilde{\bf A}^{cv,v}_{m,j}\lad v_0\rad_mc_0+\widetilde{\bf A}^{ch,v}_{m,j}\lad h_0\rad_mc_0$$
\begI
\left.\left.+\sum_{k=1}^2\lpar
\widetilde{\bf A}^{vv,v}_{m,k,j}\lad v_0\rad_k\lad v_0\rad_m
+\widetilde{\bf A}^{vh,v}_{m,k,j}\lad v_0\rad_k\lad h_0\rad_m
+\widetilde{\bf A}^{hh,v}_{m,k,j}\lad h_0\rad_k\lad h_0\rad_m\rpar\right)\right).
\endL{eq89p2}

\vspace*{1mm}\noindent
The coefficients in \rf{eq89p2} are defined by the modified formulae
(\theeddyvisc), (\theeddyadv), (\theAlfav) and (\thepitchcov) for the
respective (without tildes) coefficients of the equation for the mean
perturbation of the flow \rf{eq83}, where the following change is implemented:
A coefficient in \rf{eq83} is a spatio-temporal mean of the horizontal
component of a certain vector field, say, $\bzeta({\bf x},t)$; then
\pagebreak
the respective coefficient in \rf{eq89p2} is equal to
$\lbd\int_0^t\lbd\la\bzeta\ra_h\rbd\,{\rm d}t\rbd$.
If the symmetry of the CHM state ${\bf V}_0,{\bf H}_0,\Theta_0$
is without a time shift, then $\bf\Phi=0$; otherwise the symmetry
properties of various vector fields assumed in this section
imply that $\bf\Phi$ is $\widetilde{T}-$periodic in the fast time.

Consider the operator $\widetilde{\cal M}$, the restriction of ${\cal M}'$
onto the subspace, defined by the condition $\lad{\bf h}\rad_h=0$. The adjoint
operator is
$\widetilde{\cal M}^*=(({\cal M}'^*)^\omega,\lbd({\cal M}'^*)^h\rbd_h,({\cal M}'^*)^\theta)$
(its domain is the same as that of $\widetilde{\cal M}$).
If ${\bf S}_*^\cdot({\bf x},t)\in\ker{\cal M}^*$, then
$$\widetilde{\bf S}_*^\cdot=\left(\!{\bf S}^\omega_*,
{\bf S}^h_*\!-\lbdb\int\lbd\la(\nabla\times{\bf S}^\omega_*)\times(\nabla\times{\bf H})
-\!{\bf V}\times(\nabla\times{\bf S}^h_*)\ra_h\rbd\,dt\rbdb,S^\theta_*\right)\!\in\ker\widetilde{\cal M}^*.$$
It is more convenient to employ $\widetilde{\bf S}_*^\cdot$, than ${\bf S}_*^\cdot$.
We normalise\footnote{Suppose for simplicity that ${\cal M}^*$ is an elliptic
operator not having Jordan cells of size $2\times2$ or more,
${\cal M}^*{\bf s}_{*p}\!=\lambda_p{\bf s}_{*p}$.~The eigenvectors ${\bf s}_{*p}$
constitute a complete set. Let us expand ${\bf S}^\cdot$:
$${\bf S}^\cdot=\sum_{p=1}^\infty\sigma_p{\bf s}_{*p}\quad\Rightarrow\quad
\lad|{\bf S}^\cdot|^2\rad=\sum_{p=1}^\infty\,\sigma_p\lad{\bf s}_{*p}\cdot{\bf S}^\cdot\rad
\quad\Rightarrow\quad\lad{\bf s}_{*p}\cdot{\bf S}^\cdot\rad\ne0\mbox{ for some }p.$$
However, scalar
multiplication of the eigenvalue equation for ${\bf s}_{*p}$ by ${\bf S}^\cdot$ yields
$0=\lambda_p\lad{\bf s}_{*p}\cdot{\bf S}^\cdot\rad$, since ${\bf S}^\cdot\in\ker\cal M$.
Thus $\lad{\bf s}_{*p}\cdot{\bf S}^\cdot\rad\ne 0$ only for ${\bf s}_{*p}\in\ker{\cal M}^*$,
and hence $\lad\widetilde{\bf S}^\cdot_*\cdot{\bf S}^\cdot\rad=\lad{\bf S}^\cdot_*\cdot{\bf S}^\cdot\rad\ne 0$,
since $\la{\bf S}^h\ra_h=0$.} $\widetilde{\bf S}^\cdot_*$ so that
$\lad\widetilde{\bf S}^\cdot_*\cdot{\bf S}^\cdot\rad=1$. Substituting
$$(\bom_2,{\bf v}_2,{\bf h}_2,\theta_2)=(\bom'_2,{\bf v}'_2,{\bf h}'_2,\theta'_2)
+\sum_{k=1}^2({\bf S}^{v,\cdot}_k\lad v_2\rad_k+{\bf S}^{h,\cdot}_k\lad h_2\rad_k)$$
\begin{equation}
+(\nabla_{\bf X}\times\la{\bf v}_1\ra_h,\la{\bf v}_2\ra_h,\lad{\bf h}_2\rad_h,0),
\label{EE89p3}\end{equation}
\xrf{eq76}{eq77} and (\thevtwoave) into (A4)--(A6) for $n=2$ and
scalar multiplying the result by $\widetilde{\bf S}^\cdot_*$, obtain the
equation under discussion. It does not involve any unknown quantity except
the mean fields $\lad{\bf v}_0\rad_h$ and $\lad{\bf h}_0\rad_h$ and the
amplitude $c_0$ for the following reasons:
$\lad{\cal L}(\bom'_2,{\bf v}'_2,{\bf h}'_2,\theta'_2)\cdot\widetilde{\bf S}^\cdot_*\rad=0$,
since $\la\bom'_2\ra_v=\la{\bf v}'_2\ra_h=\lad{\bf h}'_2\rad_h=0$ and
$\widetilde{\bf S}_*^\cdot({\bf x},t)\in\ker\widetilde{\cal M}^*$;
$\lad v_2\rad_k$ and $\lad h_2\rad_k$ do not appear, since vector
fields ${\bf S}^{\cdot,\cdot}_k({\bf x},t)$ are solutions to auxiliary problems
of type I; all terms involving $\lad v_1\rad_k$, $\lad h_1\rad_k$,
$c_1$ and their derivatives do not contribute, since by our assumptions
$\bOm_0,{\bf V}_0,{\bf H}_0,\Theta_0$ is a symmetric set, and
$\bOm_1,{\bf V}_1,{\bf H}_1,\Theta_1$, ${\bf S}^\cdot$ and ${\bf S}^\cdot_*$
are antisymmetric ones, and hence these terms constitute symmetric sets
($\lad v_0\rad_k$, $\lad h_0\rad_k$, $c_0$ and their derivatives
enter into the expressions for $\bom_0,{\bf v}_0,{\bf h}_0,\theta_0$ via terms,
which are antisymmetric sets, and the expressions for
$\bom_1,{\bf v}_1,{\bf h}_1,\theta_1$ via terms, which are symmetric sets;
opposite to this, $\lad v_1\rad_k$, $\lad h_1\rad_k$ and $c_1$
enter into the expressions for $\bom_1,{\bf v}_1,{\bf h}_1,\theta_1$
via terms, which are antisymmetric sets); terms involving $\bxi^\cdot_n$
are absent because these vector fields and their derivatives exponentially
decay in fast time. Thus, when deriving the equation one can set
$\bxi^\cdot_n\!=\!0$, $\lad{\bf v}_1\rad_h\!=\!\lad{\bf h}_1\rad_h\!=\!\lad{\bf v}_2\rad_h
\!=\!\lad{\bf h}_2\rad_h\!=\!0$ and $c_1\!=\!0$ in \xrf{eq76}{eq77}, \rf{eq89p1} and~\rf{EE89p3}.

The resultant equation is an evolutionary equation for $c_0$.
(It is bulky, and for this reason it is presented in appendix F.)
If the CHM state ${\bf V}_0,{\bf H}_0,\Theta_0$ does not have a symmetry
without a time shift, linear and nonlinear pseudodifferential
operators involving the inverse Laplacian are present in
the equation -- for instance, all operators employed in
$\nabla_{\bf X}(\nabla^{-2}_{\bf X}(\nabla_{\bf X}\cdot\bf\Phi))$.
The terms, quadratic in ${\bf v}_0$, ${\bf v}_1$, ${\bf h}_0$ and ${\bf h}_1$
in (A4)--(A6) for $n=2$ give rise to cubic nonlinearity.

\pagebreak
\mi{\bf 12. Concluding remarks}

\mi
We have considered weakly nonlinear stability of small-scale Boussinesq CHM
regi\-mes to perturbations, involving large spatial and temporal scales.
Equations for the mean parts of the leading terms in power series expansions of
perturbations in the scale ratio have been derived, using homogenisation
techniques. In general CHM
regimes exhibit the $\alpha-$effect in the leading order; consequently, the
mean-field equations turn out to be linear first-order partial differential
equations, which generically have exponentially growing solutions. If the
$\alpha-$effect in the leading order is absent (e.g. if the perturbed CHM regime
is symmetric about the vertical axis or parity-invariant with or without a time
shift), then the resultant mean-field equations \rf{eq62} and \rf{eq59} are
nonlinear second-order partial differential equations, generalising
the standard
equations of magnetohydrodynamics. They possess new terms, describing eddy
diffusivity and eddy advection, both anisotropic. If the CHM regime is
non-steady and lacks $\alpha-$effect in the leading order not due to a spatial
symmetry without a time shift, new physical phenomena emerge, which are
described by non-local pseudodifferential operators. If the regime is close
to a symmetric one, e.g., it is near a symmetry breaking bifurcation, the
$\alpha-$effect emerges in the mean-field equations as well.
Close to the point of a Hopf symmetry-breaking bifurcation the mean-field
equations take the form \rf{eq71} and \rf{eq73}. Near a pitchfork
symmetry-breaking bifurcation, the mean-field equations \rf{eq83} and \rf{eq86}
involve a new scalar amplitude of an additional mean small-scale neutral mode
of the operator of linearisation. An equation for the evolution of this
amplitude, presented in Appendix F, has a cubic nonlinearity. If, in addition,
the perturbed CHM state is non-steady and the symmetry in the system is
spatio-temporal (with a non-zero time shift $\widetilde{T}$), new terms
involving a non-local operator (the inverse
Laplacian) acting on quadratic products of the mean perturbation components
emerge in the amplitude equation.

No boundary conditions have been set for the mean fields of perturbation
in the horizontal directions, since they do not affect our derivation.
They can be imposed at infinity, or on the boundary of a finite region in
slow coordinates (e.g., of a region, the size of which grows as
$\varepsilon^{-1}$ in the original fast coordinates); they must be chosen
to serve specific applications of the theory that has been developed here.
An important requirement is that the perturbation
remains globally bounded -- otherwise the basic assumption on the smallness
of the perturbation and the asymptotic nature of the expansion become broken.

In general, the mean-field equations do not satisfy the usual equations
of energy balance, suggesting that their solutions are not uniformly bounded
in time, and may, perhaps, collapse in a finite time. Such a growth, of course,
makes the equations unsuitable for description of the subsequent evolution of
the perturbation. For instance, the evolution of a perturbation of a CHM
regime emerging in a pitchfork bifurcation is significantly affected
by the sign of the coefficient $C^c$ in the term $Cc_0^3$
in the equation for the amplitude of the neutral mode:
$C^c<0$ damps the growth of the amplitude and stabilises the perturbation,
and $C^c>0$ supports its infinite superexponential growth.

We have derived the mean field equations for a perturbation of a CHM system
in a horizontal plane layer, translation invariance in which is broken by the presence
of source terms in the governing equations (\theCHMB)--\rf{eq3}.
Amplitude equations
for a large-scale perturbation of a source-free CHM system will be derived
in a sequel to the present paper. In the absence
of the sources, or if an unsteady CHM state emerging under the action of steady sources is
considered, the structure of the null space of the operator of linearisation
$\cal M$ is significantly different from the generic one, assumed here,
resulting in a different set of amplitude equations (not all of which will
be evolutionary). In a future study we will also analyse numerically solutions
to the amplitude equations for steady and periodic in time CHM regimes,
symmetric about a vertical axis. In particular, this will reveal the nature of
physical effects described by the new operators emerging in the amplitude
equations.

A similar derivation can be performed for a perturbation of a CHM regime in
a spherical shell by considering
the equations of hydromagnetic convection in spherical coordinates. These
equations will be more appropriate for geophysical applications, but their
structure is significantly more complex.

\mi{\bf Acknowledgments}

\mi
I am grateful to Dr. O.~Podvigina and Profs.~U.Frisch and M.Proctor for
stimulating discussions.
Part of this research has been carried out during my visit to the School of
Engineering, Computer Science and Mathematics, University of Exeter, UK, in
January -- May 2006. I~am grateful to the Royal Society and the University of
Exeter for their financial support. My research visits to Observatoire de la
C\^ote d'Azur were supported by the French Ministry of Education.

\mi{\bf Appendix A. The hierarchies of equations\\ for weakly nonlinear perturbations}

\mi
The following equations arise at order $\varepsilon^n$ after substitution
of the series (\thepertser) for the perturbation into \xrf{eq5p1}{eq5p3} and
expansion in power series, for the perturbed CHM state $\bOm,{\bf V,H},\Theta$
independent of~$\varepsilon$:
$${\cal L}^\omega(\bom_n,{\bf v}_n,{\bf h}_n,\theta_n)-{\partial\bom_{n-2}\over\partial T}
+\nu\lpar 2(\nabla_{\bf x}\cdot\nabla_{\bf X})\lb\bom_{n-1}\rb_v+\nabla^2_{\bf X}\bom_{n-2}\rpar$$
$$+\nabla_{\bf X}\times\lpar{\bf V}\times\bom_{n-1}+{\bf v}_{n-1}\times\bOm
-{\bf H}\times(\nabla_{\bf X}\times{\bf h}_{n-2}+\nabla_{\bf x}\times{\bf h}_{n-1})
-{\bf h}_{n-1}\times(\nabla_{\bf x}\times{\bf H})$$
$$+\sum_{k=0}^{n-2}({\bf v}_k\times\bom_{n-2-k}
-{\bf h}_k\times(\nabla_{\bf x}\times{\bf h}_{n-2-k}+\nabla_{\bf X}\times{\bf h}_{n-3-k}))\rpar$$
$$+\nabla_{\bf x}\times\lpar\sum_{k=0}^{n-1}({\bf v}_k\times\bom_{n-1-k}
-{\bf h}_k\times(\nabla_{\bf x}\times{\bf h}_{n-1-k}
+\nabla_{\bf X}\times{\bf h}_{n-2-k}))$$
$$-{\bf H}\times(\nabla_{\bf X}\times{\bf h}_{n-1})\rpar
+\beta\nabla_{\bf X}\theta_{n-1}\times{\bf e}_3=0,\eqn{A1}$$
\pagebreak
$${\cal L}^h({\bf v}_n,{\bf h}_n)-{\partial{\bf h}_{n-2}\over\partial T}
+\eta\lpar 2(\nabla_{\bf x}\cdot\nabla_{\bf X})\lb{\bf h}_{n-1}\rb_h+\nabla^2_{\bf X}{\bf h}_{n-2}\rpar$$
$$+\nabla_{\bf X}\times\lpar{\bf v}_{n-1}\times{\bf H}+{\bf V}\times{\bf h}_{n-1}+\sum_{k=0}^{n-2}{\bf v}_k\times{\bf h}_{n-2-k}\rpar
+\nabla_{\bf x}\times\sum_{k=0}^{n-1}{\bf v}_k\times{\bf h}_{n-1-k}=0,\eqn{A2}$$

$${\cal L}^\theta({\bf v}_n,\theta_n)-{\partial\theta_{n-2}\over\partial T}
+\kappa\lpar 2(\nabla_{\bf x}\cdot\nabla_{\bf X})\theta_{n-1}+\nabla^2_{\bf X}\theta_{n-2}\rpar$$
$$-({\bf V}\cdot\nabla_{\bf X})\theta_{n-1}
-\sum_{k=0}^{n-1}({\bf v}_k\cdot\nabla_{\bf x})\theta_{n-1-k}
-\sum_{k=0}^{n-2}({\bf v}_k\cdot\nabla_{\bf X})\theta_{n-2-k}=0.\eqn{A3}$$

The following equations result from \xrf{eq5p1}{eq5p3} at order
$\varepsilon^n$ for the perturbed CHM state \rf{eq65} for the parameter
dependence \rf{eq66}:
$${\cal L}^\omega(\bom_n,{\bf v}_n,{\bf h}_n,\theta_n)-{\partial\bom_{n-2}\over\partial T}
+\nu\lpar 2(\nabla_{\bf x}\cdot\nabla_{\bf X})\lb\bom_{n-1}\rb_v+\nabla^2_{\bf X}\bom_{n-2}\rpar$$
$$+\nabla_{\bf X}\times\lpar
\sum_{k=0}^{n-1}({\bf V}_k\times\bom_{n-1-k}+{\bf v}_{n-1-k}\times\bOm_k
-{\bf H}_k\times(\nabla_{\bf x}\times{\bf h}_{n-1-k})
-{\bf h}_{n-1-k}\times(\nabla_{\bf x}\times{\bf H}_k))$$
$$+\sum_{k=0}^{n-2}({\bf v}_k\times\bom_{n-2-k}
-{\bf H}_k\times(\nabla_{\bf X}\times{\bf h}_{n-2-k})
-{\bf h}_k\times(\nabla_{\bf x}\times{\bf h}_{n-2-k}+\nabla_{\bf X}\times{\bf h}_{n-3-k}))\rpar$$
$$+\nabla_{\bf x}\times\lpar
\sum_{k=1}^n({\bf V}_k\times\bom_{n-k}+{\bf v}_{n-k}\times\bOm_k
-{\bf H}_k\times(\nabla_{\bf x}\times{\bf h}_{n-k})
-{\bf h}_{n-k}\times(\nabla_{\bf x}\times{\bf H}_k))$$
$$+\sum_{k=0}^{n-1}({\bf v}_k\times\bom_{n-1-k}
-{\bf H}_k\times(\nabla_{\bf X}\times{\bf h}_{n-1-k})
-{\bf h}_k\times(\nabla_{\bf x}\times{\bf h}_{n-1-k}
+\nabla_{\bf X}\times{\bf h}_{n-2-k}))\rpar$$
$$+(\beta_2\nabla_{\bf x}\theta_{n-2}+\beta_2\nabla_{\bf X}\theta_{n-3}
+\beta_0\nabla_{\bf X}\theta_{n-1})\times{\bf e}_3=0,\eqn{A4}$$

$${\cal L}^h({\bf v}_n,{\bf h}_n)-{\partial{\bf h}_{n-2}\over\partial T}
+\eta\lpar 2(\nabla_{\bf x}\cdot\nabla_{\bf X})\lb{\bf h}_{n-1}\rb_h+\nabla^2_{\bf X}{\bf h}_{n-2}\rpar
+\nabla_{\bf x}\times\sum_{k=1}^n\lpar{\bf v}_{n-k}\times({\bf H}_k+{\bf h}_{k-1})$$
$$+{\bf V}_k\times{\bf h}_{n-k}\rpar
+\nabla_{\bf X}\times\sum_{k=0}^{n-1}\lpar{\bf v}_{n-1-k}\times({\bf H}_k+{\bf h}_{k-1})
+{\bf V}_k\times{\bf h}_{n-1-k}\rpar=0;\eqn{A5}$$

$${\cal L}^\theta({\bf v}_n,\theta_n)-{\partial\theta_{n-2}\over\partial T}
+\kappa\lpar 2(\nabla_{\bf x}\cdot\nabla_{\bf X})\theta_{n-1}+\nabla^2_{\bf X}\theta_{n-2}\rpar
-\sum_{k=1}^n\lpar({\bf V}_k\cdot\nabla_{\bf x})\theta_{n-k}$$
$$+({\bf v}_{n-k}\cdot\nabla_{\bf x})(\Theta_k+\theta_{k-1})
+({\bf V}_{n-k}\cdot\nabla_{\bf X})\Theta_{k-1}
+({\bf v}_{n-1-k}\cdot\nabla_{\bf X})\theta_{k-1}\rpar=0.\eqn{A6}$$

Any quantity with a negative subscript is assumed in this paper to be zero.

\pagebreak
\mi{\bf Appendix B. Bounds for a solution to the problem (\theMsys)\\
for a linearly stable CHM state}

\mi Suppose a CHM
state ${\bf V,H},\Theta$ is linearly stable to small-scale perturbations, which
is understood here as an exponential decay of any solution to the problem
$${\cal M}({\bf w}^\cdot)=0,\eqn{B1}$$
$$\nabla_{\bf x}\cdot{\bf w}^\omega=\nabla_{\bf x}\cdot{\bf w}^h=0,\quad
\la{\bf w}^\omega\ra_v=\la{\bf w}^h\ra_h=0\eqn{B2}$$
in any suitable norm $\|\cdot\|$, e.g., in the norm of the space of $N$ times
differentiable functions $C^N$. We show in this appendix, that for such
a CHM state any solution $\bf w=(\bom',{\bf h}',\theta)$ to the problem
(\theMsys) is globally bounded in this norm in time for any permissible initial
conditions and right-hand sides ${\bf f}=({\bf f}'^\omega,{\bf f}'^h,f'^\theta)$,
and, moreover, it decays exponentially in fast time, if $\bf f$ does so. (Slow
variables play the r\^ole of passive parameters here, and we do not explicitly
refer to them.)

By the assumption on linear stability of the CHM state,
any solution to the problem (B1)--(B2) satisfies the inequality
$$\|{\bf w}({\bf x},t_1)\|\le Ce^{-\sigma(t_1-t_2)}\|{\bf w}({\bf x},t_2)\|\eqn{B3}$$
for any $t_1\ge t_2\ge 0$ and any initial conditions satisfying (B2),
where $C$ and $\sigma>0$ are constants, independent of $t_1,\,t_2$ and $\bf w$.

By linearity, split a solution of (\theMsys) into the sum ${\bf w}={\bf w}_I+{\bf w}_{II}$,
where ${\bf w}_I$ is a solution to (B1)--(B2) with the initial condition
${\bf w}_I|_{t=0}={\bf w}({\bf x},0)$, and ${\bf w}_{II}$ is a solution to
(\theMsys) with ${\bf w}_{II}|_{t=0}=0$. By virtue of (B3),
$$\|{\bf w}_I({\bf x},t)\|\le Ce^{-\sigma t}\|{\bf w}({\bf x},0)\|.\eqn{B4}$$
By the Duhamel principle (see, e.g., Polyanin, 2001)
$${\bf w}_{II}=\int_0^t\widetilde{\bf w}({\bf x},t,\tau)\,d\tau,\eqn{B5}$$
where $\widetilde{\bf w}({\bf x},t,\tau)$ is a solution to
(B1)--(B2) for $t>\tau$ with the initial condition
$\widetilde{\bf w}({\bf x},t,\tau)|_{t=\tau}={\bf f(x},\tau)$.
Hence, by (B3),
$$\|\widetilde{\bf w}({\bf x},t,\tau)\|\le Ce^{-\sigma(t-\tau)}\|{\bf f}({\bf x},\tau)\|,$$
which implies
$$\|{\bf w}_{II}({\bf x},t)\|\le{C\over\sigma}\max_{\tau\le t}\|{\bf f}({\bf x},\tau)\|.\eqn{B6}$$
Combination of (B4) and (B6) shows, that ${\bf w}={\bf w}_I+{\bf w}_{II}$
is globally bounded in time in the norm $\|\cdot\|$.

Now, suppose $\bf f$ satisfies the inequality
$$\|{\bf f}({\bf x},t)\|\le C_{\bf f}e^{-\sigma t},$$
where $C_{\bf f}$ is a time-independent constant (without any loss of
generality we assume the same exponent $\sigma>0$ in this inequality and
in (B3)~). Then (B3) implies
$$\|\widetilde{\bf w}({\bf x},t,\tau)\|\le Ce^{-\sigma(t-\tau)}C_{\bf f}e^{-\sigma\tau}
=CC_{\bf f}e^{-\sigma t},$$
and thus from (B5)
$$\|{\bf w}_{II}({\bf x},t)\|\le CC_{\bf f}\,te^{-\sigma t}\le C'e^{-\sigma't}$$
for any $\sigma'<\sigma$, i.e. $\|{\bf w}_{II}({\bf x},t)\|$ and hence
$\|{\bf w}\|$ decay exponentially.

\mi{\bf Appendix C. A bound for a solenoidal field}

\mi We demonstrate here that if $\nabla_{\bf x}\cdot{\bf v}=0$, then
$\la\,\lb{\bf v(x,X)}\rb_h|_{{\bf X}=\varepsilon(x_1,x_2)}\ra$ is asymptotically
smaller than any power of $\varepsilon$.

Let $\lsb f\rsb$ denote in this appendix the average
of $f$ over the plane of the fast horizontal spatial variables $x_1,x_2$.
By the definition of $\lb\cdot\rb_h$ the horizontal component of
$\int_{-L/2}^{L/2}\lsb\lb{\bf v}\rb_h\rsb\,dx_3$ vanishes for all $\bf X$.
Averaging the solenoidality condition $\nabla_{\bf x}\cdot\lb{\bf v}\rb_h=0$
over the fast horizontal variables and integrating it in
the vertical direction, find that also $\lsb(\lb{\bf v}\rb_h)_3\rsb=0$ for all
$\bf X$. Thus it suffices to show that $\la(\lb{\bf v}\rb_h
-\lsb\lb{\bf v}\rb_h\rsb)|_{{\bf X}=\varepsilon(x_1,x_2)}\ra$ is asymptotically
smaller than any power of $\varepsilon$.

The equation
$$\left({\partial^2\over\partial x_1^2}+{\partial^2\over\partial x_2^2}\right)
\bphi=\lb{\bf v}\rb_h-\lsb\lb{\bf v}\rb_h\rsb$$
has a mean-free solution, globally bounded together with its derivatives, since
the average $\lsb\cdot\rsb$ of its left-hand side vanishes. Denote
$\bphi_i=\partial\bphi/\partial x_i$. Then by the chain rule
$$\left\la(\lb{\bf v}\rb_h-\lsb\lb{\bf v}\rb_h\rsb)|_{{\bf X}=\varepsilon(x_1,x_2)}\right\ra
=\left\la\left({\partial\bphi_1\over\partial x_1}
+\left.{\partial\bphi_2\over\partial x_2}\right)\right|_{{\bf X}=\varepsilon(x_1,x_2)}\right\ra$$
$$=\left\la{d\bphi_1\over dx_1}+{d\bphi_2\over dx_2}
-\varepsilon\left({\partial\bphi_1\over\partial X_1}
+\left.{\partial\bphi_2\over\partial X_2}\right)\right|_{{\bf X}=\varepsilon(x_1,x_2)}\right\ra
=-\varepsilon\left\la\left({\partial\bphi_1\over\partial X_1}
+\left.{\partial\bphi_2\over\partial X_2}\right)\right|_{{\bf X}=\varepsilon(x_1,x_2)}\right\ra.$$
(Here $d/dx_k$ denotes ``the complete partial derivative in $x_k$''.)
Since the partial derivatives have a zero mean over the plane of the
fast horizontal variables, the same operation can be applied to them
an arbitrary number of times. This concludes the demonstration.

A similar statement follows from the same arguments: if $f({\bf x},t,{\bf X},T)$
is smooth and globally bounded with its derivatives, and $\la f\ra=0$, then
$\la f|_{{\bf X}=\varepsilon(x_1,x_2)}\ra=O(\varepsilon^n)$ for any $n>0$.

\mi{\bf Appendix D. Exponential decay of the fluctuating part\\
of an integral in time.}

\mi
Let us show that $\lbd\int_0^t\bxi(t')dt'\rbd$ exponentially decays
in time in any norm $\|\cdot\|$, in which $\bxi(t)$ does. Suppose
$$\|\bxi(t)\|\le Ce^{-\sigma t},$$
where $\sigma>0$ and $C$ are some constants. Then
\pagebreak
$$\lbdb\int_0^t\bxi(t')dt'\rbdb=\int_0^t\bxi(t')dt'-\lim_{\hat{t}\to\infty}{1\over\hat{t}}
\int_0^{\hat{t}}\int_0^{t''}\bxi(t')dt'dt''$$
$$=\lim_{\hat{t}\to\infty}{1\over\hat{t}}
\int_0^{\hat{t}}\int_{t''}^t\bxi(t')dt'dt''
=-\lim_{\hat{t}\to\infty}{1\over\hat{t}}\int_t^{\hat{t}}\int_t^{t''}\bxi(t')dt'dt''$$
$$\Rightarrow\quad\left\|\lbdb\int_0^t\bxi(t')dt'\rbdb\right\|\le\lim_{\hat{t}\to\infty}
{1\over\hat{t}}\int_t^{\hat{t}}\int_t^{t''}Ce^{-\sigma t'}dt'dt''
\le\int_t^\infty Ce^{-\sigma t'}dt'=(C/\sigma)e^{-\sigma t}.$$

\mi{\bf Appendix E. A simplified form of the combined eddy diffusivity operator}

\mi
A simplified expression for the eddy diffusivity operator can be
derived for solenoidal $\lad{\bf v}_0\rad_h$ and $\lad{\bf h}_0\rad_h$:
$$\nabla_{\bf X}\times\sum_{j=1}^2\sum_{m=1}^2\sum_{k=1}^2{\partial^2\over\partial X_j\partial X_m}
\left({\bf D}^{v,v}_{m,k,j}\lad v_0\rad_k+{\bf D}^{h,v}_{m,k,j}\lad h_0\rad_k\right)$$
$$=\nabla_{\bf X}\times\sum_{k=1}^2\left(\left(\widehat{D}^{v,v}_{1,k}{\partial^2\over\partial X_k^2}
+\widehat{D}^{v,v}_{2,k}{\partial^2\over\partial X_{3-k}^2}
+\widehat{D}^{v,v}_{3,k}{\partial^2\over\partial X_1\partial X_2}
\right)\lad v_0\rad_k\right.$$
$$+\left.\left(\widehat{D}^{h,v}_{1,k}{\partial^2\over\partial X_k^2}
+\widehat{D}^{h,v}_{2,k}{\partial^2\over\partial X_{3-k}^2}
+\widehat{D}^{h,v}_{3,k}{\partial^2\over\partial X_1\partial X_2}
\right)\lad h_0\rad_k\right){\bf e}_k,$$
where, for $k=1,2$ and $n=3-k$,
$$\widehat{D}^{v,v}_{1,k}={1\over2}\left(
(D^{v,v}_{1,1,1})_1-(D^{v,v}_{1,2,2})_1-(D^{v,v}_{2,2,1})_1
+(D^{v,v}_{2,2,2})_2-(D^{v,v}_{1,1,2})_2-(D^{v,v}_{2,1,1})_2\right),$$
$$\widehat{D}^{v,v}_{2,k}=(D^{v,v}_{n,k,n})_k,\qquad
\widehat{D}^{v,v}_{3,k}=(D^{v,v}_{1,k,2})_k+(D^{v,v}_{2,k,1})_k
-(D^{v,v}_{n,n,n})_k-(D^{v,v}_{n,k,n})_n;$$
changing here the first superscript ``$v$'' to ``$h$'', obtain the expressions
for $\widehat{D}^{h,v}_{p,k}$.

Similarly,
$$\nabla_{\bf X}\times\sum_{m=1}^2\sum_{k=1}^2{\partial\over\partial X_m}\left(
{\bf D}^{v,h}_{m,k}\lad v_0\rad_k+{\bf D}^{h,h}_{m,k}\lad h_0\rad_k\right)$$
$$=\nabla_{\bf X}\times\sum_{m=1}^2\sum_{k=1}^2{\partial\over\partial X_m}\left(
\widehat{D}^{v,h}_{m,k}\lad v_0\rad_k+\widehat{D}^{h,h}_{m,k}\lad h_0\rad_k\right){\bf e}_k,$$
where, again denoting $n=3-k$,
$$\widehat{D}^{v,h}_{n,k}=(D^{v,h}_{n,k})_k,\qquad
\widehat{D}^{v,h}_{k,k}=(D^{v,h}_{k,k})_k-(D^{v,h}_{n,n})_k-(D^{v,h}_{n,k})_n,$$
and expressions for $\widehat{D}^{h,h}_{p,k}$ are obtained changing here
the first superscript ``$v$'' to ``$h$''.

\mi{\bf Appendix F. Equation for the amplitude of the mean-free neutral
mode}

\mi
In this appendix one of the main results of the paper is presented, the
equation for the amplitude of the mean-free neutral mode, $c_0$ (see \rf{eq76}):
\pagebreak
$${\partial\over\partial T}\left(c_0+\sum_{k=1}^2
\left(P^v_k\lad v_0\rad_k+P^h_k\lad h_0\rad_k\right)\right)
=\left.\sum_{j=1}^2{\partial\over\partial X_j}\right({\cal A}^c_jc_0+A^c_jc_0^2$$
$$+\sum_{k=1}^2\left({\cal A}^{v,c}_{k,j}\lad v_0\rad_k
+{\cal A}^{h,c}_{k,j}\lad h_0\rad_k+D^{c,c}_{k,j}{\partial c_0\over\partial X_k}
\left.+\sum_{m=1}^2\lpar D^{v,c}_{m,k,j}{\partial\lad v_0\rad_k\over\partial X_m}
+D^{h,c}_{m,k,j}{\partial\lad h_0\rad_k\over\partial X_m}\rpar\right)\right)$$
$$+q^cc_0+q^{cc}c_0^2+\left.\sum_{k=1}^2\right(q^v_k\lad v_0\rad_k+q^h_k\lad h_0\rad_k
+q^{cv}_k\lad v_0\rad_k c_0+q^{ch}_k\lad h_0\rad_k c_0$$
$$\left.+\sum_{m=1}^2\lpar q^{vv}_{m,k}\lad v_0\rad_k\lad v_0\rad_m
+q^{vh}_{m,k}\lad v_0\rad_k\lad h_0\rad_m
+q^{hh}_{m,k}\lad h_0\rad_k\lad h_0\rad_m\rpar\right)$$
$$+\sum_{j=1}^2\sum_{k=1}^2\left(
{\partial c_0\over\partial X_j}\left(A^{cv,c}_{k,j}\lad v_0\rad_k+A^{ch,c}_{k,j}\lad h_0\rad_k\right)
+c_0{\partial\over\partial X_j}\left(A^{vc,c}_{k,j}\lad v_0\rad_k+A^{hc,c}_{k,j}\lad h_0\rad_k\right)\right.$$
$$+\sum_{m=1}^2\!\!\left.\left({\partial\lad v_0\rad_k\over\partial X_j}
\left(A^{vv,c}_{m,k,j}\lad v_0\rad_m\!\!+\!A^{vh,c}_{m,k,j}\lad h_0\rad_m\right)
\!+{\partial\lad h_0\rad_k\over\partial X_j}
\left(A^{hv,c}_{m,k,j}\lad v_0\rad_m\!\!+\!A^{hh,c}_{m,k,j}\lad h_0\rad_m\right)\right)\!\right)$$
$$+\,C^cc_0^3\,+\sum_{k=1}^2\!\left(c_0^2\lpar C^v_k\lad v_0\rad_k+C^h_k\lad h_0\rad_k\rpar
\!+\!\sum_{m=1}^2\right(\!c_0\lpar C^{vv}_{m,k}\lad v_0\rad_k\lad v_0\rad_m
+C^{vh}_{m,k}\lad v_0\rad_k\lad h_0\rad_m$$
$$+\,C^{hh}_{m,k}\lad h_0\rad_k\lad h_0\rad_m\rpar
+\sum_{j=1}^2\lpar C^{vvv}_{m,k,j}\lad v_0\rad_k\lad v_0\rad_m\lad v_0\rad_j
+C^{vvh}_{m,k,j}\lad v_0\rad_k\lad v_0\rad_m\lad h_0\rad_j$$
$$+\,C^{vhh}_{m,k,j}\lad v_0\rad_k\lad h_0\rad_m\lad h_0\rad_j
+C^{hhh}_{m,k,j}\lad h_0\rad_k\lad h_0\rad_m\lad h_0\rad_j\rpar
\left)\left)\phantom{\sum_k^n\!\!\!\!}\right.\right.$$
$$+\sum_{i=1}^2\sum_{j=1}^2\sum_{n=1}^2\left.
{\partial^3\over\partial X_i\partial X_j\partial X_n}\nabla^{-2}_{\bf X}
\right(N^c_{i,n,j}c_0+N^{cc}_{i,n,j}c_0^2$$
$$+\sum_{m=1}^2\left.\left({\partial\over\partial X_m}\right(N^c_{i,n,m,j}c_0
+\sum_{k=1}^2\lpar N^v_{i,n,m,k,j}\lad v_0\rad_k+N^h_{i,n,m,k,j}\lad h_0\rad_k\rpar\right)$$
$$+N^v_{i,n,m,j}\lad v_0\rad_m+N^h_{i,n,m,j}\lad h_0\rad_m
+N^{cv}_{i,n,m,j}\lad v_0\rad_mc_0+N^{ch}_{i,n,m,j}\lad h_0\rad_mc_0$$
$$\left.+\sum_{k=1}^2\right(
N^{vv}_{i,n,m,k,j}\lad v_0\rad_k\lad v_0\rad_m
+N^{vh}_{i,n,m,k,j}\lad v_0\rad_k\lad h_0\rad_m
+N^{hh}_{i,n,m,k,j}\lad h_0\rad_k\lad h_0\rad_m$$
$$+{\partial^2\over\partial X_m\partial X_k}\nabla^{-2}_{\bf X}
\left.\left.\left.\left({\partial\over\partial X_1}
\left(K^v_{i,n,m,k,j}\lad v_0\rad_1+K^h_{i,n,m,k,j}\lad h_0\rad_1\right)
+\sum_{p=1}^2K^c_{i,n,m,k,j,p}{\partial c_0\over\partial X_p}\right)\right)\right)\right)$$
$$+\sum_{i=1}^2\sum_{m=1}^2\left({\partial^3\nabla^{-2}_{\bf X}\lad v_0\rad_1\over\partial X_i\partial X_m\partial X_1}
\left(M^{cv}_{i,m}c_0+\sum_{j=1}^2\left(M^{vv}_{i,m,j}\lad v_0\rad_j
+M^{hv}_{i,m,j}\lad h_0\rad_j\right)\right)\right.$$
$$+{\partial^3\nabla^{-2}_{\bf X}\lad h_0\rad_1\over\partial X_i\partial X_m\partial X_1}
\left(M^{ch}_{i,m}c_0+\sum_{j=1}^2\left(M^{vh}_{i,m,j}\lad v_0\rad_j
+M^{hh}_{i,m,j}\lad h_0\rad_j\right)\right)$$
$$+\left.\sum_{k=1}^2{\partial^3\nabla^{-2}_{\bf X}c_0\over\partial X_i\partial X_m\partial X_k}
\left(M^{cc}_{i,m,k}c_0+\sum_{j=1}^2\left(M^{cv}_{i,m,k,j}\lad v_0\rad_j
+M^{ch}_{i,m,k,j}\lad h_0\rad_j\right)\right)\right).$$

\pagebreak\noindent
It is convenient to represent its coefficients in the terms of the
10-dimensional fields
$${\bf W}_i=(\bOm_i,{\bf V}_i,{\bf H}_i,\Theta_i),$$
$$\widetilde{\bf S}^{v,\cdot}_k=({\bf S}^{v,\omega}_k,{\bf S}^{v,v}_k+{\bf e}_k,
{\bf S}^{v,h}_k,S^{v\theta}_k),\qquad
\widetilde{\bf S}^{h,\cdot}_k=({\bf S}^{h,\omega}_k,{\bf S}^{h,v}_k,
{\bf S}^{h,h}_k+{\bf e}_k,S^{h,\theta}_k),$$
$$\widetilde{\bf G}^{v,\cdot}_{m,k}=\lpar{\bf G}_{m,k}^{v,\omega}+\epsilon_{m,k,3}{\bf e}_3,
\widetilde{\bf G}_{m,k}^{v,v},{\bf G}_{m,k}^{v,h},G_{m,k}^{v,\theta}\rpar,$$
$$\widetilde{\bf G}^{h,\cdot}_{m,k}=\lpar{\bf G}_{m,k}^{v,\omega},
\widetilde{\bf G}_{m,k}^{h,v},{\bf G}_{m,k}^{h,h},G_{m,k}^{h,\theta}\rpar,\qquad
\widetilde{\bf G}^{c,\cdot}_k=\lpar{\bf G}_k^{c,\omega},
\widetilde{\bf G}_k^{c,v},{\bf G}_k^{c,h},G_k^{c,\theta}\rpar,$$
$$\widetilde{\bf Y}^{v,\cdot}_{m,k}=\lpar{\bf Y}_{m,k}^{v,\omega},
\widetilde{\bf Y}_{m,k}^{v,v},{\bf Y}_{m,k}^{v,h},Y_{m,k}^{v,\theta}\rpar,\qquad
\widetilde{\bf Y}^{h,\cdot}_{m,k}=\lpar{\bf Y}_{m,k}^{h,\omega},
\widetilde{\bf Y}_{m,k}^{h,v},{\bf Y}_{m,k}^{h,h},Y_{m,k}^{h,\theta}\rpar,$$
$$\widetilde{\bf Y}^{c,\cdot}_{m,k,j}=\lpar{\bf Y}_{m,k,j}^{c,\omega},
\widetilde{\bf Y}_{m,k,j}^{c,v},{\bf Y}_{m,k,j}^{c,h},Y_{m,k,j}^{c,\theta}\rpar,$$
bilinear forms
$${\cal B}({\bf p}^\cdot,{\bf q}^\cdot)\equiv
\lad\left({\bf p}^v\times{\bf q}^\omega+{\bf q}^v\times{\bf p}^\omega
-{\bf p}^h\times(\nabla_{\bf x}\times{\bf q}^h)
-{\bf q}^h\times(\nabla_{\bf x}\times{\bf p}^h)\right)\cdot
(\nabla_{\bf x}\times{\bf S}^\omega_*)$$
$$+\left({\bf p}^v\times{\bf q}^h+{\bf q}^v\times{\bf p}^h\right)\cdot
(\nabla_{\bf x}\times{\bf S}^h_*)-\left(({\bf p}^v\cdot\nabla_{\bf x})q^\theta
+({\bf q}^v\cdot\nabla_{\bf x})p^\theta\right)S^\theta_*\rad,$$
$${\cal C}_j({\bf p}^\cdot,{\bf q}^\cdot)\equiv
\lad\lpar{\bf p}^v\times{\bf q}^\omega+{\bf q}^v\times{\bf p}^\omega
-{\bf p}^h\times(\nabla_{\bf x}\times{\bf q}^h)
-{\bf q}^h\times(\nabla_{\bf x}\times{\bf p}^h)\rpar\times\widetilde{\bf S}^\omega_*$$
$$+((\nabla_{\bf x}\times\widetilde{\bf S}^\omega_*)\times{\bf p}^h)\times{\bf q}^h
+\lpar{\bf q}^v\times{\bf p}^h+{\bf p}^v\times{\bf q}^h\rpar\times\widetilde{\bf S}^h_*
-{\bf p}^vq^\theta\widetilde{S}^\theta_*\rad_j$$
of vector fields ${\bf p}^\cdot=({\bf p}^\omega,{\bf p}^v,{\bf p}^h,p^\theta)$
and ${\bf q}^\cdot=({\bf q}^\omega,{\bf q}^v,{\bf q}^h,q^\theta)$,
and of linear forms
$${\cal E}({\bf f})\equiv
\lad({\bf f}\times\bOm_0)\cdot(\nabla_{\bf x}\times{\bf S}^\omega_*)
+({\bf f}\times{\bf H}_0)\cdot(\nabla_{\bf x}\times{\bf S}^h_*)
-({\bf f}\cdot\nabla_{\bf x})\Theta_0S^\theta_*\rad,$$
$${\cal F}_j({\bf p}^\cdot)\equiv 2\ladb\nu{\partial{\bf p}^\omega\over\partial x_j}\!\cdot{\bf S}^\omega_*
\!+\!\eta{\partial{\bf p}^h\over\partial x_j}\!\cdot{\bf S}^h_*
\!+\!\kappa{\partial p^\theta\over\partial x_j}S^\theta_*\radb
+{\cal C}_j\left({\bf W}_0,{\bf p}^\cdot\right)
+\beta_0\lad p^\theta{\bf e}_3\times{\bf S}^\omega_*\rad_j.$$
In this notation,
$$P^v_k=\lad{\bf S}_k^{v,\omega}\cdot{\bf S}^\omega_*
+{\bf S}_k^{v,h}\cdot{\bf S}^h_*+S_k^{v,\theta}S^\theta_*\rad,\qquad
P^h_k=\lad{\bf S}_k^{h,\omega}\cdot{\bf S}^\omega_*
+{\bf S}_k^{h,h}\cdot{\bf S}^h_*+S_k^{h,\theta}S^\theta_*\rad,$$
$${\cal A}^c_j={\cal F}_j(\widehat{\bf S}^\cdot)+{\cal E}(\widetilde{\Alfa}^{c,v}_j)
+{\cal B}({\bf W}_1,\widetilde{\bf G}_j^{c,\cdot})+{\cal C}_j({\bf W}_1,{\bf S}^\cdot),$$
$${\cal A}^{v,c}_{k,j}\!=\!{\cal F}_j(\widehat{\bf S}^{v,\cdot}_k)+{\cal E}(\widetilde{\Alfa}^{v,v}_{k,j})
+{\cal B}({\bf W}_1,\widetilde{\bf G}_{j,k}^{v,\cdot})
+{\cal C}_j({\bf W}_1,\widetilde{\bf S}^{v,\cdot}_k),$$
$${\cal A}^{h,c}_{k,j}={\cal F}_j(\widehat{\bf S}^{h,\cdot}_k)+{\cal E}(\widetilde{\Alfa}^{h,v}_{k,j})
+{\cal B}({\bf W}_1,\widetilde{\bf G}_{j,k}^{h,\cdot})
+{\cal C}_j({\bf W}_1,\widetilde{\bf S}^{h,\cdot}_k),$$
$$A^c_j={\cal F}_j({\bf Q}^{cc,\cdot})+{\cal E}(\widetilde{\bf A}^v_j)
+\left({\cal C}_j({\bf S}^\cdot,{\bf S}^\cdot)
+{\cal B}({\bf S}^\cdot,\widetilde{\bf G}_j^{c,\cdot})\right)/2,$$
$$A^{cv,c}_{k,j}={\cal F}_j({\bf Q}^{cv,\cdot}_k)+{\cal E}(\widetilde{\bf A}^{cv,v}_{k,j})
+{\cal C}_j(\widetilde{\bf S}^{v,\cdot}_k,{\bf S}^\cdot)
+{\cal B}(\widetilde{\bf S}^{v,\cdot}_k,\widetilde{\bf G}_j^{c,\cdot}),$$
$$A^{ch,c}_{k,j}={\cal F}_j({\bf Q}^{ch,\cdot}_k)+{\cal E}(\widetilde{\bf A}^{ch,v}_{k,j})
+{\cal C}_j(\widetilde{\bf S}^{h,\cdot}_k,{\bf S}^\cdot)
+{\cal B}(\widetilde{\bf S}^{h,\cdot}_k,\widetilde{\bf G}_j^{c,\cdot}),$$
$$A^{vc,c}_{k,j}={\cal F}_j({\bf Q}^{cv,\cdot}_k)+{\cal E}(\widetilde{\bf A}^{cv,v}_{k,j})
+{\cal C}_j({\bf S}^\cdot,\widetilde{\bf S}^{v,\cdot}_k)
+{\cal B}({\bf S}^\cdot,\widetilde{\bf G}_{j,k}^{v,\cdot}),$$
$$A^{hc,c}_{k,j}={\cal F}_j({\bf Q}^{ch,\cdot}_k)+{\cal E}(\widetilde{\bf A}^{ch,v}_{k,j})
+{\cal C}_j({\bf S}^\cdot,\widetilde{\bf S}^{h,\cdot}_k)
+{\cal B}({\bf S}^\cdot,\widetilde{\bf G}_{j,k}^{h,\cdot}),$$
$$A^{vv,c}_{m,k,j}={\cal F}_j({\bf Q}^{vv,\cdot}_{m,k}+{\bf Q}^{vv,\cdot}_{k,m})
+{\cal E}(\widetilde{\bf A}^{vv,v}_{m,k,j}+\widetilde{\bf A}^{vv,v}_{k,m,j})
+{\cal C}_j(\widetilde{\bf S}^{v,\cdot}_m,\widetilde{\bf S}^{v,\cdot}_k)
+{\cal B}(\widetilde{\bf S}^{v,\cdot}_m,\widetilde{\bf G}_{j,k}^{v,\cdot}),$$
$$A^{vh,c}_{m,k,j}={\cal F}_j({\bf Q}^{vh,\cdot}_{m,k})+{\cal E}(\widetilde{\bf A}^{vh,v}_{m,k,j})
+{\cal C}_j(\widetilde{\bf S}^{h,\cdot}_m,\widetilde{\bf S}^{v,\cdot}_k)
+{\cal B}(\widetilde{\bf S}^{h,\cdot}_m,\widetilde{\bf G}_{j,k}^{v,\cdot}),$$
$$A^{hv,c}_{m,k,j}={\cal F}_j({\bf Q}^{vh,\cdot}_{k,m})+{\cal E}(\widetilde{\bf A}^{vh,v}_{k,m,j})
+{\cal C}_j(\widetilde{\bf S}^{v,\cdot}_m,\widetilde{\bf S}^{h,\cdot}_k)
+{\cal B}(\widetilde{\bf S}^{v,\cdot}_m,\widetilde{\bf G}_{j,k}^{h,\cdot}),$$
$$A^{hh,c}_{m,k,j}={\cal F}_j({\bf Q}^{hh,\cdot}_{m,k}+{\bf Q}^{hh,\cdot}_{k,m})
+{\cal E}(\widetilde{\bf A}^{hh,v}_{m,k,j}+\widetilde{\bf A}^{hh,v}_{k,m,j})
+{\cal C}_j(\widetilde{\bf S}^{h,\cdot}_m,\widetilde{\bf S}^{h,\cdot}_k)
+{\cal B}(\widetilde{\bf S}^{h,\cdot}_m,\widetilde{\bf G}_{j,k}^{h,\cdot}),$$
\pagebreak
$$D^{c,c}_{k,j}={\cal F}_j(\widetilde{\bf G}^{c,\cdot}_k)+{\cal E}(\widetilde{\bf D}^{c,v}_{k,j})
+\lad(H_0)_j{\bf S}^h\times{\bf S}^\omega_*\rad_k
+\delta^k_j\lad\nu{\bf S}^\omega\cdot{\bf S}^\omega_*
+\eta{\bf S}^h\cdot{\bf S}^h_*+\kappa S^\theta S^\theta_*\rad,$$
$$D^{v,c}_{m,k,j}={\cal F}_j(\widetilde{\bf G}^{v,\cdot}_{m,k})+{\cal E}(\widetilde{\bf D}^{v,v}_{m,k,j})
+\lad(H_0)_j{\bf S}_k^{v,h}\times{\bf S}^\omega_*\rad_m
\!+\delta^m_j\lad\nu{\bf S}_k^{v,\omega}\cdot{\bf S}^\omega_*
+\eta{\bf S}_k^{v,h}\cdot{\bf S}^h_*+\kappa S_k^{v,\theta}S^\theta_*\rad,$$
$$D^{h,c}_{m,k,j}={\cal F}_j(\widetilde{\bf G}^{h,\cdot}_{m,k})+{\cal E}(\widetilde{\bf D}^{h,v}_{m,k,j})
+\lad(H_0)_j({\bf S}_k^{h,h}+{\bf e}_k)\times{\bf S}^\omega_*\rad_m$$
$$+\delta^m_j\lad\nu{\bf S}_k^{h,\omega}\cdot{\bf S}^\omega_*
+\eta{\bf S}_k^{h,h}\cdot{\bf S}^h_*+\kappa S_k^{h,\theta}S^\theta_*\rad$$
(where $\delta^m_j$ is the Kronecker symbol),
$$q^c={\cal B}({\bf W}_1,\widehat{\bf S}^\cdot)+{\cal B}({\bf W}_2,{\bf S}^\cdot)
-\beta_2\lad\nabla_{\bf x}S^\theta\times{\bf S}^\omega_*\rad_3,$$
$$q^v_k={\cal B}({\bf W}_1,\widehat{\bf S}^{v,\cdot}_k)
+{\cal B}({\bf W}_2,\widetilde{\bf S}^{v,\cdot}_k)
-\beta_2\lad\nabla_{\bf x}S^{v,\theta}_k\times{\bf S}^\omega_*\rad_3,$$
$$q^h_k={\cal B}({\bf W}_1,\widehat{\bf S}^{h,\cdot}_k)
+{\cal B}({\bf W}_2,\widetilde{\bf S}^{h,\cdot}_k)
-\beta_2\lad\nabla_{\bf x}S^{h,\theta}_k\times{\bf S}^\omega_*\rad_3,$$
$$q^{cc}={\cal B}({\bf W}_1,{\bf Q}^{cc,\cdot})
+{\cal B}(\widehat{\bf S}^\cdot,{\bf S}^\cdot),$$
$$q^{cv}_k={\cal B}({\bf W}_1,{\bf Q}^{cv,\cdot}_k)
+{\cal B}(\widehat{\bf S}^\cdot,\widetilde{\bf S}^{v,\cdot}_k)
+{\cal B}({\bf S}^\cdot,\widehat{\bf S}^{v,\cdot}_k),$$
$$q^{ch}_k={\cal B}({\bf W}_1,{\bf Q}^{ch,\cdot}_k)
+{\cal B}(\widehat{\bf S}^\cdot,\widetilde{\bf S}^{h,\cdot}_k)
+{\cal B}({\bf S}^\cdot,\widehat{\bf S}^{h,\cdot}_k),$$
$$q^{vv}_{m,k}\!={\cal B}({\bf W}_1,{\bf Q}^{vv,\cdot}_{m,k})
+{\cal B}(\widetilde{\bf S}^{v,\cdot}_m,\widehat{\bf S}^{v,\cdot}_k),\quad
q^{vh}_{m,k}\!={\cal B}({\bf W}_1,{\bf Q}^{vh,\cdot}_{m,k})
+{\cal B}(\widetilde{\bf S}^{v,\cdot}_k,\widehat{\bf S}^{h,\cdot}_m)
+{\cal B}(\widetilde{\bf S}^{h,\cdot}_m,\widehat{\bf S}^{v,\cdot}_k),$$
$$q^{hh}_{m,k}={\cal B}({\bf W}_1,{\bf Q}^{hh,\cdot}_{m,k})
+{\cal B}(\widetilde{\bf S}^{h,\cdot}_m,\widehat{\bf S}^{h,\cdot}_k),$$
$$C^c={\cal B}({\bf Q}^{cc,\cdot},{\bf S}^\cdot),$$
$$C^v_k={\cal B}({\bf Q}^{cv,\cdot}_k,{\bf S}^\cdot)
+{\cal B}({\bf Q}^{cc,\cdot},\widetilde{\bf S}^{v,\cdot}_k),\qquad
C^h_k={\cal B}({\bf Q}^{ch,\cdot}_k,{\bf S}^\cdot)
+{\cal B}({\bf Q}^{cc,\cdot},\widetilde{\bf S}^{h,\cdot}_k),$$
$$C^{vv}_{m,k}={\cal B}({\bf Q}^{vv,\cdot}_{m,k},{\bf S}^\cdot)
+{\cal B}({\bf Q}^{cv,\cdot}_k,\widetilde{\bf S}^{v,\cdot}_m),\qquad
C^{hh}_{m,k}={\cal B}({\bf Q}^{hh,\cdot}_{m,k},{\bf S}^\cdot)
+{\cal B}({\bf Q}^{ch,\cdot}_k,\widetilde{\bf S}^{h,\cdot}_m),$$
$$C^{vh}_{m,k}={\cal B}({\bf Q}^{vh,\cdot}_{m,k},{\bf S}^\cdot)
+{\cal B}({\bf Q}^{cv,\cdot}_k,\widetilde{\bf S}^{h,\cdot}_m)
+{\cal B}({\bf Q}^{ch,\cdot}_m,\widetilde{\bf S}^{v,\cdot}_k),$$
$$C^{vvh}_{m,k,j}={\cal B}({\bf Q}^{vv,\cdot}_{m,k},\widetilde{\bf S}^{h,\cdot}_j)
+{\cal B}({\bf Q}^{vh,\cdot}_{j,m},\widetilde{\bf S}^{v,\cdot}_k),\qquad
C^{vhh}_{m,k,j}={\cal B}({\bf Q}^{vh,\cdot}_{m,k},\widetilde{\bf S}^{h,\cdot}_j)
+{\cal B}({\bf Q}^{hh,\cdot}_{j,m},\widetilde{\bf S}^{v,\cdot}_k),$$
$$C^{vvv}_{m,k,j}={\cal B}({\bf Q}^{vv,\cdot}_{m,k},\widetilde{\bf S}^{v,\cdot}_j),\qquad
C^{hhh}_{m,k,j}={\cal B}({\bf Q}^{hh,\cdot}_{m,k},\widetilde{\bf S}^{h,\cdot}_j),$$
$$N^c_{i,n,m,j}=-{\cal E}((\widetilde{D}^{c,v}_{m,j})_n{\bf e}_i)
+{\cal E}(\widetilde{\bf d}^{c,v}_{i,n,m,j})
+{\cal F}_j(\widetilde{\bf Y}^{c,\cdot}_{i,n,m}),$$
$$N^v_{i,n,m,k,j}=-{\cal E}((\widetilde{D}^{v,v}_{m,k,j})_n{\bf e}_i)
+\delta^1_k\delta^1_m({\cal E}(\widetilde{\bf d}^{v,v}_{i,n,j})
+{\cal F}_j(\widetilde{\bf Y}^{v,\cdot}_{i,n})),$$
$$N^h_{i,n,m,k,j}=-{\cal E}((\widetilde{D}^{h,v}_{m,k,j})_n{\bf e}_i)
+\delta^1_k\delta^1_m({\cal E}(\widetilde{\bf d}^{h,v}_{i,n,j})
+{\cal F}_j(\widetilde{\bf Y}^{h,\cdot}_{i,n})),$$
$$N^c_{i,n,j}=-{\cal E}((\widetilde{\cal A}^{c,v}_j)_n{\bf e}_i)
+{\cal B}({\bf W}_1,\widetilde{\bf Y}^{c,\cdot}_{i,n,j}),$$
$$N^v_{i,n,m,j}=-{\cal E}((\widetilde{\cal A}^{v,v}_{m,j})_n{\bf e}_i)
+\delta^1_m\delta^1_j{\cal B}({\bf W}_1,\widetilde{\bf Y}^{v,\cdot}_{i,n}),$$
$$N^h_{i,n,m,j}=-{\cal E}((\widetilde{\cal A}^{h,v}_{m,j})_n{\bf e}_i)
+\delta^1_m\delta^1_j{\cal B}({\bf W}_1,\widetilde{\bf Y}^{h,\cdot}_{i,n}),$$
$$N^{cc}_{i,n,j}=-{\cal E}((\widetilde{A}^v_j)_n{\bf e}_i),\quad
N^{cv}_{i,n,m,j}=-{\cal E}((\widetilde{A}^{cv,v}_{m,j})_n{\bf e}_i),\quad
N^{ch}_{i,n,m,j}=-{\cal E}((\widetilde{A}^{ch,v}_{m,j})_n{\bf e}_i),$$
$$N^{vv}_{i,n,m,k,j}=-{\cal E}((\widetilde{A}^{vv,v}_{m,k,j})_n{\bf e}_i),\quad
N^{vh}_{i,n,m,k,j}=-{\cal E}((\widetilde{A}^{vh,v}_{m,k,j})_n{\bf e}_i),$$
$$N^{hh}_{i,n,m,k,j}=-{\cal E}((\widetilde{A}^{hh,v}_{m,k,j})_n{\bf e}_i),$$
$$K^v_{i,n,m,k,j}=-{\cal E}((\widetilde{d}^{v,v}_{m,k,j})_n{\bf e}_i),\quad
K^h_{i,n,m,k,j}=-{\cal E}((\widetilde{d}^{h,v}_{m,k,j})_n{\bf e}_i),$$
$$K^c_{i,n,m,k,j,p}=-{\cal E}((\widetilde{d}^{c,v}_{m,k,j,p})_n{\bf e}_i),$$
$$M^{cv}_{i,m}={\cal B}(\widetilde{\bf Y}^{v,\cdot}_{i,m},{\bf S}^\cdot),\quad
M^{vv}_{i,m,j}={\cal B}(\widetilde{\bf Y}^{v,\cdot}_{i,m},\widetilde{\bf S}^{v,\cdot}_j),\quad
M^{hv}_{i,m,j}={\cal B}(\widetilde{\bf Y}^{v,\cdot}_{i,m},\widetilde{\bf S}^{h,\cdot}_j),$$
$$M^{ch}_{i,m}={\cal B}(\widetilde{\bf Y}^{h,\cdot}_{i,m},{\bf S}^\cdot),\quad
M^{vh}_{i,m,j}={\cal B}(\widetilde{\bf Y}^{h,\cdot}_{i,m},\widetilde{\bf S}^{v,\cdot}_j),\quad
M^{hh}_{i,m,j}={\cal B}(\widetilde{\bf Y}^{h,\cdot}_{i,m},\widetilde{\bf S}^{h,\cdot}_j),$$
$$M^{cc}_{i,m,k}={\cal B}(\widetilde{\bf Y}^{c,\cdot}_{i,m,k},{\bf S}^\cdot),\quad
M^{vc}_{i,m,k,j}={\cal B}(\widetilde{\bf Y}^{c,\cdot}_{i,m,k},\widetilde{\bf S}^{v,\cdot}_j),\quad
M^{hc}_{i,m,k,j}={\cal B}(\widetilde{\bf Y}^{c,\cdot}_{i,m,k},\widetilde{\bf S}^{h,\cdot}_j).$$
All the coefficients $K,M$ and $N$ vanish, if the CHM state
${\bf V}_0,{\bf H}_0,\Theta_0$ possesses a symmetry without a time shift
($\widetilde{T}=0$).

\mi{\bf Appendix G. Evaluation of coefficients of the eddy correction terms}

\mi It suffices to solve 28 auxiliary problems (4, 8, 10 and 6 of types I--IV,
respectively) to evaluate the coefficients ${\bf D}^{\cdot,\cdot},\ {\bf d}^{\cdot,\cdot}$
and ${\bf A}^{\cdot\cdot,\cdot}$ in the mean-field equations \rf{eq59} and \rf{eq62}
for the weakly nonlinear stability problem
(or these coefficients in \rf{eq71}, \rf{eq73}, \rf{eq83} and \rf{eq86}~). However, it is
possible to exploit the fact that solutions to the auxiliary problems
(${\bf G}^{\cdot,v}_{m,k},\ {\bf G}^{\cdot,h}_{m,h},\ G^{\cdot,\theta}_{m,h})$,
(${\bf Q}^{\cdot\cdot,v}_{m,k},\ {\bf Q}^{\cdot\cdot,h}_{m,k},\ Q^{\cdot\cdot,\theta}_{m,k}$)
and (${\bf Y}^{\cdot,v}_{m,k},\ {\bf Y}^{\cdot,h}_{m,h},\ Y^{\cdot,\theta}_{m,h})$
enter the averages \rf{eq63p2} and (\theeddyadh) only in scalar products
with the vector field $({\bf H\times e}_3,-{\bf V\times e}_3,0)$, and in (\theeddyvisc)
and (\theeddyadv) with ${\bf W}_{n,j}\equiv(-V^j{\bf e}_n-V^n{\bf e}_j,
H^j{\bf e}_n+H^n{\bf e}_j,0)$ (4 vector fields for \hbox{$n,j=1,2$}).
We refer to the scalar product $\lad{\bgamma\cdot\bf W}\rad$ of 7-dimensional
vector fields $\bgamma=(\bgamma^v,\bgamma^h,\gamma^\theta)$ and
${\bf W}=({\bf W}^v,{\bf W}^h,W^\theta)$ in the layer.

Since in \rf{eq59} gradients, on which the curl acts, can be neglected,
it suffices to evaluate in (\theeddyvisc) and (\theeddyadv) scalar
products with ${\bf W}_{1,2},\ {\bf W}_{2,1}$ and ${\bf W}_{1,1}-{\bf W}_{2,2}$.
The number of auxiliary problems to be solved for evaluation of coefficients
in \rf{eq59} and \rf{eq62} is reduced to 8, if {\it auxiliary problems for the adjoint
operator} are considered, following Zheligovsky (2005, 2006a,b). They have
the same numerical complexity as the auxiliary problems considered above.
Then evaluation of coefficients ${\bf D}^{c,\cdot}$, ${\bf d}^{c,\cdot}$,
$\Alfa^{\cdot,\cdot}$, ${\bf A}^\cdot$ and ${\bf A}^{c\cdot,\cdot}$
(see (\theAlfav), (\theAlfah), (\thepitchcov), (\thepitchcoh)~) in new terms
in equations \rf{eq71}, \rf{eq73}, \rf{eq83} and \rf{eq86} also does not
require solving auxiliary problems of types V and II$'$--V$'$: it is only
necessary to compute the mean-free eigenvector ${\bf S}^\cdot\in\ker\cal M$.

Let
$${\bf W}^v=\la{\bf W}^v\ra_h+\la{\bf W}^v\ra_h+\widehat{\bf W}^v+\nabla W^v,\quad
{\bf W}^h=\la{\bf W}^h\ra_h+\widehat{\bf W}^h+\nabla W^h,$$
$$\bgamma^v=\la\bgamma^v\ra_v+\widehat{\bgamma}^v+\nabla\gamma^v,\quad
\bgamma^h=\la\bgamma^h\ra_h+\widehat{\bgamma}^h+\nabla\gamma^h\eqn{G1}$$
(in this appendix all differential operators are in fast variables) be
decompositions of three-dimensional vector components of $\bf W$ and $\bgamma$
into spatial mean, solenoidal and potential parts. Suppose $\widehat{\bgamma}$
and $\bf Z$ satisfy the equations
$${\cal L}^v(\widehat{\bgamma}^v,\widehat{\bgamma}^h,\widehat{\gamma}^\theta,\widehat{\gamma}^p)=\widehat{\bf F}^v,\quad
{\cal L}^h(\widehat{\bgamma}^v,\widehat{\bgamma}^h)=\widehat{\bf F}^h,\quad
{\cal L}^\theta(\widehat{\bgamma}^v,\widehat{\gamma}^\theta)=\widehat{F}^\theta,\eqn{G2}$$
$$\nabla\cdot\widehat{\bgamma}^v=\nabla\cdot\widehat{\bgamma}^h=0,\quad
\la\widehat{\bgamma}^v\ra_h=\la\widehat{\bgamma}^h\ra_h=0;$$
$$({\cal L}^\dagger)^v({\bf Z}^v,{\bf Z}^h,Z^\theta)=\widehat{\bf W}^v,\quad
({\cal L}^\dagger)^h({\bf Z}^v,{\bf Z}^h)=\widehat{\bf W}^h,\quad
({\cal L}^\dagger)^\theta({\bf Z}^v,Z^\theta)=0,\eqn{G3}$$
$$\nabla\cdot{\bf Z}^v=\nabla\cdot{\bf Z}^h=0,\quad\la{\bf Z}^v\ra_h=\la{\bf Z}^h\ra_h=0\eqn{G4}$$
and the boundary conditions for vector fields in the domain of the adjoint
operator (in our case coinciding with the boundary conditions for the
domain of $\cal L$, (\theBCs) and \rf{eq4}~). Here ${\cal L}^\dagger$
is the adjoint operator to the restriction of $\widetilde{\cal L}$
(defined in section 4) onto the subspace of its domain, where
horizontal magnetic components have zero spatial means:
$${\cal L}^\dagger({\bf v,h},\theta)=((\widetilde{\cal L}^*)^v({\bf v,h},\theta),
\lb(\widetilde{\cal L}^*)^h({\bf v,h},\theta)\rb_h,(\widetilde{\cal L}^*)^\theta({\bf v,h},\theta)).$$
Then
$$\lad\bgamma^v\cdot{\bf W}^v+\bgamma^h\cdot{\bf W}^h\rad\eqn{G5}$$
$$=\lad\la{\bf W}^v\ra_h\cdot\la\bgamma^v\ra_h+\la{\bf W}^h\ra_h\cdot\la\bgamma^h\ra_h
+\nabla W^v\cdot\nabla\gamma^v+\nabla W^h\cdot\nabla\gamma^h+
\widehat{\bf W}^v\cdot\widehat{\bgamma}^v+\widehat{\bf W}^h\cdot\widehat{\bgamma}^h\rad$$
$$=\lad\la{\bf W}^v\ra_h\cdot\la\bgamma^v\ra_h+\la{\bf W}^h\ra_h\cdot\la\bgamma^h\ra_h
-W^v\nabla\cdot\bgamma^v-W^h\nabla\cdot\bgamma^h$$
$$+(\widehat{\bgamma}^v,\widehat{\bgamma}^h,\widehat{\gamma}^\theta)
\cdot\left(({\cal L}^\dagger)^v({\bf Z}^v,{\bf Z}^h,Z^\theta),
({\cal L}^\dagger)^h({\bf Z}^v,{\bf Z}^h),({\cal L}^\dagger)^\theta({\bf Z}^v,Z^\theta)\right)\rad$$
$$=\lad\la{\bf W}^v\ra_h\cdot\la\bgamma^v\ra_h+\la{\bf W}^h\ra_h\cdot\la\bgamma^h\ra_h
-W^v\nabla\cdot\bgamma^v-W^h\nabla\cdot\bgamma^h+
\widehat{\bf F}^v\cdot{\bf Z}^v+\widehat{\bf F}^h\cdot{\bf Z}^h+\widehat{F}^\theta Z^\theta\rad.$$

Thus, for evaluation of the eddy coefficients it is enough to solve the
auxiliary problems for the adjoint operator (G3)--(G4) and to modify the
statements of auxiliary problems in order to match the form (G2) by making
substitutions (G1) and considering vector potentials of the vorticity equations.
Only solutions to auxiliary problems of types II and II$'$ are not solenoidal
and have non-vanishing spatial means of horizontal magnetic components; their
divergences are determined by \rf{eq44sol}, \rf{eq45sol} and \rf{eq44p2x}
and the spatial means by \rf{EQbegin0}; for the problem
(\theauxIIc) the spatial mean can be
found applying \rf{solvealpha}. Equations in the form of vector potentials
of the vorticity equations for problems of type III are presented
by \xrf{EQ26p1p}{EQ28p1p}. They can be also easily obtained for problems of
all other types, except for types II and II$'$; for these equations,
``uncurling'' can be performed numerically in the Fourier series representation.

If the perturbed CHM state ${\bf V,H},\Theta$ is steady, then a steady
solution to (G3)--(G4) is sought. Numerical solution of the evolutionary
problem (G3)--(G4) for time-dependent data can be problematic, since
the operator ${\cal L}^\dagger$ ceases to be parabolic. However, it becomes
parabolic, if time is reversed. Thus, it is possible to overcome this
difficulty, setting initial conditions for ${\bf Z}^{\cdot,\cdot}$ at
$t=\hat{t}>0$, and solving (G3)--(G4) for decreasing $t$ to $t=0$.
This transformation suffices, if the CHM state ${\bf V,H},\Theta$ is periodic
or quasiperiodic, and a periodic or quasiperiodic, respectively, solution
is sought. (Alternatively, such a solution can be found numerically
in the form of Fourier series in time.) If time dependence is more complex,
the spatio-temporal averaging of the scalar products involving
${\bf Z}^{\cdot,\cdot}$ in (G5) can be implemented as
$$\lad{\bf F\cdot Z}\rad=\lim_{\hat{t}\to\infty}
\lim_{\ell\to\infty}{1\over\hat{t}L\ell^2}\int_0^{\hat{t}}\int_{-L/2}^{L/2}
\int_{-\ell/2}^{\ell/2}\int_{-\ell/2}^{\ell/2}
{\bf F}({\bf x},t)\cdot{\bf Z}(\hat{t};{\bf x},t)\,dx_1\,dx_2\,dx_3\,dt$$
(if this limit exists), where ${\bf Z}(\hat{t};{\bf x},t)$ denotes the solution,
obtained with the time reversal.

\pagebreak
\mi{\bf References}

\mi Anufriev, A.P. and Braginsky, S.I. Influence of irregularities of the boundary
of the Earth's core on the velocity of the liquid and on the magnetic field.
{\it Geomagn. Aeron.}, 1975, {\bf 15}, 754--757.

\mi Anufriev, A.P. and Braginsky, S.I. Influence of irregularities of the boundary
of the Earth's core on the velocity of the liquid and on the magnetic field. II.
{\it Geomagn. Aeron.}, 1977a, {\bf 17}, 78--82.

\mi Anufriev, A.P. and Braginsky, S.I. Effect of irregularities of the boundary of
the Earth's core on the speed of the fluid and on the magnetic field. III.
{\it Geomagn. Aeron.}, 1977b, {\bf 17}, 492--496.

\mi Bowin, C. Topography at the core-mantle boundary. {\it Geophys. Res. Lett.}
1986, {\bf 13}, 1513--1516.

\mi Braginsky, S.I. Self excitation of a magnetic field during the
motion of a highly conducting fluid, {\it JETP}, 1964a, {\bf 47}, 1084--1098
(Engl. transl.: {\it Sov. Phys. JETP}, 1965, {\bf 20}, 726--735.)

\mi Braginsky, S.I. Theory of the hydromagnetic dynamo, {\it JETP}, 1964b,
{\bf 47}, 2178--2193. (Engl. transl.: {\it Sov. Phys. JETP}, 1965, {\bf 20},
1462--1471.)

\mi Braginsky, S.I. Kinematic models of the Earth's hydromagnetic dynamo,
{\it Geomagn. Aeron.}, 1964c, {\bf 4} (4), 732--747. (Engl. transl.:
{\it Geomagn. Aeron.}, 1964, {\bf 4}, 572--583.)

\mi Braginsky, S.I. Magnetohydrodynamics of the Earth's core,
{\it Geomagn. Aeron.}, 1964d, {\bf 4} (5), 898--916. (Engl. transl.:
{\it Geomagn. Aeron.}, 1964, {\bf 4}, 698--712.)

\mi Braginsky, S.I. Magnetic waves in the Earth's core,
{\it Geomagn. Aeron.}, 1967, {\bf 7} (6), 1050--1060. (Engl. transl.:
{\it Geomagn. Aeron.}, 1967, {\bf 7}, 851--859.)

\mi Braginsky, S.I. An almost axially symmetric model of the hydromagnetic
dynamo of the Earth, I, {\it Geomagn. Aeron.}, 1975, {\bf 15} (1), 149--156.

\mi Baptista, M., Gama, S.M.A. and Zheligovsky, V.
Eddy diffusivity in convective hydromagnetic systems.
{\it Eur. Phys. J. B}, 2007, {\bf 60}, 337-352\newline
[http://xxx.lanl.gov/abs/nlin.CD/0511020].

\mi Bassom, A.P. and Zhang, K. Strongly nonlinear convection cells in a rapidly
rotating fluid layer. {\it Geophys. Astrophys. Fluid Dyn.}, 1994,
{\bf 76}, 223--238.

\mi Bensoussan, A., Lions, J.-L. and Papanicolaou, G. {\it Asymptotic
analysis for periodic structures}, 1978. Amsterdam, North Holland, 700 pp.

\mi Biferale, L., Crisanti, A., Vergassola, M. and Vulpiani, A. Eddy viscosity
in scalar transport, {\it Phys. Fluids}, 1995, {\bf 7} (11), 2725--2734.

\mi Bodenschatz, E., Pesch, W. and Ahlers, G. Recent developments in
Rayleigh-B\'enard convection. {\it Ann. Rev. Fluid Mech.}, 2000, {\bf 32},
709--778.

\pagebreak
\mi Cioranescu, D. and Donato, P. {\it An introduction to homogenization},
1999. Oxford Univ. Press, 262 pp.

\mi Cross, M.C. and Newell, A.C. Convection patterns in large aspect ratio
systems, {\it Physica D}, 1984, {\bf 10}, 299--328.

\mi Dubrulle, B. and Frisch, U. Eddy viscosity of parity-invariant flow,
{\it Phys. Rev. A}, 1991, {\bf 43}, 5355--5364.

\mi E, W. and Shu, C.-W. Effective equations and the inverse cascade theory
for Kolmogorov flows, {\it Phys. Fluids A}, 1993, {\bf 5}, 998--1010.

\mi Frisch, U. {\it Turbulence, The Legacy of A.N. Kolmogorov}, 1995.
Cambridge Univ. Press, NY.

\mi Frisch, U., Legras, B. and Villone, B. Large-scale Kolmogorov flow on the
beta-plane and resonant wave interactions, {\it Physica D}, 1996, {\bf 94},
36--56.

\mi Gama, S. and Chaves, M. Time evolution of the eddy viscosity in two-dimensional
Navier-Stokes flow, {\it Phys. Rev. Lett.}, 2000, {\bf 61}, 2118--2120.

\mi Gama, S., Vergassola, M. and Frisch, U. Negative eddy viscosity in
isotropically forced two-dimensional flow: linear and nonlinear dynamics,
{\it J. Fluid Mech.}, 1994, {\bf 260}, 95--126.

\mi Guckenheimer, J. and Holmes, P. Nonlinear oscillations, dynamical systems,
and bifurcations of vector fields, 1990. Springer-Verlag, NY, 453 pp.

\mi Jikov V.V., Kozlov, S.M. and Oleinik, O.A. {\it Homogenization of differential
operators and integral functionals}, 1994. Springer-Verlag, Berlin, 570 pp.

\mi Julien, K., Knobloch, E. Fully nonlinear oscillatory convection
in a rotating layer, {\it Phys. Fluids}, 1997, {\bf 9}, 1906--1913.

\mi Julien, K., Knobloch, E. Strongly nonlinear convection cells in a rapidly
rotating layer: the tilted f-plane, {\it J. Fluid Mech.}, 1998, {\bf 360}, 141--178.

\mi Julien, K., Knobloch, E. and Werne, J. A new class of equations for
rotationally constrained flows, {\it Theoret. Comput. Fluid Dynamics},
1998, {\bf 11}, 251--261.

\mi Julien, K., Knobloch, E., Fully nonlinear three-dimensional convection
in a rapidly rotating layer, {\it Phys. Fluids}, 1999, {\bf 11} (6), 1469--1483.

\mi Julien, K., Knobloch, E. and Tobias, S.M.
Strongly nonlinear magnetoconvection in three dimensions,
{\it Physica D}, 1999, {\bf 128}, 105--129.

\mi Julien, K., Knobloch, E. and Tobias, S.M.
Strongly nonlinear magnetoconvection in the presence of oblique fields,
{\it J. Fluid Mech.}, 2000, {\bf 410}, 285--322.

\mi Julien, K., Knobloch, E. and Tobias, S.M.
Highly supercritical convection in strong magnetic fields.
In {\it Advances in nonlinear dynamos},
Eds Ferriz-Mas, A. and N\'u\~nez, M., 2003, 195--223. Taylor and Francis, London.

\pagebreak
\mi Lanotte, A., Noullez, A., Vergassola, M. and Wirth, A.
Large-scale dynamo by negative magnetic eddy diffusivities,
{\it Geophys.~Astrophys.~Fluid Dyn.}, 1999, {\bf 91}, 131--146.

\mi Liusternik, L.A. and Sobolev, V.J. {\it Elements of functional analysis},
1961. Frederick Ungar Publ. Co., NY.

\mi Majda, A.J. and Kramer, P.R. Simplified models for turbulent diffusion:
theory, numerical modelling, and physical phenomena. {\it Phys. Rep.}, 1999,
{\bf 314}, 237-574.

\mi Matthews, P.C. Asymptotic solutions for nonlinear magnetoconvection.
{\it J. Fluid Mech.}, {\bf 387}, 397--409, 1999.

\mi Merrill, R.T., McEllhiny, M.W. and McFadden, Ph.L. {\it The magnetic field
of the Earth. Paleomagnetism, the core and the deep mantle}, 1996.
Academic Press, San Diego, 527 pp.

\mi Murakami, Y., Murakami, M. and Gotoh, K. Three-dimensional negative eddy
viscosity effect on the onset of instability in some planar flows.
{\it Phys. Rev. E}, 1995, {\bf 51} (5), 5128--5131.

\mi Nepomnyashchy, A.A. On the stability of the secondary flow of a viscous
fluid in an infinite domain, {\it Appl. Math. Mech.}, 1976, {\bf 40}, 886--891.

\mi Newell A.C. Two-dimensional convection patterns in large aspect
ratio systems. In Fujita, H. (Ed.), {\it Nonlinear partial differential
equations in applied science}, 1983. North-Holland, Amsterdam, 202--231.

\mi Newell, A.C., Passot, T. and Lega, J. Order parameter equations for
patterns, {\it Ann. Rev. Fluid Mech.}, 1993, {\bf 25}, 399--453.

\mi Newell, A.C., Passot, T., Bowman, C., Ercolani, N. and Indik, R.
Defects are weak and self-dual solutions of the Cross-Newell phase diffusion
equation for natural patterns, {\it Physica D}, 1996, {\bf 97}, 185--205.

\mi Newell, A.C., Passot, T. and Souli, M. Convection at finite Rayleigh numbers
in large-aspect-ratio containers {\it Phys. Rev. Lett.}, 1990a, {\bf 64} (20),
2378--2381.

\mi Newell, A.C., Passot, T. and Souli, M. The phase diffusion and mean drift
equations for convection at finite Rayleigh numbers in large containers,
{\it J. Fluid Mech.}, 1990b, {\bf 220}, 187--552.

\mi Novikov, A. Eddy viscosity of cellular flows by upscaling.
{\it J. Comp. Phys.}, 2004, {\bf 195}, 341--354.

\mi Novikov, A. and Papanicolaou, G. Eddy viscosity of cellular flows.
{\it J. Fluid Mech.}, 2001, {\bf 446}, 173--198.

\mi Oleinik, O.A., Shamaev, A.S. and Yosifian, G.A. {\it Mathematical problems
in elasticity and homogenization}, 1992. Elsevier Science Publishers,
Amsterdam, 398 pp.

\mi Podvigina, O.M. Magnetic field generation by convective flows in a plane
layer, {\it Eur. Phys. J. B}, 2006, {\bf 50}, 639--652.

\pagebreak
\mi Polyanin, A.D. A handbook of linear equations of mathematical physics, 2001.
Fizmatlit, Moscow, 576 pp.

\mi Ponty, Y., Passot, T. and Sulem, P.L. Pattern dynamics in rotating convection
at finite Prandtl number, {\it Phys. Rev. E}, 1997, {\bf 56} (4), 4162--4178.

\mi Ponty, Y., Gilbert, A.D. and Soward A.M. Kinematic dynamo action
in flows driven by shear and convection, {\it J. Fluid Mech.}, 2001, {\bf 435},
261--287.

\mi Ponty, Y., Gilbert, A.D. and Soward, A.M. Dynamo action due to Ekman layer
instability. In Chossat, P., Armbruster, D. and Oprea, I. (Eds.), {\it Dynamo
and dynamics, a mathematical challenge}, 2001b. Kluwer, Boston, 75--82.

\mi Ponty, Y., Gilbert, A.D. and Soward, A.M. The onset of thermal convection in
Ekman--Couette shear flow with oblique rotation, {\it J. Fluid Mech.}, 2003,
{\bf 487}, 91--123.

\mi Rotvig, J. and Jones, C.A. Rotating convection driven dynamos at low Ekman
number, {\it Phys. Rev. E}, 2002, {\bf 66}, 056308, 15 pp.\newline
[http://link.aps.org/abstract/PRE/v66/e056308].

\mi She, Z.S. Metastability and vortex pairing in the Kolmogorov flow,
{\it Phys. Lett. A}, 1987, {\bf 124}, 161--164.

\mi Sivashinsky, G.I. and Yakhot, V. Negative viscosity effect in large-scale
flows, {\it Phys. Fluids}, 1985, {\bf 28}, 1040--1042.

\mi Sivashinsky, G.I. and Frenkel A.L. On negative eddy viscosity under conditions
of isotropy, {\it Phys. Fluids A}, 1992, {\bf 4} (8), 1608--1610.

\mi Soward, A.M. A kinematic theory of large magnetic Reynolds number
dynamos, {\it Phil. Trans. Roy. Soc. A}, 1972, {\bf 272}, 431--462.

\mi Soward, A.M. A convection driven dynamo I. The weak field case,
{\it Phil. Trans. Roy. Soc. A}, 1974, {\bf 275}, 611--651.

\mi Starr, V.P. {\it Physics of negative viscosity phenomena}, 1968.
McGraw-Hill, NY.

\mi Vergassola, M. and Avellaneda, M. Scalar transport in compressible
flow, {\it Physica D}, 1997, {\bf 106}, 148--166.

\mi Wirth, A. Complex eddy-viscosity: a three-dimensional effect,
{\it Physica D}, 1994, {\bf 76}, 312--317.

\mi Wirth, A., Gama, S. and Frisch, U. Eddy viscosity of
three-dimensional flow, {\it J. Fluid Mech.}, 1995, {\bf 288}, 249--264.

\mi Zheligovsky, V.A. On the linear stability of spatially periodic steady
magnetohydrodynamic systems with respect to long-period perturbations,
{\it Izvestiya, Physics of the Solid Earth}, 2003, {\bf 39} (5), 409--418
[http://arxiv.org/abs/nlin/0512076 -- in Russian].

\pagebreak
\mi Zheligovsky, V.A. Convective plan-form two-scale dynamos in a plane layer,
{\it Geophys. Astrophys. Fluid Dyn.}, 2005, {\bf 99}, 151--175
[http://arxiv.org/abs/physics/0405045].

\mi Zheligovsky, V.A. Weakly nonlinear stability of centrally symmetric
magnetohydrodynamic systems to perturbations involving large scales,
{\it Izvestiya, Physics of the Solid Earth}, 2006a, {\bf 42} (3), 244-253
[http://arxiv.org/abs/nlin.CD/0601012 -- in Russian].

\mi Zheligovsky V.A. Weakly nonlinear stability to large-scale
perturbations in convective magnetohydrodynamic systems without
the $\alpha$-effect. {\it Izvestiya, Physics of the Solid Earth}, 2006b,
{\bf 42} (12), 1051-1068 [http://arxiv.org/abs/nlin.CD/0601013 -- in Russian].

\mi Zheligovsky, V.A. and Podvigina, O.M. Generation of multiscale magnetic field
by parity-invariant time-periodic flows, {\it Geophys. Astrophys. Fluid Dyn.},
2003, {\bf 97}, 225--248 [http://xxx.lanl.gov/abs/physics/0207112].

\mi Zheligovsky, V.A., Podvigina, O.M. and Frisch, U.
Dynamo effect in parity-invariant flow with large and moderate
separation of scales, {\it Geophys. Astrophys. Fluid Dyn.}, 2001, {\bf 95},
227--268 {[http://xxx.lanl.gov/abs/nlin.CD/0012005]}.
\end{document}